%% file: main.tex
\documentclass[
12pt,		
openright,	
twoside,  
a4paper,			
chapter=TITLE,		
english,			
french,				
spanish,			
brazil				
]{USPSC}
\usepackage[T1]{fontenc}		
\usepackage[utf8]{inputenc}		
\usepackage{lmodern}			
\usepackage{lastpage}			
\usepackage{indentfirst}		
\usepackage{color}				
\usepackage{graphicx}			
\usepackage{float} 				
\usepackage{chemfig,chemmacros} 
\usepackage{tikz}				
\usetikzlibrary{positioning}
\usepackage{pdfpages}
\usepackage{makeidx}            
\usepackage{hyphenat}          
\usepackage[absolute]{textpos} 
\usepackage{eso-pic}           
\usepackage{makebox}           

\newcommand{\doi}[1]{DOI: \href{https://doi.org/#1}{#1}}
\usepackage{caption}
\usepackage{subcaption}
\usepackage{amsfonts}
\usepackage{mathrsfs}

\usepackage{cite}              
\usepackage[num, abnt-emphasize=bf, abnt-thesis-year=both, abnt-repeated-author-omit=no, abnt-last-names=abnt, abnt-etal-cite, abnt-etal-list=3, abnt-etal-text=it, abnt-and-type=e, abnt-doi=doi, abnt-url-package=none, abnt-verbatim-entry=no]{abntex2cite} 
\usepackage{alphalph}

\renewcommand{\footnotesize}{\small} 

\usepackage{lipsum}				

\usepackage{multicol}	
\usepackage{multirow}	
\usepackage{longtable}	
\usepackage{threeparttablex}    
\usepackage{array}

\usepackage{ABNT6023-2018}

\siglaunidade{IFSC}
\programa{DFBe}
\definecolor{blue}{RGB}{41,5,195}

\makeatletter
\hypersetup{
	pdftitle={\@title}, 
	pdfauthor={\@author},
	pdfsubject={\imprimirpreambulo},
	pdfcreator={LaTeX with abnTeX2},
	pdfkeywords={abnt}{latex}{abntex}{USPSC}{trabalho acadêmico}, 
	colorlinks=true,       		
	linkcolor=blue,            	
	citecolor=blue,        		
	filecolor=magenta,      		
	urlcolor=blue,
	bookmarksdepth=4	
}
\makeatother

\makeindex

\begin{document}

\selectlanguage{english}

\frenchspacing 

\renewcommand{\ABNTEXchapterfontsize}{\fontsize{12}{12}\bfseries}
\renewcommand{\ABNTEXsectionfontsize}{\fontsize{12}{12}\bfseries}
\renewcommand{\ABNTEXsubsectionfontsize}{\fontsize{12}{12}\normalfont}
\renewcommand{\ABNTEXsubsubsectionfontsize}{\fontsize{12}{12}\normalfont}
\renewcommand{\ABNTEXsubsubsubsectionfontsize}{\fontsize{12}{12}\normalfont}

\imprimircapa
\imprimirfolhaderosto*

\includepdf{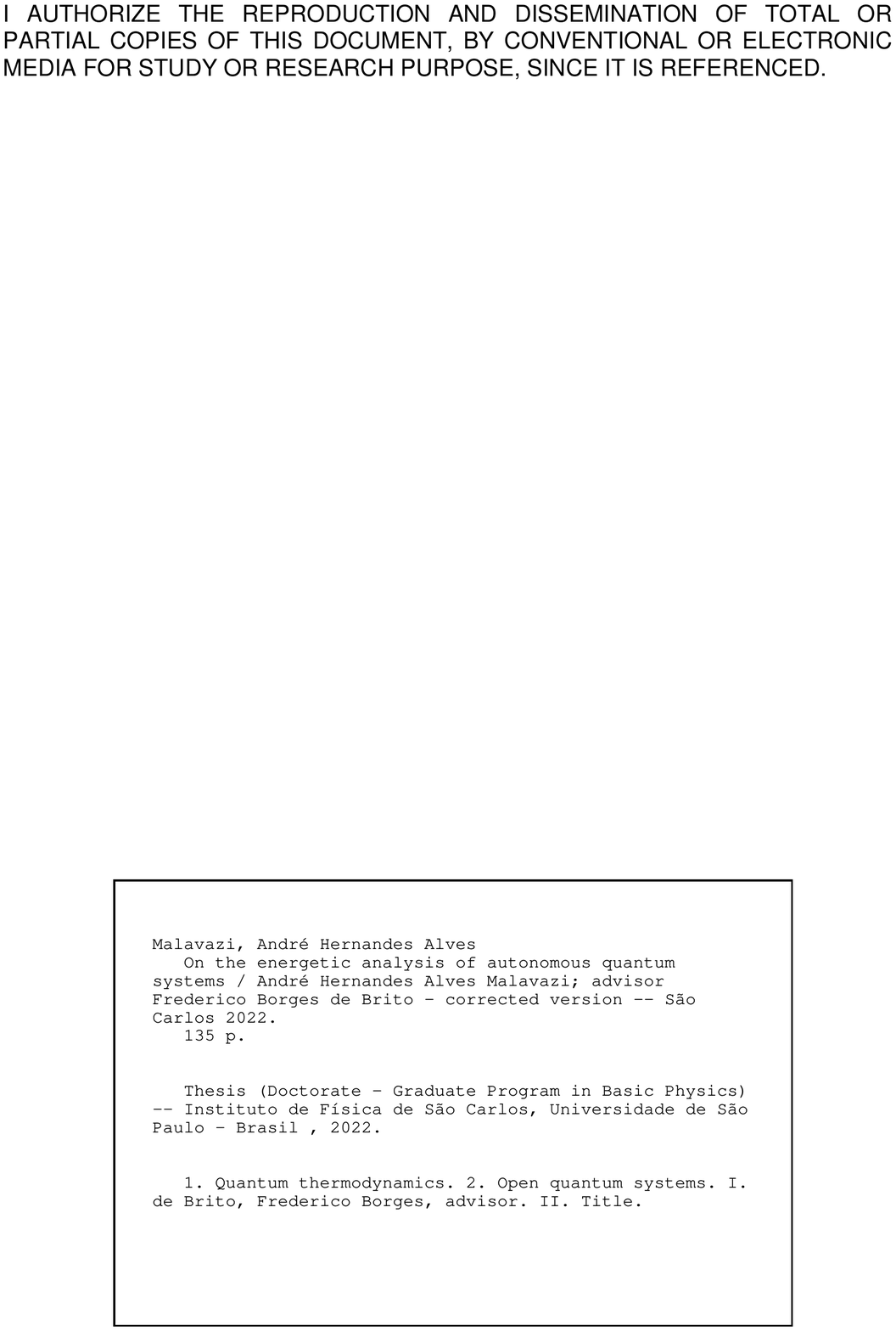}










\include{Outros/USPSC-Dedicatoria}

\include{Chapters/Acknowledgements}

\include{Outros/Epigrafe}

\include{Chapters/Abstract}
%
\include{Chapters/Resumo}

\pdfbookmark[0]{\listfigurename}{lof}
\listoffigures*
\cleardoublepage




\pdfbookmark[0]{\contentsname}{toc}
\tableofcontents*
\cleardoublepage
\textual

\include{Chapters/Chapter1/Chapter1}
\include{Chapters/Chapter2/Chapter2}

\include{Chapters/Chapter3/Chapter3}

\include{Chapters/Chapter4/Chapter4}

\postextual

\bibliography{References}

%
%





\end{document}

%% file: Outros/USPSC-Dedicatoria.tex
\begin{dedicatoria}
   \vspace*{\fill}
   \centering
   \noindent
   \textit{ This thesis is dedicated to all people who work hard against misinformation \\ and to developing science in Brazil.\\ Most importantly, it is dedicated to all, direct and indirect, victims of this pandemic.} \vspace*{\fill}
\end{dedicatoria}

%% file: Chapters/Acknowledgements.tex
\begin{agradecimentos}

A countless number of people pass through our lives. Most of them are with us just for brief moments, while some become essential and irreplaceable constants. Yet, all of them represents a piece of our totality, in one way or another. I am the sum of several crucial parts, and for that, I would like to thank and acknowledge you all:

I am very grateful to my loving family: my parents, Lilian and Jefferson, my brother, Arthur, my grandmother, Beatriz, and my younger family members, Minnie (RIP), Bella and Mia. You represent my foundations and the most special constant behind me and my achievements. Thank you for your unconditional support and love throughout my life, especially during the last years. The path I chose is uncertain and tortuous, but I am very fortunate for having you by my side. In fact, there is no André without you.

I also would like to thank my advisor, Prof. Dr. Frederico Brito, for sharing his knowledge, providing support during the execution of this project and, most importantly, all the trust and patience. I have grown in many ways during this time, and I am sure I have become a better person and, without a doubt, a better scientist after his guidance.

Many great friends also joined me during this trajectory, both on and offline: "I am glad you are here with me, at the end of all things...". A special thanks to my "roommates" Gabriela and Lais for sharing much more than a simple office; to Cleverson and all the members of the "Jovens quânticos" for all the passionate scientific discussions; to Rodrigo and Pedro for existing and making life more bearable; to Dimitrios and Madeira for the several coffees, advice and funny moments; to Diogo, Jessica, Guilherme, Raul, Edmilson and Cesar for welcoming me when I moved to IFSC; to Adriane for all our talks and mutual support; to Letícia for all the help and assistance; to Paula, Fernando, Maria Luiza and Artur for sharing a story and being a second loving family for many years; to my hometown friends, José, Rodrigo and Ikeda for also being important constants during my life. Last but not least, I am grateful to Thalyta for all her patience, care and shared memories. Thank you for being supportive and staying by my side during these arduous times.

Finally, I would like to acknowledge all the staff from IFSC, especially Ricardo, for supporting me since day one. This study was financed in part by the Coordenação de Aperfeiçoamento de Pessoal de Nível Superior – Brasil (CAPES) – Finance Code 001, and I am very greatfull for the financial support.

\end{agradecimentos}

%% file: Outros/Epigrafe.tex
\begin{epigrafe}
    \vspace*{\fill}
	\begin{flushright}
		\textit{``The unexamined life is not worth living.''\\
		Socrates}
	\end{flushright}
\end{epigrafe}

%% file: Chapters/Abstract.tex
\begin{resumo}[Abstract]
 \begin{otherlanguage*}{english}
	\begin{flushleft} 
		\setlength{\absparsep}{0pt} 
 		\SingleSpacing  		\imprimirautorabr~~\textbf{\imprimirtitleabstract}.	\imprimirdata.  \pageref{LastPage}p. 
		\imprimirtipotrabalhoabs~-~\imprimirinstituicao, \imprimirlocal, 	\imprimirdata. 
 	\end{flushleft}
	\OnehalfSpacing 
   During the last decades, there have been many theoretical and experimental
advances both in the extension of thermodynamics to comprise microscopic
systems out-of-equilibrium and in the understanding of quantum mechanics.
Along with the state-of-the-art capability of controlling fragile
quantum systems in a wide variety of physical platforms, this context
has paved the way for the current strategic efforts to develop a thermodynamic
theory of quantum systems. In this sense, the research field coined
as quantum thermodynamics (QT) already plays a key role in the design
and development of future quantum-based technologies. More specifically,
QT aims both to apply the usual thermodynamic concepts and notions
to describe arbitrary non-equilibrium quantum systems and to understand
the emergence of classical thermodynamic behaviour from the underlying
fundamentally quantum dynamics. However, despite all current progress,
there is still no consolidated formalism for a general thermodynamic
description of fully autonomous quantum objects. Besides, the lack
of consensus on some central aspects, such as the definitions of quantum
counterparts of thermodynamic quantities, is particularly notorious. In this thesis, we focus on the energetic analysis within autonomous
quantum systems. To this aim, we propose a novel and general formalism
for a dynamic description of the energy exchanges between interacting
subsystems. From the Schmidt decomposition approach, we identify effective
Hamiltonians as the representative operators for characterizing the
local internal energies, whose expectation values satisfy the usual thermodynamic
notion of energy additivity. In contrast to the currently used methodologies,
such procedure treats the subsystems with equal footing and do not
rely on any sort of approximations and additional hypotheses, e.g.,
semi-classical description, weak-coupling regime, strict energy conservation
and Markovian dynamics. In short, our proposal contributes to the
development of QT by providing a new formalism that does not suffer
from the usual restrictive shortcomings and establishes a new and
exact route for defining other general thermodynamic quantities to
the quantum regime.

   \vspace{\onelineskip}
 
   \noindent 
   \textbf{Keywords}: Quantum thermodynamics. Open quantum systems. 
 \end{otherlanguage*}
\end{resumo}

%% file: Chapters/Resumo.tex
\setlength{\absparsep}{18pt} 
\begin{resumo}
	\begin{flushleft} 
			\setlength{\absparsep}{0pt} 
			\SingleSpacing 
			\imprimirautorabr~~\textbf{\imprimirtituloresumo}.	\imprimirdata. \pageref{LastPage}p. 
			\imprimirtipotrabalho~-~\imprimirinstituicao, \imprimirlocal, \imprimirdata. 
 	\end{flushleft}
\OnehalfSpacing 			
 Durante as últimas décadas, houve muitos avanços teóricos e experimentais
tanto na extensão da termodinâmica para abranger sistemas microscópicos
fora de equilíbrio quanto na compreensão da mecânica quântica. Somada
a capacidade de última geração de controlar sistemas quânticos frágeis
em uma ampla variedade de plataformas físicas, esse contexto abriu
caminho para os atuais esforços estratégicos para desenvolver uma
teoria termodinâmica de sistemas quânticos. Nesse sentido, o campo
de pesquisa cunhado como termodinâmica quântica (TQ) já desempenha
um papel fundamental no projeto e desenvolvimento de futuras tecnologias
baseadas em fenômenos quânticos. Mais especificamente, a TQ visa tanto
aplicar os conceitos e as noções termodinâmicas usuais para descrever
sistemas quânticos arbitrários fora do equilíbrio quanto entender
o surgimento do comportamento termodinâmico clássico a partir da dinâmica
fundamentalmente quântica subjacente. No entanto, apesar de todo o
progresso atual, ainda não existe um formalismo consolidado para uma
descrição termodinâmica geral de objetos quânticos totalmente autônomos.
Além disso, é particularmente notória a falta de consenso em relação
a alguns aspectos centrais, como as definições de análogos quânticos
de grandezas termodinâmicas. Nesta tese, focamos na análise energética
em sistemas quânticos autônomos. Para isso, propomos um novo formalismo
geral para uma descrição dinâmica das trocas energéticas entre subsistemas
interagentes. A partir da abordagem da decomposição de Schmidt, identificamos
Hamiltonianos efetivos como os operadores representativos para caracterização
das energias internas locais, cujos valores esperados satisfazem a
noção termodinâmica usual da aditividade de energia. Ao contrário
das metodologias atualmente utilizadas, tal procedimento trata os
subsistemas em pé de igualdade e não depende de nenhum tipo de aproximações
e hipóteses adicionais, por exemplo, descrição semiclássica, regime
de acoplamento fraco, conservação de energia estrita e dinâmica Markoviana.
Em suma, nossa proposta contribui para o desenvolvimento da TQ fornecendo
um novo formalismo que não sofre das restrições usuais e estabelece
uma nova e exata rota para definir outras grandezas termodinâmicas
gerais para o regime quântico.

 \textbf{Palavras-chave}: Termodinâmica quântica. Sistemas quânticos abertos.
\end{resumo}

%% file: Chapters/Chapter1/Chapter1.tex
\chapter{Introduction}
\label{}

Quantum thermodynamics is a rapidly growing and promising field of research, accumulating efforts from several distinct perspectives. Its flourishing is inextricably linked to recent technological advances and current independent theoretical progress in the fields of nonequilibrium thermodynamics and quantum mechanics. On the one hand, state-of-the-art experiments allow the precise control of small, and possibly quantum, systems in several platforms. On the other, yet the laws of thermodynamics have been successfully verified to microscopic scales when considering ensemble realizations of the experiment, and studies on quantum information elucidated the technological potential of harnessing genuine quantum phenomena. However, the task of developing a consistent and somehow meaningful thermodynamic theory of quantum systems proved to be non-trivial and far from being straightforward. In this sense, the most current approaches are restricted to specific scenarios of approximative descriptions and semi-classical treatments, i.e., it is commonly assumed externally driven systems (by means of classical agents), weak-coupling regime and Markovian dynamics. Along these lines, there is no unifying formalism for characterizing the thermodynamics \textit{within} general autonomous quantum systems, and many fundamental open questions remain unanswered. Remarkably, it is unclear how to define quantum counterparts of the most basic thermodynamic quantities. This context highlights the importance of further research, especially on conceptual aspects of the theory. The work in question falls within this scope\footnote{The main discussions and results introduced in this thesis were presented in \cite{Malavazi} after its defense.}.

This thesis is organized as follows: Chapter (\ref{part:Chapter-2--})
contextualizes the field of quantum thermodynamics. It initially introduces
the basic concepts and formalism of thermodynamics and quantum mechanics separately.
Then it provides a brief overview of quantum thermodynamics and discusses
its current status and open problems; Chapter (\ref{part:Chapter-3--})
contains the main sections of the thesis. It develops and proposes
a novel framework for analyzing the energetics within autonomous pure
bipartite quantum systems. This proposal is exact, symmetrical and
based on a well-established mathematical tool: the Schmidt decomposition procedure. More
specifically, it introduces time-dependent local effective Hamiltonians
interpreted as the representative operators for characterizing the
physical local internal energies. This identification allows the description
of the subsystems effective dynamics within general interacting subsystems, regardless
of their particular properties and coupling regimes, and also recovers
the usual thermodynamic property of energy additivity. Besides, it
briefly discusses the current approaches and difficulties for defining
general quantum versions of core thermodynamic quantities. After that,
the introduced formalism is extended for mixed states; Finally, the conclusions of this thesis are addressed in Chapter
(\ref{part:Chapter-4--}). It summarizes its
main messages and provides a perspective for future work considering
the general context of quantum thermodynamics.

%% file: Chapters/Chapter2/Chapter2.tex
\chapter{Contextualization}
\label{part:Chapter-2--}

This chapter aims to contextualize the field of research in which
this work is inserted, known as \textit{quantum thermodynamics}. To
this end, a brief detour is necessary to introduce its main ingredients,
namely \textit{thermodynamics} and \textit{quantum mechanics}. Of
course, this is not an extensive review of both areas. Instead, these
notes intend to provide an overview of the context and main motivations
for the development of a thermodynamic theory of quantum systems.

In this sense, Section (\ref{sec:Thermodynamics}) presents the basic
concepts of classical equilibrium thermodynamics and the recently
established research area of stochastic thermodynamics. Then, Section
(\ref{sec:Quantum-mechanics}) discusses the formalism of quantum
mechanics. Finally, Section (\ref{sec:Quantum-thermodynamics}) formally
introduces quantum thermodynamics and its current status.

\input{Chapters/Chapter2/Section1}
\input{Chapters/Chapter2/Section2}
\input{Chapters/Chapter2/Section3}

%% file: Chapters/Chapter2/Section1.tex
\section{Thermodynamics\label{sec:Thermodynamics}}

Thermodynamics is, undoubtedly, one of the most successful physical
theories in the history of science. It is no overstating to say that
the foundations of modern society were significantly influenced by
the technological innovations brought by it. Its range of predictions
and applicability is surprisingly far-reaching, especially if considering
its restrictions and the circumstances of its developments. In this
sense, all branches of modern science and engineering function accordingly
to the laws of thermodynamics. Nevertheless, even after centuries
of discussions, its results and statements continue to inspire and
instigate novel research and surprising connections.

Despite earlier developments, the theory of thermodynamics was born
during the 17th century and flourished as an influential branch of
physics during the 18th and 19th centuries\footnote{For a historical perspective on the development of thermodynamics,
see. \cite{Evangelista}}. Since its early days, thermodynamics was developed as a phenomenological
discipline and approached semi-empirically. It initially progressed
through scientists' efforts to understand better the behaviour of
gases and the \textquotedbl abhors\textquotedbl{} nature of the vacuum\footnote{\textquotedbl\textit{Natura abhorret vacuum}\textquotedbl}.
Then, it proceeded as a fundamental way to characterize different
aspects of energy exchange through work and heat, along with the notions
of irreversibility and entropy. As often happens in science, these
investigations were partially motivated by the technological advancements
of the time: the development of new instruments, such as vacuum pumps,
the thermometer and steam engines, provided the necessary means to
perform novel and better measurements and push even further the progress
of the theory. Surprisingly, this context also triggered the unprecedented
technological transformation that led western society toward the Industrial
Revolution and, eventually, the thriving of modern machinery. In fact,
the progress of thermodynamics was driven considerably by the attempts
to understand and develop efficient machines. In this sense, despite
the pragmatical and sober motivation, it is remarkable how these studies
guided scientists toward the establishment of general and fundamental
laws of nature.

Seeing from the privilege of modern science, the phenomenological
approach to thermodynamics could hardly have been different. The microscopic
behaviour of matter remained hidden from scientists' eyes for a long
time, and the tools of analytical mechanics, along with its mechanistic
view of nature, were being developed in parallel to thermodynamics.
In this sense, its history is also intrinsically connected to the
concomitant developments of chemistry and classical mechanics. Along
these lines, at that time, the empirical observation that mass\footnote{Strictly speaking, mass does not conserve due to the mass-energy equivalence.
However, in the present thermodynamic context, this difference is
negligible.}, momentum and energy were conserved quantities in isolated systems
was still under scrutiny and far from consensus. In fact, even these
individual notions - including \textquotedbl isolated systems\textquotedbl{}
- were not entirely established. Heat, for instance, was not directly
associated with the mechanical motions of particles or the \textquotedbl vis
viva\textquotedbl , the earlier version of kinetic energy, until
the efforts of Joule, when he established the relationship between
energy, work and heat. Instead, the orthodox explanation was given
by the so-called caloric theory. Interestingly, this was also the
context for the development of heat engines and the theoretical studies
concerning their performances, which set the stage for the works of
Sadi Carnot (1796 - 1832) - and later Rudolf Clausius (1822 - 1888)
- on what we know today as the second law of thermodynamics and the
concept of entropy.

After that, names like James Clerk Maxwell (1831 - 1879), Ludwig Boltzmann
(1844 - 1906) and Josiah Willard Gibbs (1839 - 1903) were responsible
for providing the microscopic and mechanical basis of phenomenological
equilibrium thermodynamics. Such progress is depicted by the kinetic
theory of gases and the following foundation of statistical mechanics.
Most importantly, these developments represent a crucial paradigm
shift concerning the understanding of the macroscopic nature, i.e.,
the introduction of probabilistic reasoning and thermodynamic ensembles
provided the necessary tools for - at least partially - bridging the
gap between the underlying microscopic motion and the observed macroscopic
behaviour. More recently, during the 20th century, physicists, such
as Lars Onsager (1903 - 1976) and Ilya Prigogine (1917 - 2003),
began to extrapolate thermodynamics to the uncharted territory of
non-equilibrium regimes. After all, most natural processes occur under
these conditions. Along these lines, the analysis of systems close
to equilibrium, known as the linear regime, provided the formalism
for investigating several physical phenomena and represented the starting
point of the characterization of non-equilibrium states.

As mentioned earlier, thermodynamics has still been an active field
of research. During the last decades, one can find great efforts to
extend the theory and its results to even broader scenarios, namely
microscopic and far-from-equilibrium settings. In such regimes, thermal
fluctuations play a critical role in the dynamics of the desired physical
system, and the usual thermodynamic quantities are updated to random
variables. In this sense, stochastic thermodynamics and its fluctuation
theorems have been providing seminal results concerning non-equilibrium
processes and irreversibility.

\input{Chapters/Chapter2/Subsection_1_1}
\input{Chapters/Chapter2/Subsection_1_2}

%% file: Chapters/Chapter2/Subsection_1_1.tex
\subsection{Classical equilibrium thermodynamics\label{subsec:Classical-equilibrium-thermodyna}}

Let us briefly present and review the classical equilibrium thermodynamics
and its main statements. Of course, there are several traditional
textbooks written on the subject. For a comprehensive discussion of
thermodynamics see. \cite{zemansky,Callen,termo,Solr-32024,Prigogine}

Classical thermodynamics restricts itself to a specific, though meaningful,
physical regime and scope. Essentially, it was developed to describe
the behaviour of large scale properties of classical \textit{macroscopic}
objects and their general relationships. Since it is a phenomenological
description, it does not require any microscopic consideration concerning
the structure of matter. Instead, it provides a macroscopic and fully
general characterization in terms of a few measurable gross variables
$\{X_{j}\}$, known as thermodynamic coordinates, such as volume ($V$),
pressure ($P$) and temperature ($T$). In this sense, once obtained
the relevant coordinates for describing a system, one has specified
its physical state and, along with it, has access to all thermodynamic
properties that are functions of it, referred to as state functions,
i.e., $F(\{X_{j}\})$. Such a procedure is only possible because classical
thermodynamics revolves around the concept of states in \textit{thermodynamic
equilibrium}. These states are characterized by their static, or time-invariant,
nature, such that all possible thermodynamic coordinates remain fixed\footnote{One can also refer to equilibrium states relative to specific macroscopic
coordinates, such as mechanical, chemical and thermal.}. In contrast, non-equilibrium states might not even have specified
values for them. It is an empirical fact that many physical systems
naturally (spontaneously) evolve in time from non-equilibrium states towards these
constant states. It is worth mentioning that everything that happens
during this dynamic process is outside the scope of classical equilibrium
thermodynamics.

Along these lines, it is also possible to define a succession of changes
connecting different equilibrium states, which establishes a \textit{thermodynamic
process}. If the initial and final states are the same, the whole
process is known as a thermodynamic cycle, which represents the building
blocks of any heat engine. Visually, thermodynamic processes and cycles
are usually depicted by trajectories in the space of equilibrium states
defined in terms of the thermodynamic coordinates $\{X_{j}\}$. Of
course, any change presupposes some degree of variation of a macroscopic
variable that, in general, would perturb the system and kick it out
of equilibrium until it reaches a novel one in terms of the new setting.
Nevertheless, as long as the changes are kept infinitesimal, the system
is weakly perturbed and maintained close to equilibrium. Such a convenient
condition characterizes a so-called \textit{quasi-static process}.
Notice that any ideal thermodynamic process would require an infinite
amount of time to be performed. However, in many realistic scenarios,
this theoretical abstraction represents a satisfactory approximation.
Interestingly, the rate of processes is not addressed by classical
equilibrium thermodynamics, which highlights that the theory is not
a truly dynamic one, at least not in the usual sense.

\subsubsection{The laws of thermodynamics}

The physical object under scrutiny, or the thermodynamics system,
is usually classified according to the ways it interacts with its
exterior, especially concerning the possibility of exchanging matter
or energy with its surroundings. In this sense, isolated systems,
as the name suggests, do not interact with other objects and, therefore,
exchange neither things. Closed systems only exchange energy, while
open systems exchange both energy and matter. Naturally, these interactions
are accompanied by thermodynamic processes, both from the system and
its surroundings. The possible state transformations are described,
and constrained, by a set of empirical and mathematical statements
known as the \textit{laws of thermodynamics}. Let us now briefly introduce
them.

\paragraph{Zeroth law of thermodynamics}

As mentioned earlier, thermodynamic systems eventually reach equilibrium
states. It implies that if two independent systems are allowed to
interact, the single entity constituted by both individuals will also
equilibrate. Thus, it is an experimental fact that interacting bodies
that are specified by distinct temperatures will reach thermal equilibrium.
Along these lines, if systems $A$ and $B$ are individually in thermal
equilibrium with system $C$, then $A$ and $B$ are also in equilibrium.
Such transitivity property is known as the \textit{zeroth law of thermodynamics}.

\paragraph{First law of thermodynamics}

Essentially, the first law of thermodynamics refers to the conservation
of energy within thermodynamic systems. However, instead of being
written in terms of the sum of all the kinetic and potential contributions
of its microscopic constituents, the internal energy relative to state
$k$, $U_{k}$, is a state function specified by the relevant macroscopic
variables $\{X_{j}^{k}\}$, i.e.,
\begin{equation}
U_{k}\equiv U(\{X_{j}^{k}\})+U_{0},
\end{equation}
where $U_{0}$ is the energy of a reference state. Notice that it
means that the internal energy change $U_{b}-U_{a}$ from a state
$a$ to $b$ is independent of the thermodynamic process, $\gamma$,
connecting them, such that
\begin{equation}
\int_{\gamma}\,dU=\int_{\gamma^{\prime}}\,dU=U_{b}-U_{a},
\end{equation}
where $dU$ is an exact differential, and
\begin{equation}
\oint\,dU=0\label{eq: interal cyclic energy}
\end{equation}
for any cyclic transformation.

It is clear that isolated system, by construction, maintains its internal
energy fixed since any energetic change requires the system to be
submitted to a thermodynamic process. Thus, let us now consider closed
systems. Thermodynamics, and the first law specifically, not only
state energy conservation but also splits its possible changes into
two different \textquotedbl flavours\textquotedbl : heat, $Q$,
is the type of energy exchanged once systems of different temperatures
are interacting or, as commonly said, put into thermal contact; work,
$W$, in contrast, is the energy transferred (not stochastically) by the external change
of the thermodynamic coordinates. Along these lines, in the simplest
scenario consisting of two interacting bodies reaching thermal equilibrium,
the energetic exchange of the system under consideration is completely
due to heat, or
\begin{equation}
\Delta U=Q.
\end{equation}
However, for a thermally insulated system, one can perform work and
modify the system's state simply by externally controlling its macroscopic
variables, such that
\begin{equation}
\Delta U=W.
\end{equation}
This kind of procedure where no heat is involved, is characterized
as an adiabatic process. Nevertheless, considering general interactions,
both contributions might be present for arbitrary processes. Thus,
for an infinitesimal energy change, $dU$, the \textit{first law of
thermodynamics} is stated as
\begin{equation}
dU\equiv\delta Q+\delta W,
\end{equation}
where $\delta W$ and $\delta Q$ are the infinitesimal work performed
on the system and the heat transferred into it, respectively. In contrast
with internal energy, both individual quantities are not state functions.
Instead, they are intrinsic to thermodynamic processes. Such path
dependency is mathematically represented by the inexact differentials
$\delta W$ and $\delta Q$. Thus, given two possible trajectories,
$\gamma$ and $\gamma^{\prime}$, connecting different states $a$
and $b$, we have
\begin{align}
W_{\gamma} & \equiv\int_{\gamma}\,\delta W\neq\int_{\gamma^{\prime}}\,\delta W\equiv W_{\gamma^{\prime}},\\
Q_{\gamma} & \equiv\int_{\gamma}\,\delta Q\neq\int_{\gamma^{\prime}}\,\delta Q\equiv Q_{\gamma^{\prime}},
\end{align}
even though
\begin{equation}
U_{b}-U_{a}=W_{\gamma}+Q_{\gamma}=W_{\gamma^{\prime}}+Q_{\gamma^{\prime}}.
\end{equation}
It is worth mentioning that, for open systems, one just must consider
an additional term $dU_{matter}$ for the energy flow due to matter
exchange.

\paragraph{Second law of thermodynamics}

Interestingly, there are several distinct, though ultimately equivalent,
ways for formulating the second law of thermodynamics. In this sense,
according to Lord Kelvin (William Thomson, 1824 - 1907) \cite{zemansky}:
\begin{quotation}
\textquotedbl\textit{It is impossible by means of inanimate material
agency to derive mechanical effect from any portion of matter by cooling
it below the temperature of the coldest of the surrounding objects.}\textquotedbl{}
\end{quotation}
On the same matter, Max Planck (1858 - 1947) wrote:
\begin{quotation}
\textquotedbl\textit{It is impossible to construct an engine which,
working in a complete cycle, will produce no effect other than the
raising of a weight and the cooling of a heat reservoir.}\textquotedbl{}
\end{quotation}
However, both statements above are often combined into the so-called
Kelvin-Planck statement of the second law: 
\begin{quotation}
\textquotedbl\textit{It is impossible to construct an engine that,
operating in a cycle, will produce no effect other than the extraction
of heat from a reservoir and the performance of an equivalent amount
of work}.\textquotedbl{}
\end{quotation}
Also, in the words of Clausius:
\begin{quotation}
\textquotedbl\textit{It is impossible to construct a device that,
operating in a cycle, will produce no effect other than the transference
of heat from a cooler to a hotter body.}\textquotedbl{}
\end{quotation}
Finally, Carnot's theorem states:
\begin{quotation}
\textquotedbl\textit{No heat engine operating between two heat reservoirs
can be more efficient than a reversible heat engine operating between
the same two reservoirs.}\textquotedbl{}
\end{quotation}
Thus, according to these words, the second law of thermodynamics is
a statement of what is impossible to achieve in the attempt of interchanging
heat and work. However, these statements transcend the context of
heat engines and refrigerators and have surprisingly far-reaching
and less down-to-earth consequences. Essentially, the second law captures
a fundamental feature of natural phenomena not covered by the first
law, i.e., there are classes of processes not prohibitive from the
perspective of the conservation of energy that still does not happen
in nature. In this sense, the second law is an independent and complementary
statement that also portrays a fundamental natural restriction. Moreover,
it constrains even further the set of possible thermodynamic processes.

At the core of the second law also lies the concepts of reversibility
and irreversibility. Suppose a thermodynamic system that interacts
with its surroundings. Naturally, both work and heat may be exchanged
during any change of the macroscopic variables. Nevertheless, a reversible
process is, by definition, a process that, once realized, the system
and its surroundings must be returned to their previous states, without
any other change, by restoring the settings to their initial conditions.
However, this is only achievable - a priori - by requiring processes
executed by slow infinitesimal transformations (quasi-statically)\footnote{In fact, every reversible process is necessarily quasi-static, while
the converse is not true.} and in the absence of dissipative effects, such as friction. Along
these lines, it is clear that reversible processes are theoretical
idealizations and, consequently, every natural process occurs in a
finite time and is irreversible to some extent.

Along with (ir)reversibility, entropy is a cornerstone of the second
law. By introducing it, Clausius provided the means for a consistent
mathematical formulation of the previous statements. Interestingly,
the notion of entropy transcended thermodynamics and soon achieved
the status of one of the most fundamental quantities of modern physics.
As a direct extension of the work of Carnot on heat engines and cyclic
thermodynamic processes \cite{Carnot}, Clausius showed what we know today as Clausius's
theorem \cite{clausius1854},
\begin{equation}
\oint\,\frac{\delta Q}{T}=0,\label{eq:clausius theorem}
\end{equation}
where $T$ is the temperature at which the heat exchange took place\footnote{This equality is obtained by noticing that any thermodynamic cycle
can be seen as a composition of infinitesimal Carnot cycles and summing
up all the ratios $\frac{Q_{i}}{T_{i}}$.}. Eq. (\ref{eq:clausius theorem}) is valid for arbitrary cyclic and
reversible processes and also hints at the existence of a state function
specified by the macroscopic variables $\{X_{j}\}$, i.e., $\frac{\delta Q}{T}$ which
depends only on the thermodynamic coordinates and its integral between
two different states $a$ and $b$ is independent of the path (similar
to Eq. (\ref{eq: interal cyclic energy})). Thus, Clausius proposed
a new quantity $S$ and named it \textit{entropy}\footnote{Entropy means \textit{transformation}, from the Greek word $\tau\rho\pi\eta$.
\cite{Prigogine}}, such that
\begin{equation}
dS\equiv\frac{\delta Q}{T}
\end{equation}
and, therefore, the entropic change is
\begin{equation}
\Delta S\equiv S_{b}-S_{a}=\int_{a}^{b}\frac{\delta Q}{T}.
\end{equation}
Notice that, despite $\delta Q$ being a path-dependent quantity,
the ratio $\frac{\delta Q}{T}$ is an exact differential, $dS$. It
is imperative to highlight both the close relationship between entropy
and heat and the importance of the reversibility hypotheses. Let us
suppose a system of interest interacting with a reservoir at temperature
$T$. During any reversible process the heat exchanged, $Q$, between
the system and reservoir happens at the same temperature, in such
a way that their local entropic changes are the opposite, i.e.,
\begin{equation}
\Delta S_{system}=-\Delta S_{reservoir}.
\end{equation}
The equality above is true for every reversible process and sketches
an even more general result: The sum of the entropies of the system
of interest and all its surroundings, also known as the \textit{universe}\footnote{In the context of thermodynamics, the word \textit{universe} does
not have a cosmological sense.}, is invariant over reversible processes, i.e., given $S_{universe}\equiv S_{system}+S_{surroundings}$,
we have
\begin{equation}
\Delta S_{universe}=0.
\end{equation}

Of course, as mentioned earlier, reversible processes do not represent
the typical behaviour in nature. Since entropy is a state function,
given a system at equilibrium, any possible cyclic process, reversible
or not, would still satisfy $\oint dS_{system}=0$. However, if
any irreversibility is taken into account, such as friction and other
dissipative effects, Eq. (\ref{eq:clausius theorem}) should be updated
to the following inequality
\begin{equation}
\oint\,\frac{\delta Q}{T}\leq0,
\end{equation}
which also implies that
\begin{equation}
dS_{system}\geq\frac{\delta Q}{T}.\label{eq: inequality}
\end{equation}
Besides, in these cases, the entropic change from the system of interest
is not completely compensated by the changes that occur in the surroundings.
In fact, in general, one has for the\textit{ second law of thermodynamics}
\begin{equation}
\Delta S_{universe}\geq0.\label{eq: entropy change universe}
\end{equation}
Clearly, these expressions automatically include the previous ones:
if no irreversible processes take place, then the equalities are recovered.
Eq. (\ref{eq: entropy change universe}) above is the most common
mathematical statement of the second law: the entropy of the universe
always increases or remains constant. It is also directly related
to the notion of the \textit{arrow of time}, i.e., since every natural
process inexorably increases the entropy, one can distinguish the
past from the future.

Note that the entropic change of an irreversible process is higher
than a reversible one with the same exchanged heat, $Q$, and temperature
$T$. More modern approaches often rewrite the inequality from Eq.(\ref{eq: inequality})
into the following equality: 
\begin{equation}
dS_{system}=d\varPhi+d\Sigma,
\end{equation}
where $d\varPhi\equiv\frac{\delta Q}{T}$ is recognized as the entropy
change due to the external energy flow\footnote{If the system is open, there is also an energetic contribution due
to matter flow.}, known as \textit{entropy flux}, and $d\Sigma$, or the \textit{entropy
production}, is the change due to the irreversible processes. Along
these lines, it is clear that entropy $dS_{system}$ and $d\varPhi$
might acquire positive or negative values, while entropy production
is always a positive quantity,
\begin{equation}
d\Sigma\geq0.\label{eq:entropy production}
\end{equation}
Interestingly, while the first law states that the energy of an isolated
system is conserved, the second implies that its entropy is not and,
more importantly, can only increase during natural processes. Of course,
both fundamental quantities are intrinsically connected, and many
applications of thermodynamics intend to characterize their changes.

\paragraph{Third law of thermodynamic}

Finally, the \textit{third law of thermodynamics} is mathematically
stated as
\begin{equation}
T\rightarrow0_{+}\,\Rightarrow\,S\rightarrow S_{0},
\end{equation}
i.e., when the temperature $T$ approaches the limiting absolute zero,
the entropy approaches a constant value, $S_{0}$, independent of
all macroscopic variables.

\paragraph{Brief remarks}

In the context of thermodynamics, physical properties are commonly
separated into two distinct categories: On the one hand, \textit{intensive}
quantities are independent of the mass or the size and number of constituents of the system;
On the other, \textit{extensive} ones are those proportional to the
system's mass, number of constituents, or size. Along these lines, let us consider a thermodynamic
system in equilibrium described by the macroscopic variables temperature
$T$, volume $V$ and pressure $P$. Naturally, one can also prescribe
an internal energy $U$ and entropy $S$. Then, suppose one can split
the whole system into two equal smaller subsystems. Both partitions
will maintain their previous temperature $T$ and pressure $P$, while
their new volume, internal energy and entropy will be divided by two
($V/2$, $U/2$ and $S/2$). Thus, it is clear that $T$ and $P$
are intensive quantities, while $V$, $U$ and $S$ are extensive
ones. In particular, notice that extensivity also means that these
properties are \textit{additive}, i.e., the whole energy and entropy,
for instance, are the sum of the subsystem's energy and entropy, respectively.

Finally, it is worth emphasizing - once again - that the laws of thermodynamics
were developed and established for macroscopic systems in equilibrium.
It was not necessary any considerations concerning the microscopic
behaviour of matter. Surprisingly, however, it does not mean that
these laws are not valid in other regimes.

%% file: Chapters/Chapter2/Subsection_1_2.tex
\subsection{Stochastic thermodynamics\label{subsec:Stochastic-thermodynamics}}

The microscopic foundations of classical equilibrium thermodynamics
were provided by the advent of statistical mechanics. It also brought
a new probabilistic perspective to the understanding of natural phenomena
at the molecular level. These efforts allowed the field to shift from
purely phenomenological reasoning to more solid ground. Later, linear
response theory contributed to the first expeditions outside the equilibrium
context. More recently, however, the scenarios are even more extreme:
the current high degree of control of systems and devices below the
microscopic scale urged a thermodynamic description of small systems
far from equilibrium. In such circumstances, essentially, one is interested
in characterizing the energetics of finite objects at finite times.
Biological systems at the cellular level, molecular machines, and
colloidal particles trapped by optical tweezers, for instance, are
paradigmatic examples of physical objects operating under these conditions.
In contrast with macroscopic scenarios, the energy within small systems
is of the order of $k_{B}T$ and its fluctuations are not negligible.
In fact, deviations from the average behaviour are typical features
of such regimes and carry valuable thermodynamic information.

In general, these discussions belong to the recent field of \textit{stochastic
thermodynamics} \cite{seifert2008stochastic,sekimoto2010stochastic,seifert2012stochastic,schuster2013nonequilibrium,VANDENBROECK20156,ciliberto2017experiments}, whose developments has been successfully
helping to bridge the gap between the well known macroscopic laws
and the small-scale behaviour of matter. In this framework, thermodynamic
systems are both driven by external classical forces and constantly influenced
by the fluctuations induced by their environments. Along these lines,
thermodynamic quantities, such as work, heat and entropy production,
are understood as fluctuating entities specified at the phase space
trajectory level and characterized by stochastic dynamical processes,
often depicted by Markovian master equations. Despite the shared probabilistic
spirit, stochastic thermodynamics exceeds the scope of equilibrium
statistical mechanics and the linear response regime, i.e., it allows
the description of general microscopic scenarios and the consideration
of arbitrary non-equilibrium processes, in such a way that the connection
with the well established macroscopic laws and conclusions are directly
obtained once considered the statistical ensemble level. Furthermore,
this perspective provided novel insights on the relationship between
irreversibility and the second law of thermodynamics. Interestingly,
although predicting the positivity of the \textit{average} of entropy
production, stochastic thermodynamics highlighted the existence of
trajectories with negative entropy production and provided the mathematical
formalism for quantifying their probabilities of occurrence.

\subsubsection{Fluctuations theorems}

Along with a more refined view of thermodynamics, the so-called \textbf{\textit{f}}\textit{luctuation
}\textbf{\textit{t}}\textit{heorems} (FTs) represent the most influential
results of the field. Initially discovered during the '90s \cite{evans1993probability,evans1994equilibrium,gallavotti1995dynamical,jarzynski1997nonequilibrium,jarzynski1997equilibrium,crooks1998nonequilibrium,kurchan1998fluctuation,crooks1999entropy,lebowitz1999gallavotti}
and later experimentally verified in several setups \cite{hummer2001free,liphardt2002equilibrium,feitosa2004fluidized,ciliberto2004experimental,collin2005verification,Douarche_2005,garnier2005nonequilibrium,douarche2006work,PhysRevLett.96.070603,mossa2009dynamic,naert2012experimental,ciliberto2017experiments}, the
FTs were instrumental in triggering further studies on non-equilibrium
systems and fostering the foundation of stochastic thermodynamics.
Essentially, the FTs are general mathematical statements concerning
the probability distributions of the relevant thermodynamic quantities
over an ensemble of identically prepared systems. They are usually
expressed according to their \textbf{\textit{i}}\textit{ntegral} (IFTs)
or \textbf{\textit{d}}\textit{etaile}d (DFTs) forms, given, respectively,
by \cite{seifert2012stochastic}
\begin{equation}
\langle e^{-\zeta}\rangle=\int d\zeta\,P(\zeta)e^{-\zeta}=1\label{eq:integral FTs}
\end{equation}
 and 
\begin{equation}
\frac{P(\zeta)}{P(-\zeta)}=e^{-\zeta},
\end{equation}
where $\zeta$ is a functional of the stochastic trajectory, e.g.,
fluctuating work and heat, and $P(\zeta)$ is its respective distribution.
From Eq. (\ref{eq:integral FTs}), one can easily use Jensen's inequality\footnote{For the exponential function, Jensen's inequality becomes $e^{-\langle\zeta\rangle}\geq\langle e^{-\zeta}\rangle$.}
and unveil the hidden inequality
\begin{equation}
\langle\zeta\rangle\geq0.
\end{equation}

For instance, suppose a system initially prepared in thermal equilibrium
at temperature $T$ and then decoupled with its surroundings. If the
system's Hamiltonian is parametrized as $H(\lambda)$, where $\lambda$
is an externally controlled parameter, work $w$ can be performed
by simply changing it from $\lambda_{i}$ to $\lambda_{f}$ since
$w\equiv\int\frac{\partial H(\lambda)}{\partial\lambda}d\lambda$.
By defining the dissipated work along with a specific work protocol
$\lambda_{i}\rightarrow\lambda_{f}$ as $w_{diss}\equiv w-\Delta F_{fi}$,
where $\Delta F\equiv F_{f}-F_{i}$ is the Helmholtz free energy difference
for the equilibrium states associated with $H(\lambda_{i})$ and $H(\lambda_{f})$,
one can show the following IFT
\begin{equation}
\langle e^{-\beta w_{diss}}\rangle=1\:\Longleftrightarrow\:\langle e^{-\beta w}\rangle=e^{-\beta\Delta F}\label{eq:Jarzinski}
\end{equation}
 and, therefore,
\begin{equation}
\langle w_{diss}\rangle\geq0\:\Longleftrightarrow\:\langle w\rangle\geq\Delta F_{fi},\label{eq: second law jarzynski}
\end{equation}
where $\beta^{-1}\equiv k_{B}T$ with $k_{B}$ the Boltzmann constant
and the ensemble averages $\langle.\rangle$ are computed after performing
the same protocol several - ideally infinite - times. Interestingly,
these expressions make no mention of the protocol's execution speed
or its specific details. In fact, it just depends on the final and
initial values $\lambda_{f}$ and $\lambda_{i}$. Eq. (\ref{eq:Jarzinski})
is the emblematic - and surprising - \textit{Jarzynski equality} \cite{jarzynski1997nonequilibrium}
relating arbitrary non-equilibrium processes, $\lambda_{i}\rightarrow\lambda_{f}$,
with genuine equilibrium quantities $\Delta F_{fi}$, while the Eq.
(\ref{eq: second law jarzynski}) implied by Eq. (\ref{eq:Jarzinski})
is one of the well-known forms of the second law of thermodynamics.
Notice, however, that $\langle w\rangle\geq\Delta F_{fi}$ is even
more revealing than the classical thermodynamic one since it is the
\textit{average} value of work that is higher than the difference
of free energy, which also means that for individual elements of the
ensemble such inequality may be violated. Of course, this conclusion
does not represent a contradiction with the classical result: for
macroscopic systems and their energy scales, the fluctuations are
not easily observed, in a way that the work performed by the execution
of the same protocol are essentially equal and, consequently, their
distributions are mathematically depicted by delta functions. Additionally,
if $P_{\Lambda}(w)$ is the probability distribution of performing
some work $w$ along the trajectory defined by the protocol $\Lambda:\lambda_{i}\rightarrow\lambda_{f}$
and $P_{\tilde{\Lambda}}(w)$ is the distribution related to the trajectory
obtained by the time-reversed protocol, $\tilde{\Lambda}:\lambda_{f}\rightarrow\lambda_{i}$,
under the assumption of Markovian dynamics and initial equilibrium
states for both protocols, one can show the following DFTs-like expression
\begin{equation}
\frac{P_{\Lambda}(w)}{P_{\tilde{\Lambda}}(-w)}=e^{-\beta w_{diss}}=e^{-\beta(w-\Delta F_{fi})},\label{eq:Crooks FT}
\end{equation}
also known as the \textit{Crooks theorem}. \cite{crooks1999entropy} Interestingly,
Eq. (\ref{eq:Crooks FT}) means the probability of observing trajectories
whose work spent along with $\Lambda$ is given by $w$ is exponentially
more likely than the probability of observing work $-w$ along with
the reverse protocol $\tilde{\Lambda}$. In fact, their probabilities
are equal iff the work performed is equal to the free energy difference,
$w=\Delta F_{fi}\Longleftrightarrow P_{\Lambda}(w)=P_{\tilde{\Lambda}}(-w)$.
Moreover, it is easy to see that Eq. (\ref{eq:Crooks FT}) directly
implies the Jarzynski equality, Eq. (\ref{eq:Jarzinski})\footnote{$\int dw\,P_{\tilde{\Lambda}}(-w)e^{-\beta w}=e^{-\beta\Delta F}\int dw\,P_{\Lambda}(w)\Rightarrow\langle e^{-\beta w}\rangle=e^{-\beta\Delta F}$.}.

Finally, by properly identifying entropy production, $\Sigma$, along
with stochastic trajectories, one can also show the following IFT
for arbitrary protocols and initial conditions \cite{seifert2012stochastic,schuster2013nonequilibrium}
\begin{equation}
\langle e^{-\frac{\Sigma}{k_{B}}}\rangle=1.
\end{equation}
 Again, it also implies what can be seen as a more refined version of the second law of
thermodynamics depicted by Eq. (\ref{eq:entropy production}) concerning
the average value of entropy production, such that
\begin{equation}
\langle\Sigma\rangle\geq0.
\end{equation}

In short, these theorems are simple equalities concerning the distributions
of stochastic thermodynamic quantities. Surprisingly, they are valid
even for \textit{general} processes far from equilibrium. In this sense, despite their
simplicity, they represented generalizations of the classical macroscopic
thermodynamic results and provided novel insights on the underlying
statistical nature of the laws of thermodynamics.

%% file: Chapters/Chapter2/Section2.tex
\section{Quantum mechanics\label{sec:Quantum-mechanics}}

Quantum mechanics (QM) is one of the main pillars of modern science, not only because of its remarkable accuracy and outstanding success in predicting physical phenomena but also because it represents, along with Einstein's theory of general relativity and Maxwell's of electrodynamics, our best attempt to understand the very fundamental aspects of nature. Since its early days, at the beginning of the 20th century, it became clear that the development of QM would require a novel and courageous change in the traditional way of thinking. After all, physics - as was commonly believed at the time - was not complete, and the failure of classical theories to explain the contemporary experimental observations was undeniable evidence of this defeat. Hence, despite centuries of immense accomplishment, physicists were forced to adapt in the face of empirical confrontation. Along these lines, the so-called ultraviolet catastrophe is a paradigmatic example of how nature's behaviour contradicted classical intuition and, in contrast, radical new hypotheses were suddenly necessary. This interesting episode is commonly seen as the birth of the quantum theory, and Max Planck (1858 - 1947) is regarded as one of its founding fathers. Latter, seminal works by Albert Einstein (1879 - 1955), Niels Bohr (1885 - 1962), Louis de Broglie (1892 - 1987), Erwin Schrödinger (1887 - 1961), Werner Heisenberg (1901 - 1976) and many others helped develop and establish the basis and mathematical structure of the theory.

During the first decades of the last century, QM flourished both from
a theoretical perspective and experimental validation. The revolutionary
ideas of energy quantization, particle-wave duality and the intrinsic
probabilistic behaviour at the atomic scale, although highly non-intuitive,
seemed to be a superior way to describe how nature works. This early
progress was of fundamental importance to pave the way for the later
development of quantum field theory and the conception of modern technologies,
such as the groundbreaking developments of the transistor, the laser
and, more recently, magnetic resonance machines. However, despite
this great success, QM remained challenging conceptually. On the one
hand, it was unclear how to physically interpret some fundamental
aspects of the theory, such as the wave function and non-local correlations.
On the other, there was a debate to understand if it was indeed a
complete description of reality. Interestingly, even after one century
of development and crucial progress, some of these foundational questions
remain open, which, in a sense, preserve these old debates echoing
throughout modern research. This, however, did not prevent the field
from expanding and developing. Applications of QM to specific questions
gave rise to new and exciting research fields. The birth of quantum
information and computation in the '80s and their first progress during
the '90s, for instance, are recent and fascinating ramifications of
the quantum theory in the context of information processing.

Nowadays, QM maintains its influence as one of the most fundamental
theories of modern physics and keeps changing the world consistently
in several aspects ranging from the production of new materials to
the development of novel - and unique - technologies. Along these
lines, this section will be dedicated to briefly introducing - or
reviewing - the elementary aspects of the theory and its particular
mathematical formalism. Of course, it does not aim for an extensive
presentation of the subject and its most modern tools. A comprehensive
discussion of QM can be found in the excellent textbooks that also
guided the following introduction. \cite{Sakurai,Piza,Griffiths,nielsen00,BRE02}

\input{Chapters/Chapter2/Subsection_2_1}
\input{Chapters/Chapter2/Subsection_2_2}

%% file: Chapters/Chapter2/Subsection_2_1.tex
\subsection{Mathematical description of quantum systems}

Quantum theory, essentially, is mathematically written and stated
with the vocabulary provided by linear algebra. In fact, the whole
formalism of QM, and its basic elements, can be described by vectors
and linear transformations. In this sense, complex vector spaces with
inner product, known as Hilbert spaces, are the most fundamental underlying
structures setting the stage for describing quantum systems, whose
physical states are simply represented by vectors within these abstract
entities. 

Along these lines, for every quantum system, there is an associated
Hilbert space $\mathcal{H}$ with the appropriate dimensionality $d:=dim(\mathcal{H})$
- possibly infinity - such that its vector state is fully characterized
and labelled by, according to Dirac's notation, the so-called \textit{ket}
$|\Psi\rangle\in\mathcal{H}$. From now on, let us assume only finite-dimensional
Hilbert spaces. These vectors, or kets, encode all possible information
concerning the physical system in question and are the quantum analogous
to the usual notion of classical states in the phase space, given
by $(\overrightarrow{\boldsymbol{p}},\overrightarrow{\boldsymbol{q}})$\footnote{Of course, the analogy only corresponds to their similar descriptive
roles inside their particular theoretical frameworks.}. Just like any regular vector, every ket $|\Psi\rangle$ can be represented
in terms of a basis, i.e., given a set $\{|b_{j}\rangle,j=1,...,d\}$
of $d$ linearly independent kets spanning $\mathcal{H}$, one can
describe the state as the following \textit{superposition} (linear
combination)
\begin{equation}
|\Psi\rangle=\sum_{j=1}^{d}\alpha_{j}|b_{j}\rangle,\label{eq:state}
\end{equation}
where the coefficients $\{\alpha_{j},j=1,...,d\}$ are complex numbers.
Besides, it is worth mentioning that Hilbert spaces inherit all properties
and basic operations defined for arbitrary vector spaces. Along with
$\mathcal{H}$, we should also define the so-called dual vector space
$\mathcal{H}^{*}$, whose for every possible element $|\Psi\rangle$
of $\mathcal{H}$, there is a single component $\langle\Psi|$ - known
as \textit{bra} - from $\mathcal{H}^{*}$, such that
\begin{equation}
|\Psi\rangle=\sum_{j=1}^{d}\alpha_{j}|b_{j}\rangle\longleftrightarrow\langle\Psi|=\sum_{j=1}^{d}\alpha_{j}^{*}\langle b_{j}|,
\end{equation}
where $\alpha_{j}^{*}$ denotes the complex conjugate of $\alpha_{j}$.
Now we have all the necessary ingredients to properly define the inner
product between two elements $|\psi\rangle$ and $|\phi\rangle$ of
$\mathcal{H}$ as simply as 
\begin{equation}
\langle\psi|\phi\rangle,\label{eq:inner product}
\end{equation}
where $\langle\psi|\phi\rangle=(\langle\phi|\psi\rangle)^{*}$ and
$\langle\psi|\psi\rangle\geq0$. Thus, two non-null kets are orthogonal,
relative to Eq. (\ref{eq:inner product}), iff $\langle\psi|\phi\rangle=0$.
Also, it allows the identification of the norm of a ket $|\psi\rangle$
as $|||\psi\rangle||\equiv\sqrt{\langle\psi|\psi\rangle}$. In this
sense, every possible basis $\{|b_{j}\rangle,j=1,...,d\}$ of $\mathcal{H}$
should contain orthogonal and normalized (unit norm) elements, i.e.,
they should satisfy the orthonormality condition stated by $\langle b_{j}|b_{k}\rangle=\delta_{jk}$,
where $\delta_{jk}$ is the usual Kronecker delta.

In QM, the system's state is encoded only in the \textquotedbl direction\textquotedbl{}
of $|\psi\rangle$. In fact, any ket from the set $\{c|\Psi\rangle\in\mathcal{H}|c\in\boldsymbol{C}_{\neq0}\}$
represents the same physical state. For this reason, and latter interpretative
convenience, it is commonly assumed normalized kets for describing
quantum systems, $\langle\Psi|\Psi\rangle=1$, which also implies
that, given Eq. (\ref{eq:state}), 
\begin{equation}
\sum_{j=1}^{d}\alpha_{j}^{*}\alpha_{j}=\sum_{j=1}^{d}|\alpha_{j}|^{2}=1.
\end{equation}
Interestingly, one could add a complex phase $c=e^{i\theta}$ with
unit modulus such that $|\Psi\rangle\rightarrow e^{i\theta}|\Psi\rangle$,
in such a way that $\langle\Psi|\Psi\rangle=1$ is still satisfied.
Nevertheless, this phase, often called as global phase, does not affect
any physical prediction obtained from $|\Psi\rangle$.

\subsubsection{Operators and observables\label{subsec:Operators-and-observables}}

In the following sections, we denote operators by $\hat{(.)}$. In
this context, operators refer to linear transformations $\hat{T}:\mathcal{H}\rightarrow\mathcal{H}^{\prime}$
mapping one element $|\Psi\rangle$ from a particular Hilbert space
$\mathcal{H}$ to another $|\Phi\rangle$ from $\mathcal{H}^{\prime}$,
such that $\hat{T}|\Psi\rangle=|\Phi\rangle$ and
\begin{equation}
\hat{T}\left(\sum_{j=1}^{d}\alpha_{j}|b_{j}\rangle\right)=\sum_{j=1}^{d}\alpha_{j}(\hat{T}|b_{j}\rangle).
\end{equation}
Nevertheless, most of the time, we are interested in operators $\hat{O}$
acting within a specific Hilbert space, such that $\hat{O}:\mathcal{H}\rightarrow\mathcal{H}$.
From now on, the set of linear operators satisfying this is denominated
by $\mathcal{L}(\mathcal{H})$. Without any loss of generality, let
us restrain our discussion for such cases.

As shown earlier, kets can be represented in terms of any basis $\{|b_{j}\rangle,j=1,...,d\}$
of $\mathcal{H}$. Similarly, operators might be depicted by $d\times d$
matrices. In this sense, any operator $\hat{O}\in\mathcal{L}(\mathcal{H})$
is fully characterized by the complex matrix elements $o_{kj}\equiv\langle b_{k}|\hat{O}|b_{j}\rangle$
and written in terms of the outer products defined by $|b_{k}\rangle\langle b_{j}|$,
such that
\begin{equation}
\hat{O}=\sum_{k=1}^{d}\sum_{j=1}^{d}o_{kj}|b_{k}\rangle\langle b_{j}|.\label{eq: operator}
\end{equation}
It is worth mentioning that any orthonormal basis of $\mathcal{H}$
is a valid and equivalent basis of description, in the sense that
all possible representations are directly connected via similarity
transformations. This fact gives us the freedom to perform calculations
in the most suitable one. Thus, given Eqs. (\ref{eq:state}, \ref{eq: operator}),
we have - in general - $\hat{O}|\Psi\rangle=\sum_{k=1}^{d}\sum_{j=1}^{d}\alpha_{j}o_{kj}|b_{k}\rangle$.
In particular, the so-called identity operator $\hat{1}\in\mathcal{L}(\mathcal{H})$
that maps any ket to itself, such that $\hat{1}|\Psi\rangle=|\Psi\rangle$,
is clearly obtained iff $o_{kj}=\delta_{kj}$ in the equation above.
Hence, 
\begin{equation}
\hat{1}=\sum_{j=1}^{d}|b_{j}\rangle\langle b_{j}|
\end{equation}
 This important equality is true for any possible basis, and it is
known as the \textit{completeness relation}. Often, we are interested
in the applications of different operators in a given ket, such as
$\hat{O}_{2}\hat{O}_{1}|\Psi\rangle$. Since they are depicted by
matrices, the general non-commutative property is, naturally, inherited,
i.e., $\hat{O}_{1}\hat{O}_{2}\neq\hat{O}_{2}\hat{O}_{1}$. Sometimes,
however, operators might commute and, for this reason, it is convenient
to define the \textit{commutator} $[,]$ between two operators as
\begin{equation}
[\hat{O}_{1},\hat{O}_{2}]\equiv\hat{O}_{1}\hat{O}_{2}-\hat{O}_{2}\hat{O}_{1}.
\end{equation}

Let us present some import classes of linear operators. The Hermitian
conjugate of $\hat{O}$ - or simply adjoint - defined by $\hat{O}^{\dagger}$,
is the unique operator which guarantees that the bra $(\langle\Psi|\hat{O}^{\dagger})$
is the dual vector of the ket $(\hat{O}|\Psi\rangle)$. Operationally,
the matrix $\hat{O}^{\dagger}$ is the conjugate transpose of $\hat{O}$,
which means that $\langle b_{k}|\hat{O}|b_{j}\rangle=(\langle b_{j}|\hat{O}^{\dagger}|b_{k}\rangle)^{*}$.
Nevertheless, if 
\begin{equation}
\hat{O}=\hat{O}^{\dagger},\label{eq:hermiticity}
\end{equation}
then $\hat{O}$ is denominated as a \textit{Hermitian} operator. Additionally,
\textit{Unitary} operators $\hat{U}$ are those who satisfy the following
equality 
\begin{equation}
\hat{U}\hat{U}^{\dagger}=\hat{U}^{\dagger}\hat{U}=\hat{1},
\end{equation}
which also implies that they preserve the inner product, such that
$\langle\psi|\phi\rangle=(\langle\psi|\hat{U}^{\dagger})(\hat{U}|\phi\rangle)$.
Both Hermitian and Unitary linear operators are examples of a broader
class of operators, known as \textit{Normal} ones. A Normal operators
$\hat{N}$, by definition, commutes with its own Hermitian conjugate,
i.e.,
\begin{equation}
[\hat{N},\hat{N}^{\dagger}]=0
\end{equation}
or $\hat{N}\hat{N}^{\dagger}=\hat{N}^{\dagger}\hat{N}$. Along these
lines, this class of operators are accompanied by the \textit{spectral
theorem}, which states that \cite{nielsen00}: if $\hat{N}\in\mathcal{L}(\mathcal{H})$
is normal, then it is unitarily diagonalizable, i.e., there is a basis
$\{|N_{j}\rangle,j=1,...,d\}$ of $\mathcal{H}$ such that
\begin{equation}
\hat{N}=\sum_{j=1}^{d}n_{j}|N_{j}\rangle\langle N_{j}|,\label{eq:spectral decomposition}
\end{equation}
where $\{n_{j},j=1,...,d\}$ and $\{|N_{j}\rangle,j=1,...,d\}$ are
the respective eigenvalues and eigenvectors of $\hat{N}$, since $\hat{N}|N_{j}\rangle=n_{j}|N_{j}\rangle$.
The form of Eq. (\ref{eq:spectral decomposition}) above is commonly
referred to as the \textit{spectral decomposition} of $\hat{N}$.
Interestingly, if two different normal operators $\hat{N}_{1}$ and
$\hat{N}_{2}$ have spectral decompositions according to the same
basis, they must commute, i.e., if $\hat{N}_{k}=\sum_{j=1}^{d}n_{k,j}|N_{j}\rangle\langle N_{j}|$
for $k=1,2$ share the same eigenvectors, then it is guaranteed that
$[\hat{N}_{1},\hat{N}_{2}]=0$.

In QM, meaningful physical quantities, such as position, momentum
and spin, are known as observables and mathematically described by
hermitian linear operators and the machinery presented above. Along
these lines, given an observable $\hat{O}\in\mathcal{L}(\mathcal{H})$,
its eigenvalues $\{o_{j}\}_{j=1,...,d}$ are understood as the possible
outcomes of an eventual measurement performed on the quantum system.
The hermiticity condition, stated by Eq. (\ref{eq:hermiticity}),
guarantee that these values are real numbers\footnote{Given Eq. (\ref{eq:spectral decomposition}), if $\hat{N}=\hat{N}^{\dagger}$,
then $\{n_{j}\}_{j}\in\mathbb{R}$.}.

\subsubsection{Projective measurements}

Suppose that $\hat{B}\in\mathcal{L}(\mathcal{H})$ is the observable
we are interested in measuring, with the following spectral decomposition
\begin{equation}
\hat{B}=\sum_{j=1}^{d}b_{j}|b_{j}\rangle\langle b_{j}|.
\end{equation}
If $|\Psi\rangle$ is the system's state, and since $\{|b_{j}\rangle,j=1,...,d\}$
constitute a valid basis of $\mathcal{H}$, let us write it in the
following representation $|\Psi\rangle=\sum_{j=1}^{d}\alpha_{j}|b_{j}\rangle$,
where $\alpha_{k}\equiv\langle b_{k}|\Psi\rangle$. Notice that, in
general, $|\Psi\rangle$ is in a superposition of the eigenvectors
of $\hat{B}$. Once a measurement is performed, two things happen:
\textbf{i)} as a result of the process, the measurement apparatus
obtains one of the possible eigenvalues of $\hat{B}$; \textbf{ii)}
the system's state immediately collapses into one of the superposition
elements. Such state update corresponds to the eigenvector relative
to the measurement outcome, i.e., if $b_{k}$ is measured, then
\begin{equation}
|\Psi\rangle\rightarrow|\Psi^{\prime}\rangle\equiv|b_{k}\rangle.
\end{equation}
This whole event - known as a projective measurement - is seen as
fundamentally probabilistic, i.e., in the orthodox presentation of
QM, all information encoded by $|\Psi\rangle$ are the statistics
of the outcomes, which are characterized by the coefficients $\{\alpha_{j},j=1,...,d\}$
playing the role of probability amplitudes. In this sense, by defining
the so-called projectors $\hat{\Pi}_{k}\equiv|b_{k}\rangle\langle b_{k}|\in\mathcal{L}(\mathcal{H})$,
the probability of measuring $b_{k}$ and the system collapsing to
$|b_{k}\rangle$ is given by 
\begin{equation}
Prob(b_{k})\equiv\langle\Psi|\Pi_{k}|\Psi\rangle=|\alpha_{k}|^{2}\geq0,\label{eq:postulate2}
\end{equation}
while the collapse itself is mathematically written as
\begin{equation}
|\Psi^{\prime}\rangle=\frac{\Pi_{k}|\Psi\rangle}{\sqrt{\langle\Psi|\Pi_{k}|\Psi\rangle}}.
\end{equation}
Notice that, it means that if any posterior measurement of $\hat{B}$
is performed right after the first one, the system would still be
found at the same state $|\Psi^{\prime}\rangle=|b_{k}\rangle$ with
$Prob(b_{k})=1$. Also, as expected, the normalization $\langle\Psi|\Psi\rangle=1$
guarantees that
\begin{equation}
\sum_{k=1}^{d}Prob(b_{k})=1.
\end{equation}
It is clear that the physical act of measuring is not being modelled
at all, and the treatment above is, essentially, phenomenological.
In fact, Eq. (\ref{eq:postulate2}) is a fundamental \textit{postulate}
of QM and, therefore, it is not demonstrable from first principles\footnote{Still today, there are several discussions on this matter, especially
on foundational questions of QM. \cite{Neto}}.

From the measurement statistics of $\hat{B}$ above, one can compute
its average value simply as $\sum_{k=1}^{d}b_{k}Prob(b_{k})$. Thus,
it is easy to see that the expectation value of any given observable
$\hat{O}\in\mathcal{L}(\mathcal{H})$ relative to the state $|\Psi\rangle$
can be defined as
\begin{equation}
\langle\hat{O}\rangle\equiv\langle\Psi|\hat{O}|\Psi\rangle.
\end{equation}

It is worth mentioning that, even though $|\Psi\rangle$ might represent
a single physical quantum system, in practice, such probabilities
are only accessible at an ensemble level analysis, in a way that it
is implicitly assumed several identical copies of the same physical
ket. This identical copy setting is known as a \textit{pure} ensemble.

\subsubsection{Composite systems}

Let us now generalize the previous formalism to include the description
of - possibly - several quantum systems\footnote{Notice that the majority of the machinery presented here also works for a single quantum systems that has several degrees of freedom, such as orbital motion, spin, etc.}. In many cases, the quantum
system being described is composed of two or more, non necessarily
equal, subsystems. If $\mathcal{H}^{(k)}$ is the Hilbert space of
the $k$th element of a group of $N$ subsystems, the whole Hilbert
space $\mathcal{H}^{(0)}$ is constituted by the following tensor
product $\mathcal{H}^{(0)}=\mathcal{H}^{(1)}\otimes\mathcal{H}^{(2)}...\otimes\mathcal{H}^{(N)}$.
Of course, each subsystem might have its own dimensionality $d^{(k)}:=dim(\mathcal{H}^{(k)})$,
such that $d^{(0)}=\Pi_{k=1}^{N}d^{(k)}$. The representation of possible
states in $\mathcal{H}^{(0)}$ is given by the tensor product structure
$|\psi^{(1)}\rangle\otimes|\psi^{(2)}\rangle...\otimes|\psi^{(N)}\rangle=|\psi^{(1)}\psi^{(2)}...\psi^{(N)}\rangle$,
where $|\psi^{(k)}\rangle\in\mathcal{H}^{(k)}$ stands for an individual
ket belonging to the $k$th subsystem. It means that, in the particular
case in which every subsystem is prepared in the state $|\Psi^{(k)}\rangle$,
the whole system is described by $|\Psi\rangle=|\Psi^{(1)}\rangle\otimes|\Psi^{(2)}\rangle...\otimes|\Psi^{(N)}\rangle$.

For simplicity, let us focus on bipartite systems, i.e., quantum systems
composed of two individual subsystems. Thus, let $\mathcal{H}^{(0)}=\mathcal{H}^{(1)}\otimes\mathcal{H}^{(2)}$
and $\{|b_{j}^{(1,2)}\rangle,j=1,...,d^{(1,2)}\}$ be a possible basis
of $\mathcal{H}^{(1,2)}$. Then any bipartite state can be written
as
\begin{align}
|\Psi\rangle & =\sum_{j=1}^{d^{(1)}}\sum_{k=1}^{d^{(2)}}a_{jk}|b_{j}^{(1)}\rangle\otimes|b_{k}^{(2)}\rangle,\label{Estado bipartido geral}
\end{align}
where $a_{jk}\equiv\langle b_{j}^{(1)},b_{k}^{(2)}|\Psi\rangle$ is
the probability amplitude of finding each individual subsystem into
the states $|b_{j}^{(1)}\rangle$ and $|b_{k}^{(2)}\rangle$. Eq.
(\ref{Estado bipartido geral}) above is very general. In some cases,
if $a_{jk}=c_{j}d_{k}$, then the whole system can be written as 
\begin{equation}
|\Psi\rangle=|\phi^{(1)}\rangle\otimes|\phi^{(2)}\rangle,\label{eq:separable}
\end{equation}
where $|\phi^{(1)}\rangle=\sum_{j=1}^{d^{(1)}}c_{j}|b_{j}^{(1)}\rangle$
and $|\phi^{(2)}\rangle=\sum_{k=1}^{d^{(2)}}d_{k}|b_{k}^{(2)}\rangle$,
i.e., for each subsystem, one can attribute an individual local state
$|\phi^{(1,2)}\rangle$ that it is independent of the other. Whenever
the previous form of Eq. (\ref{eq:separable}) is obtained, the whole
system is denominated \textit{separable}, and we have a product state.
Otherwise, the subsystems are non-separable and, therefore, called
\textit{entangled}. Entanglement is a genuine quantum property that,
essentially, means the different quantum systems have fundamentally
correlated states, regardless of their physical spatial distance.
Interestingly, it implies that the whole entangled system is not understood
as the simple composition of individual subsystems but an indivisible
entity instead.

In this context, local operators $\hat{O}^{(k)}\in\mathcal{L}(\mathcal{H}^{(k)})$
are the ones that act on individual subsystems and, therefore, specific
Hilbert spaces. In a global perspective, however, a local operator
acting in the subsystem $(2)$ of $|\psi^{(1)}\rangle\otimes|\psi^{(2)}\rangle$,
for instance, translates to
\begin{equation}
\left(\hat{1}^{(1)}\otimes\hat{O}^{(2)}\right)|\psi^{(1)}\rangle\otimes|\psi^{(2)}\rangle=|\psi^{(1)}\rangle\otimes\hat{O}^{(2)}|\psi^{(2)}\rangle,
\end{equation}
which means that $\hat{O}^{(2)}$ acts on $|\psi^{(2)}\rangle$, while
$|\psi^{(1)}\rangle$ remains unchanged due to the identity operator
$\hat{1}^{(1)}$. Besides, one could also be interested in applying
the operators $\hat{O}^{(1)}$ and $\hat{O}^{(2)}$ simultaneously.
In this case, the whole operator is written as $\hat{O}^{(1)}\otimes\hat{O}^{(2)}$
and
\begin{equation}
\left(\hat{O}^{(1)}\otimes\hat{O}^{(2)}\right)|\psi^{(1)}\rangle\otimes|\psi^{(2)}\rangle=\hat{O}^{(1)}|\psi^{(1)}\rangle\otimes\hat{O}^{(2)}|\psi^{(2)}\rangle.
\end{equation}

Along these lines, bipartite states can be cast in a very convenient
and special form, given by the Schmidt decomposition. Since this procedure
is the basis of this work, let us discuss it in more detail.

\paragraph{The Schmidt decomposition\label{par:The-Schmidt-decomposition}}

The Schmidt decomposition is a very useful theorem concerning the
representation form of pure bipartite quantum states (or any composite
one with a bipartite separation, actually). Its convenience has far-reaching
consequences and applications in quantum theory, especially in the
context of quantum information and quantum computation. Essentially,
it claims that any pure bipartite state can be written compactly as
a single sum of correlated orthonormal basis. In fact, as briefly
sketched below, this is a direct consequence of a more general result
concerning matrix factorization, known as \textbf{S}ingular \textbf{V}alue
\textbf{D}ecomposition (SVD for short). Despite its simplicity and
aesthetical appeal, the Schmidt decomposition facilitates both the
entanglement analysis and the computation of reduced states\footnote{Formally presented in Section (\ref{par:Reduced-systems}).}
when dealing with such systems. It is also worth mentioning that this
tool is closely related to the concept of state purification.

Consider a pure bipartite quantum state $|\Psi\rangle\in\mathcal{H}^{(0)}=\mathcal{H}^{(1)}\otimes\mathcal{H}^{(2)}$,
such that $dim(\mathcal{H}^{(1,2)})=d^{(1,2)}$ and, without any loss
of generality, $d^{(1)}\leq d^{(2)}$. Given a pair of local orthonormal
basis $\{|b_{j}^{(1,2)}\rangle,j=1,...,d^{(1,2)}\}$ for $\mathcal{H}^{(1,2)}$,
we know that $|\Psi\rangle$ can be written according to Eq. (\ref{Estado bipartido geral}).
Notice that this general expression contains a double sum $\sum_{j=1}^{d^{(1)}}\sum_{k=1}^{d^{(2)}}$,
each one relative to its respective subsystem's Hilbert space. Clearly,
this represents a superposition over all possible combination of basis
elements, where $|a_{jk}|^{2}$ quantify the probability associated
with the ket $|b_{j}^{(1)}\rangle\otimes|b_{k}^{(2)}\rangle$. Alternatively,
the coefficients $a_{jk}$ can be seen as the entries of a $d^{(1)}\times d^{(2)}$
matrix $\overleftrightarrow{A}$, such that $a_{jk}=[\overleftrightarrow{A}]_{jk}$.
The SVD guarantee the following matrix factorization below
\begin{equation}
\overleftrightarrow{A}=\overleftrightarrow{L}\overleftrightarrow{\Lambda}\overleftrightarrow{R}^{\dagger},
\end{equation}
where $\overleftrightarrow{L}$ and $\overleftrightarrow{R}$ are
matrices with orthogonal columns, i.e., $\overleftrightarrow{L}^{\dagger}\overleftrightarrow{L}=\overleftrightarrow{R}^{\dagger}\overleftrightarrow{R}=\overleftrightarrow{1}$,
and $\overleftrightarrow{\Lambda}$ is a diagonal positive semi-definite
matrix. Since we are assuming $d^{(1)}\leq d^{(2)}$, both $\overleftrightarrow{L}$
and $\overleftrightarrow{\Lambda}$ are square $d^{(1)}\times d^{(1)}$
matrices, while $\overleftrightarrow{R}$ is a $d^{(2)}\times d^{(1)}$
one\footnote{If $d^{(1)}\geq d^{(2)}$, their forms would be altered: $\overleftrightarrow{R}$
would be a square matrix while $\overleftrightarrow{L}$ would be
rectangular.}. The diagonal elements $\{\lambda_{j},j=1,...,d^{(1)}\}$ of $\overleftrightarrow{\Lambda}$
are called its singular values, and the number $n\leq d^{(1)}$ of
non-zero $\lambda_{j}$ is known as the Schmidt rank. Thus,
\begin{equation}
a_{jk}=\sum_{\eta=1}^{n}L_{j\eta}\lambda_{\eta}R_{k\eta}^{*},
\end{equation}
and, therefore,
\begin{equation}
|\Psi\rangle=\sum_{\eta=1}^{n}\lambda_{\eta}|\varphi_{\eta}^{(1)}\rangle\otimes|\varphi_{\eta}^{(2)}\rangle,\label{SD}
\end{equation}
where $|\varphi_{\eta}^{(1)}\rangle\equiv\sum_{j=1}^{d^{(1)}}L_{j\eta}|b_{j}^{(1)}\rangle$
and $|\varphi_{\eta}^{(2)}\rangle\equiv\sum_{k=1}^{d^{(2)}}R_{k\eta}^{*}|b_{k}^{(2)}\rangle$.
Additionally, the normality condition of $|\Psi\rangle$ and the $\overleftrightarrow{L}(\overleftrightarrow{R})$
column's orthogonality guarantee that $\sum_{j=1}^{n}\lambda_{j}^{2}=1$
and $\langle\varphi_{m}^{(1,2)}|\varphi_{n}^{(1,2)}\rangle=\delta_{mn}$,
respectively. Eq. (\ref{SD}) is the well known Schmidt decomposition
of $|\Psi\rangle$ in its full glory, where the local kets $\{|\varphi_{j}^{(1)}\rangle,j=1,...,d^{(1)}\}$
and $\{|\varphi_{j}^{(2)}\rangle,j=1,...,d^{(2)}\}$ form an orthonormal
basis for $\mathcal{H}^{(1)}$ and $\mathcal{H}^{(2)}$, called Schmidt
basis, and the singular values $\lambda_{j}$ are also known as the
Schmidt coefficients of $|\Psi\rangle$\footnote{Depending on the reference, sometimes the square of the singular values
are denominated as the Schmidt coefficients.}. These coefficients are unambiguously defined, while the Schmidt
basis are unique up to eventual degenerate coefficients and phases.
Such form is general and far from being intuitive, and, despite the
obvious simplification compared with Eq. (\ref{Estado bipartido geral}),
it has many interesting features that justify its use: notice that
the double sum is changed for a single one bounded by the Schmidt
rank $n\leq d^{(1)}$, which means that regardless the dimension $d^{(2)}$
only an $n$-dimensional subspace of $\mathcal{H}^{(2)}$ is relevant
for the whole state description; besides, it becomes evident that
in order to have entangled bipartite systems, or non-separable $|\Psi\rangle\neq|\varphi^{(1)}\rangle\otimes|\varphi^{(2)}\rangle$,
it is required at least two non-zero Schmidt coefficients, i.e., $|\Psi\rangle$
is an entangled state iff $n>$1; plus, it gives all necessary information
for writing the reduced states (as discussed in Section (\ref{par:Reduced-systems})).

\subsubsection{Density operator}

The framework presented until now was built to describe pure states
$|\Psi\rangle\in\mathcal{H}$. As mentioned earlier, every physical
ket encodes the probabilities associated with measurements performed
on a chosen basis. Such statistics, however, is only observed at the
ensemble level, i.e., after the access of a collection of identical
copies of $|\Psi\rangle$. Although feasible, it is clear that this
is a particular case of a more general ensemble. In most realistic
scenarios, one has access to a statistical mixture of $N$ arbitrary
states $\{P_{\eta},|\Psi_{\eta}\rangle\in\mathcal{H},\eta=1,...,N\}$,
where $\{P_{\eta}\}\geq0$ with $\sum_{\eta=1}^{N}P_{\eta}=1$ characterizes
the distribution of pure states $|\Psi_{\eta}\rangle$. The mathematical
entity representing quantum ensembles is known as density operators
or density matrices and written as
\begin{equation}
\hat{\rho}\equiv\sum_{\eta=1}^{N}P_{\eta}|\Psi_{\eta}\rangle\langle\Psi_{\eta}|\in\mathcal{L}(\mathcal{H}).\label{eq:density operator}
\end{equation}
If $P_{j}=1$ and $P_{\eta}=0$ for all $\eta\neq j$, then the quantum
state $\hat{\rho}=|\Psi_{j}\rangle\langle\Psi_{j}|$ is called pure.
Otherwise, it is referred to as a mixed state. Eq. (\ref{eq:density operator})
is the most general representation of a quantum system. In fact, the
whole formalism of QM can be stated in terms of density operators
$\hat{\rho}$. Notice that, instead of being represented by kets,
states are depicted by operators according to this framework. Also,
in contrast with the probabilities mentioned in the context of pure
ensembles, the distribution of states $\{P_{\eta}\}$ is fully classical,
in the sense that it only captures our ignorance concerning the preparation
- or access - of possible quantum states. It is worth mentioning the
pure states within the set $\{|\Psi_{\eta}\rangle\}$ are not necessarily
orthogonal. Nevertheless, since $\hat{\rho}$ is clearly hermitian,
one can write it conforming to its spectral decomposition, such that
\begin{equation}
\hat{\rho}=\sum_{j=1}^{d}p_{j}|\varphi_{j}\rangle\langle\varphi_{j}|,\label{eq:density matrix spectral decomposition}
\end{equation}
where $\{|\varphi_{j}\rangle,j=1,...,d\}$ is a basis of $\mathcal{H}$
and $\{p_{j}\}\geq0$ with $\sum_{j}^{d}p_{j}=1$. Along these lines,
it is clear that density matrices correspond to a particular class
of operators within $\mathcal{L}(\mathcal{H})$, satisfying some properties
to characterize physical systems. The corresponding density matrices'
subset will be represented by $\mathscr{D}(\mathcal{H})$. Mathematically,
any $\hat{\rho}\in\mathscr{D}(\mathcal{H})$ should meet the following
conditions:
\begin{enumerate}
\item \textbf{Hermitian:} $\hat{\rho}=\hat{\rho}^{\dagger}$;
\item \textbf{Unit trace:} $tr\{\hat{\rho}\}=1$\footnote{The trace $tr\{\hat{O}\}$ of any operator $\hat{O}$ is simply the
sum of its diagonal elements. Since this quantity is independent of
the representation choice, one can use any basis to compute it.};
\item \textbf{Positive semi-definite:} $\hat{\rho}\geq0$ or $\{p_{j}\}\geq0$.
\end{enumerate}
While condition \textbf{1.} implies real eigenvalues, condition \textbf{2.}
states the normalization of $\hat{\rho}$ or $\sum_{j=1}^{d}p_{j}=1$
and condition \textbf{3.} guarantee positive probabilities, $p_{j}\geq0$
for all $j$. Besides, every density matrix $\hat{\rho}$ satisfies
the following inequality
\begin{equation}
tr\{\hat{\rho}^{2}\}\leq1.\label{eq:purity}
\end{equation}
Interestingly, by computing Eq. (\ref{eq:purity}), one can easily
verify if the state $\hat{\rho}$ is pure or mixed. Notice that $tr\{\hat{\rho}^{2}\}=1$
iff $\hat{\rho}=|\Psi\rangle\langle\Psi|$ is pure. Otherwise, it
is mixed. For this reason, Eq. (\ref{eq:purity}) is commonly known
as state purity.

\paragraph{Projective measurements}

Concerning measurements and observables. Let us suppose, again, the
observable $\hat{B}=\sum_{j=1}^{d}b_{j}|b_{j}\rangle\langle b_{j}|$.
If a projective measurement is performed in an ensemble characterized
by Eq. (\ref{eq:density matrix spectral decomposition}), the probability
of obtaining $b_{k}$ is
\begin{equation}
Prob(b_{k})\equiv tr\{\hat{\Pi}_{k}\hat{\rho}\}=\sum_{j=1}^{d}p_{j}|\langle b_{k}|\varphi_{j}\rangle|^{2}\geq0,
\end{equation}
which is, essentially, a sum over the probability of observing $b_{k}$
conditioned by each possible element of the mixture $\{p_{j},|\varphi_{j}\rangle,j=1,...,d\}$.
After the measurement, as expected, the new state is pure and given
by 
\begin{equation}
\hat{\rho}^{\prime}\equiv\frac{\hat{\Pi}_{k}\hat{\rho}\hat{\Pi}_{k}^{\dagger}}{tr\{\hat{\Pi}_{k}\hat{\rho}\}}=|b_{k}\rangle\langle b_{k}|.
\end{equation}

From the expressions above, the average value of these measurements,
$\langle\hat{B}\rangle$, can be easily computed by $\sum_{k=1}^{d}b_{k}Prob(b_{k})$.
It is straightforward to check that $\langle\hat{B}\rangle=tr\{\hat{B}\hat{\rho}\}.$
This previous relation can be generalized to any observable $\hat{O}\in\mathcal{L}(\mathcal{H})$,
such that its expectation value $\langle\hat{O}\rangle$ relative to
the ensemble $\hat{\rho}$ is defined as\footnote{Due to the cyclic property of the trace, the equality $tr\{\hat{A}\hat{B}\}=tr\{\hat{B}\hat{A}\}$
is satisfied for any pair of operators $\hat{A}$ and $\hat{B}$.}
\begin{equation}
\langle\hat{O}\rangle\equiv tr\{\hat{O}\hat{\rho}\}=tr\{\hat{\rho}\hat{O}\}.
\end{equation}

\paragraph{Reduced states\label{par:Reduced-systems}}

If the quantum system of interest is composed of different subsystems,
the density operator formalism provide a straightforward manner to
describe them. Often, one has access only to a part of a much broader
quantum system, in a way that it is impossible to obtain the whole
state description. In this sense, the reduced density operators are
the representative ones for describing local states.

Let us suppose a bipartite system, represented by $\hat{\rho}^{(0)}\in\mathscr{D}(\mathcal{H}^{(0)})$
and $\mathcal{H}^{(0)}=\mathcal{H}^{(1)}\otimes\mathcal{H}^{(2)}$.
The reduced density operators are defined by 
\begin{align}
\hat{\rho}^{(1)} & \equiv tr_{2}\{\hat{\rho}^{(0)}\}\in\mathscr{D}(\mathcal{H}^{(1)}),\\
\hat{\rho}^{(2)} & \equiv tr_{1}\{\hat{\rho}^{(0)}\}\in\mathscr{D}(\mathcal{H}^{(2)}),
\end{align}
where $tr_{k}\{(.)\}$ is the partial trace over subspace $k$. Of
course, these quantities are genuine density operators and, therefore,
inherit all desired properties. In this sense, the statistics of local
observables agrees with the expected behaviour, i.e., suppose one
perform measurements of an observable $\hat{O}^{(1)}$ in the subsystem
$(1)$, then
\begin{equation}
\langle\hat{O}^{(1)}\rangle\equiv tr\left\{ \left(\hat{O}^{(1)}\otimes\hat{1}^{(2)}\right)\hat{\rho}^{(0)}\right\} =tr_{1}\left\{ \hat{O}^{(1)}\hat{\rho}^{(1)}\right\} .
\end{equation}

If the whole system is described by $\hat{\rho}^{(0)}=\hat{\rho}^{(1)}\otimes\hat{\rho}^{(2)}$,
the state is separable, such that the reduced density matrices are
clearly $\hat{\rho}^{(1,2)}=tr_{2,1}\{\hat{\rho}^{(0)}\}$. In these
cases, access to the local descriptions is enough to characterize
the whole system. Nevertheless, if $\hat{\rho}^{(0)}\neq\hat{\rho}^{(1)}\otimes\hat{\rho}^{(2)}$,
then the subsystems are somehow correlated, in a way that even knowing
both individual local states, $\hat{\rho}^{(1,2)}$, one still cannot
infer the whole state $\hat{\rho}^{(0)}$.

Let us now consider the particular case in which the whole bipartite
system is described by a pure state, such that $\hat{\rho}^{(0)}=|\Psi\rangle\langle\Psi|$
and, without any loss of generality, $d^{(1)}\leq d^{(2)}$. Thus,
there is no classical uncertainty concerning the ensemble. Interestingly,
as will be shown below, this is not necessarily true for the subsystems.
As we saw, in general, any ket $|\Psi\rangle$ can be written as Eq.
(\ref{Estado bipartido geral}). However, the Schmidt decomposition
presented in Eq. (\ref{SD}) provides a more convenient form. Along
these lines, the whole state can be cast in the following form
\begin{equation}
\hat{\rho}^{(0)}=\sum_{\eta,\alpha=1}^{n}\lambda_{\eta}\lambda_{\alpha}|\varphi_{\eta}^{(1)}\rangle\langle\varphi_{\alpha}^{(1)}|\otimes|\varphi_{\eta}^{(2)}\rangle\langle\varphi_{\alpha}^{(2)}|,
\end{equation}
such that the local states can be immediately computed,
\begin{align}
\hat{\rho}^{(1)} & =\sum_{\eta=1}^{n}\lambda_{\eta}^{2}|\varphi_{\eta}^{(1)}\rangle\langle\varphi_{\eta}^{(1)}|,\\
\hat{\rho}^{(2)} & =\sum_{\eta=1}^{n}\lambda_{\eta}^{2}|\varphi_{\eta}^{(2)}\rangle\langle\varphi_{\eta}^{(2)}|.
\end{align}
Hence, as long the whole system is pure, its Schmidt decomposition
gives all necessary information for inferring the local spectral decompositions
above, i.e., the square of the Schmidt coefficients, $\{\lambda_{\eta}^{2}(t),\eta=1,...,n\}$,
characterize the distributions across the pure states given by the
Schmidt basis, $\{|\varphi_{\eta}^{(1,2)}(t)\rangle,\eta=1,...,n\}$.
Notice that, despite having no uncertainty from a global perspective,
the local subsystems are represented by mixed states with identical
probabilities. In particular, since by hypotheses, $d^{(2)}\geq d^{(1)}$
the subsystem $(2)$ must necessarily contain $(d^{(2)}-n)$ null
eigenvalues. Interestingly, this is true for any conceivable bipartite
system, regardless of the dimensions in question. Such a non-intuitive
result is a direct consequence of entanglement: as mentioned earlier
in Section (\ref{par:The-Schmidt-decomposition}), for $|\Psi\rangle$
to be entangled it is required at least two non-zero Schmidt coefficients,
i.e., the Schmidt rank $n>1$. In contrast, if $n=1$ such that $\lambda_{\eta}=\delta_{k\eta}$,
then $\hat{\rho}^{(1,2)}=|\varphi_{k}^{(1,2)}\rangle\langle\varphi_{k}^{(1,2)}|,$
and the whole system is clearly separable,
\begin{equation}
|\Psi\rangle=|\varphi_{k}^{(1)}\rangle\otimes|\varphi_{k}^{(2)}\rangle\rightarrow\hat{\rho}^{(0)}=|\varphi_{k}^{(1)}\rangle\langle\varphi_{k}^{(1)}|\otimes|\varphi_{k}^{(2)}\rangle\langle\varphi_{k}^{(2)}|.
\end{equation}

It is worth mentioning that the reasoning above is also commonly explored
the other way around. Given a quantum system $(1)$ described by $\hat{\rho}^{(1)}\in\mathscr{D}(\mathcal{H}^{(1)})$,
by adding an auxiliary system $(2)$, the former can be seen as a
subsystem of a bigger pure one depicted by $|\Psi\rangle$, i.e.,
$\hat{\rho}^{(1)}=tr_{2}\{|\Psi\rangle\langle\Psi|\}$. This procedure
is known as purification of $\hat{\rho}^{(1)}$.

%% file: Chapters/Chapter2/Subsection_2_2.tex
\subsection{Quantum dynamics}

In the previous subsection, we presented the mathematical tools and formalism
for describing quantum systems. Let us now briefly discuss how to
express their time evolution considering the different possible contexts.

\subsubsection{Isolated and closed system dynamics}

The most simple dynamical scenario consists of \textit{isolated} and
\textit{closed} quantum systems. By definition, isolated systems do
not interact - in any way - with any other kind of object, classical
or quantum. In these cases, the described system is fully quantized,
and no other external entity is included - implicitly or explicitly
- in the picture, i.e., the system in question represents the totality
of elements in the context under scrutiny. Consequently, an isolated
quantum system evolves \textit{autonomously} in time, uninfluenced
by anything else outside its own existence. Of course, the whole system
itself may be constituted by individual interacting parts. In contrast,
closed quantum systems represent less restrictive scenarios. It is
allowed the presence of external interaction with classical agents,
which is often justified by semi-classical reasoning. In these cases,
the considered influences are implicit, and the totality of elements
is not fully quantized. In general, such situations portray quantum
systems evolving in time according to externally controlled (deterministic)
fields driving its dynamics. 

Mathematically, both cases are differentiated by the eventual time-dependency
of their time-translation generator. On the one hand, isolated systems
have fixed Hamiltonians. On the other hand, the external interaction
considered in closed systems induces time-dependent ones. Interestingly,
despite the fundamental - and subtle - differences, their dynamic
behaviour is described by the same functional form. 

Thus, let us proceed considering a closed quantum system at time $t$
represented by $|\Psi(t)\rangle\in\mathcal{H}$. In general, given
the initial state $|\Psi(t_{0})\rangle$, we are interested in describing
$|\Psi(t)\rangle$ at any subsequent time $t\geq t_{0}$. Formally,
this task is performed by the so-called time-evolution operator $\hat{\mathcal{U}}(t,t_{0})\in\mathcal{L}(\mathcal{H})$,
such that
\begin{equation}
|\Psi(t)\rangle=\hat{\mathcal{U}}(t,t_{0})|\Psi(t_{0})\rangle.\label{eq:ket time evolution}
\end{equation}
Essentially, $\hat{\mathcal{U}}(t,t_{0})$ is the linear operator
that maps any initial physical ket to a time-evolved one. Under the
general requirements of $\underset{t\rightarrow t_{0}}{lim}\,\hat{\mathcal{U}}(t,t_{0})=\hat{1}$,
the state normalization $\langle\Psi(t)|\Psi(t)\rangle=1$ for all
$t$ and map composition $\hat{\mathcal{U}}(t_{2},t_{0})=\hat{\mathcal{U}}(t_{2},t_{1})\hat{\mathcal{U}}(t_{1},t_{0})$,
one can show that $\hat{\mathcal{U}}(t,t_{0})$ must be unitary, $\hat{\mathcal{U}}(t,t_{0})\hat{\mathcal{U}}^{\dagger}(t,t_{0})=\hat{1}$,
and satisfies the following differential equation
\begin{equation}
i\hbar\frac{d}{dt}\hat{\mathcal{U}}(t,t_{0})=\hat{H}(t)\hat{\mathcal{U}}(t,t_{0}),\label{eq:eq diff time evolution operator}
\end{equation}
where $\hbar$ is the Planck's constant, and $\hat{H}(t)=\hat{H}^{\dagger}(t)\in\mathcal{L}(\mathcal{H})$
is the - possibly - time-dependent Hamiltonian and the time-translation
generator of this dynamics. The general solution of Eq. (\ref{eq:eq diff time evolution operator})
above is given by
\begin{equation}
\hat{\mathcal{U}}(t,t_{0})=\overleftarrow{\mathcal{T}}e^{-\frac{i}{\hbar}\int_{t_{0}}^{t}ds\,\hat{H}(s)}
\end{equation}
where $\overleftarrow{\mathcal{T}}$ is the chronological time-ordering
operator. Of course, as mentioned earlier, if the system is isolated
then $\hat{H}$ is constant, and the time-evolution operator is considerably
simpler:
\begin{equation}
\hat{\mathcal{U}}(t,t_{0})=e^{-\frac{i}{\hbar}\hat{H}(t-t_{0})}.
\end{equation}

Finally, combining Eqs. (\ref{eq:ket time evolution}, \ref{eq:eq diff time evolution operator})
one obtains the usual \textit{Schrödinger equation} for describing
the state's dynamics,
\begin{equation}
i\hbar\frac{d}{dt}|\Psi(t)\rangle=\hat{H}(t)|\Psi(t)\rangle.\label{eq:Scho eq}
\end{equation}

Alternatively, if the system of interest is initially described by
a mixed state $\hat{\rho}(t_{0})=\sum_{j=1}^{d}p_{j}|\varphi_{j}(t_{0})\rangle\langle\varphi_{j}(t_{0})|\in\mathscr{D}(\mathcal{H})$,
each element of the ensemble will certainly time evolve according
to Eq. (\ref{eq:Scho eq}) above, such that $|\varphi_{j}(t)\rangle=\hat{\mathcal{U}}(t,t_{0})|\varphi_{j}(t_{0})\rangle$.
Thus, at any later time $t\geq t_{0}$, the density matrix is given
by
\begin{equation}
\hat{\rho}(t)\equiv\hat{\Phi}_{t,t_{0}}\hat{\rho}(t_{0})=\hat{\mathcal{U}}(t,t_{0})\hat{\rho}(t_{0})\hat{\mathcal{U}}^{\dagger}(t,t_{0}),\label{eq:unitary map}
\end{equation}
where 
\begin{equation}
\hat{\Phi}_{t,t_{0}}:\mathscr{D}(\mathcal{H})\rightarrow\mathscr{D}(\mathcal{H})
\end{equation}
is the so-called (linear) dynamical map representing the unitary time
evolution\footnote{Notice that $\hat{\Phi}_{t,t_{0}}$ acts on an operator instead of
a ket. For this reason, these linear operators are commonly known
as \textit{superoperators}.}. Along these lines, it is straightforward to obtain the density matrix
counterpart of Eq. (\ref{eq:Scho eq}) for the equation of motion,
given by the so-called \textit{Liouville-von Neumann equation},
\begin{equation}
i\hbar\frac{d}{dt}\hat{\rho}(t)=[\hat{H}(t),\hat{\rho}(t)].\label{eq:liou vN equation}
\end{equation}
It is important to highlight that the eigenvalues $\{p_{j},j=1,...,d\}$
of $\hat{\rho}(t)$ remain invariant for this class of time evolution.

In short, the whole dynamics of both isolated and closed systems are
fully characterized once having access to initial conditions and the
unitaries $\hat{\mathcal{U}}(t,t_{0})$, written as functionals of
the Hamiltonians, mapping states at different instants of time.

\subsubsection{Open systems dynamics}\label{subsec:Open-systems-dynamics}

This class of dynamics is fundamentally distinct from the isolated
and closed ones. As the name suggests, open quantum systems consist
of physical systems that explicitly interact with other - possibly
many - quantum objects. Differently from the other cases, as a consequence
of the interaction, the dynamics of open systems are unavoidably coupled
and correlated to their environments, which directly affect their
time evolutions and lead to the phenomena of decoherence and dissipation.
More specifically, their equation of motion is no longer unitary nor
described by Eq. (\ref{eq:liou vN equation}) in such a way that a
more general description formalism is required. Along these lines,
an individual subsystem of a larger entity is - in general - treated
as an open quantum system. As discussed earlier in Section (\ref{par:Reduced-systems}),
subsystems are characterized by reduced states. Therefore, essentially,
we are interested in finding their dynamical equations. Solving this
task is the most fundamental objective of the subfield known as open
quantum systems and is of imperative importance for understanding
the complex behaviour of quantum matter and the development of future
quantum devices. It is worth mentioning that open systems represent
the standard approach for many realistic scenarios: the intrinsic
fragility of quantum systems means that they are easily disturbed
by external influences. Nevertheless, if the interactions are sufficiently
weak, the system might be treated as approximately isolated. Let us
focus on the general aspects of open system dynamics.

Suppose the system of interest is described, at any time $t$, by
$\hat{\rho}^{(1)}(t)\in\mathscr{D}(\mathcal{H}^{(1)})$ and corresponds
to a subsystem of a larger isolated quantum system depicted by $\hat{\rho}^{(0)}\in\mathscr{D}(\mathcal{H}^{(0)})$,
such that $\mathcal{H}^{(0)}=\mathcal{H}^{(1)}\otimes\mathcal{H}^{(2)}$,
where $\mathcal{H}^{(2)}$ is the Hilbert space that encompasses the
environment of $(1)$. Notice this configuration establishes a bipartition
between the described system and \textquotedbl everything else\textquotedbl .
Since the total system is isolated - by hypotheses - given an initial
state $\hat{\rho}^{(0)}(t_{0})$, its time evolution is guaranteed
to be unitary, such that 
\begin{equation}
\hat{\rho}^{(0)}(t)=\hat{\mathcal{U}}^{(0)}(t,t_{0})\hat{\rho}^{(0)}(t_{0})\hat{\mathcal{U}}^{(0)\dagger}(t,t_{0}),
\end{equation}
where $\hat{\mathcal{U}}^{(0)}(t,t_{0})=e^{-\frac{i}{\hbar}\hat{H}^{(0)}(t-t_{0})}\in\mathcal{L}(\mathcal{H}^{(0)})$
is its time evolution operator, and $\hat{H}^{(0)}\in\mathcal{L}(\mathcal{H}^{(0)})$
is the Hamiltonian relative to the whole bipartition. This Hamiltonian
have the following structure
\begin{equation}
\hat{H}^{(0)}:=\hat{H}^{(1)}\otimes\hat{1}^{(2)}+\hat{1}^{(1)}\otimes\hat{H}^{(2)}+\hat{H}_{int},\label{eq:total hamiltonian}
\end{equation}
where $\hat{H}^{(1,2)}\in\mathcal{L}(\mathcal{H}^{(1,2)})$ are the
bare Hamiltonians for each subsystem, and $\hat{H}_{int}\in\mathcal{L}(\mathcal{H}^{(0)})$
is the term representing all the physical interactions between $(1)$
and $(2)$. It is important to highlight the fact the bare Hamiltonians
are local operators restricted to their individual subspaces while
the interaction is clearly non-local, in the sense that it acts on
the total Hilbert space.

According to Section (\ref{par:Reduced-systems}), at any time $t$,
the local states are simply
\begin{equation}
\hat{\rho}^{(1,2)}(t)=tr_{2,1}\{\hat{\mathcal{U}}^{(0)}(t,t_{0})\hat{\rho}^{(0)}(t_{0})\hat{\mathcal{U}}^{(0)\dagger}(t,t_{0})\}.\label{eq: time evolved subsystem}
\end{equation}
Thus, using Eqs. (\ref{eq:liou vN equation}, \ref{eq:total hamiltonian},
\ref{eq: time evolved subsystem}), it is straightforward to show
that their exact dynamical equations are
\begin{equation}
i\hbar\frac{d}{dt}\hat{\rho}^{(1,2)}(t)=tr_{2,1}\{[\hat{H}^{(0)},\hat{\rho}^{(0)}]\}=[\hat{H}^{(1,2)},\hat{\rho}^{(1,2)}(t)]+tr_{2,1}\{[\hat{H}_{int},\hat{\rho}^{(0)}]\},
\end{equation}
where the commutator $[\hat{H}^{(1,2)},\hat{\rho}^{(1,2)}(t)]$ resembles
the contribution that appears in Eq. (\ref{eq:liou vN equation})
and $tr_{2,1}\{[\hat{H}_{int},\hat{\rho}^{(0)}]\}$ is an extra component
due to the interaction term, $\hat{H}_{int}$. While the former is
the unitary part of the dynamics, the latter contains non-unitary
contributions\footnote{This operator might contain terms of the following form $[(.),\hat{\rho}^{(1,2)}(t)]$
as well. The remaining part, however, is non-unitary.}. It is clear that the role played by the interaction is of extreme
relevance to the local dynamics. In fact, one can easily check that
its absence would imply that both subsystems would be individually
isolated and, therefore, evolving in time independently, i.e., $\hat{H}^{(0)}=\hat{H}^{(1)}\otimes\hat{1}^{(2)}+\hat{1}^{(1)}\otimes\hat{H}^{(2)}$
and $i\hbar\frac{d}{dt}\hat{\rho}^{(1,2)}(t)=[\hat{H}^{(1,2)},\hat{\rho}^{(1,2)}(t)]$.
Also, from a global perspective, the interaction is responsible for
inducing the formation of both classical and quantum correlations
within the system in a way that $\hat{\rho}^{(0)}(t)\neq\hat{\rho}^{(1)}(t)\otimes\hat{\rho}^{(2)}(t)$
for general situations. Of course, the resulting local dynamics of
subsystem $(1)$ fundamentally depends on the nature of $(2)$ that,
in principle, could be as simple as a single qubit or a complex many-body
system. Nevertheless, in many scenarios involving open quantum systems,
it is commonly assumed that $(2)$ is a reservoir with an infinite
number of degrees of freedom or a large heat bath. These situations,
as expected, might be very complicated to be analysed by the exact
expressions above or even treated numerically. For this reason, the
usual approaches and techniques rely on approximative methods and
different hypotheses aiming to simplify the description, which also
restrains the analysis for specific regimes of validity. Along these
lines, initial uncorrelated states, weak-coupling and Markovian dynamics
represent the most common assumptions leading to the orthodox microscopic
derivations of the phenomenological master equations.

As we can see, the interaction breaks the unitary mapping given by
Eq. (\ref{eq:unitary map}). This relationship, however, might be
generalized by defining the dynamical map $\hat{\Lambda}_{t,t_{0}}^{(1)}:\mathscr{D}(\mathcal{H}^{(1)})\rightarrow\mathscr{D}(\mathcal{H}^{(1)})$,
such that
\begin{equation}
\hat{\rho}^{(1)}(t)\equiv\hat{\Lambda}_{t,t_{0}}^{(1)}\hat{\rho}^{(1)}(t_{0})=tr_{2}\{\hat{\mathcal{U}}^{(0)}(t,t_{0})\hat{\rho}^{(0)}(t_{0})\hat{\mathcal{U}}^{(0)\dagger}(t,t_{0})\}.\label{eq:dynamical map}
\end{equation}
Any physical dynamical map of this kind should be completely positive
and trace-preserving, also known as a CPTP-map, i.e.,
\begin{enumerate}
\item \textbf{Completely positive:} $\hat{\Lambda}_{t,t_{0}}^{(1)}\hat{\rho}^{(1)}(t_{0})\geq0$;
\item \textbf{Trace-preserving:} $tr\{\hat{\Lambda}_{t,t_{0}}^{(1)}\hat{\rho}^{(1)}(t_{0})\}=tr\{\hat{\rho}^{(1)}(t_{0})\}=1$.
\end{enumerate}
Similarly to the time evolution operator when considering physical
kets, these conditions are the most basic requirements that guarantee
a linear mapping between proper density matrices. Thus, the whole
dynamics is fully characterized by a set of CPTP dynamical maps, $\{\hat{\Lambda}_{t,t_{0}}^{(1)},t\geq t_{0}\}$,
describing the density matrix throughout the time evolution. In some
scenarios, the set of dynamical maps satisfies the so-called semigroup
property $\hat{\Lambda}_{t_{2},t_{0}}^{(1)}=\hat{\Lambda}_{t_{2},t_{1}}^{(1)}\hat{\Lambda}_{t_{1},t_{0}}^{(1)}$
for $t_{2}\geq t_{1}\geq t_{0}$ which also implies the usual Lindblad-like
form for master equations.

For instance, let us consider initially uncorrelated systems, such that the whole
bipartite system is depicted by a product state $\hat{\rho}^{(0)}(t_{0})=\hat{\rho}^{(1)}(t_{0})\otimes\hat{\rho}^{(2)}(t_{0})$.
If the spectral decomposition of subsystem $(2)$ at time $t_{0}$
is given by
\begin{equation}
\hat{\rho}^{(2)}(t_{0})=\sum_{\alpha=1}^{d^{(2)}}p_{\alpha}(t_{0})|\varphi_{\alpha}^{(2)}(t_{0})\rangle\langle\varphi_{\alpha}^{(2)}(t_{0})|,
\end{equation}
then Eq. (\ref{eq:dynamical map}) becomes
\begin{equation}
\hat{\rho}^{(1)}(t)=\hat{\Lambda}_{t,t_{0}}^{(1)}\hat{\rho}^{(1)}(t_{0})=\sum_{\alpha=1}^{d^{(2)}}\sum_{\beta=1}^{d^{(2)}}\hat{K}_{\beta\alpha}^{(1)}(t,t_{0})\hat{\rho}^{(1)}(t_{0})\hat{K}_{\beta\alpha}^{(1)\dagger}(t,t_{0}),\label{eq: Krauss}
\end{equation}
where 
\begin{equation}
\hat{K}_{\beta\alpha}^{(1)}(t,t_{0})\equiv\sqrt{p_{\alpha}(t_{0})}\langle\varphi_{\beta}^{(2)}(t_{0})|\hat{\mathcal{U}}^{(0)}(t,t_{0})|\varphi_{\alpha}^{(2)}(t_{0})\rangle\in\mathcal{L}(\mathcal{H}^{(1)})
\end{equation}
 and $\sum_{\alpha,\beta=1}^{d^{(2)}}\hat{K}_{\beta\alpha}^{(1)\dagger}(t,t_{0})\hat{K}_{\beta\alpha}^{(1)}(t,t_{0})=\hat{1}^{(1)}.$
The form of Eq. (\ref{eq: Krauss}) is known as the \textit{Kraus
representation} - or operator sum representation - and is the general
structure of a dynamical map, where the operators $\{\hat{K}_{\beta\alpha}^{(1)}(t,t_{0}),\beta,\alpha=1,...,d^{(2)}\}$
are the \textit{Kraus operators}. In particular, if there is only
one Kraus operator $\hat{K}_{j}^{(1)}(t,t_{0})$, then the map simplifies
to $\hat{\Lambda}_{t,t_{0}}^{(1)}\hat{\rho}^{(1)}(t_{0})=\hat{K}_{j}^{(1)}(t,t_{0})\hat{\rho}^{(1)}(t_{0})\hat{K}_{j}^{(1)\dagger}(t,t_{0})$
and the time evolution is unitary, since $\hat{K}_{j}^{(1)\dagger}(t,t_{0})\hat{K}_{j}^{(1)}(t,t_{0})=\hat{1}^{(1)}$.

In short, the dynamics of open systems are considerably more complicated
than the isolated and closed ones and require a more general approach
for treating them. As mentioned earlier, in realistic scenarios, any
object is unavoidable coupled to its external environment. Its interactions
often lead to an intricate non-unitary time evolution that is not
easily described by analytic means. In this sense, all the formalism
and approximations developed within the context of open quantum systems
are useful tools for understanding the behaviour of interacting systems.

%% file: Chapters/Chapter2/Section3.tex
\section{Quantum thermodynamics\label{sec:Quantum-thermodynamics}}

Given the previous introductions from Sections (\ref{sec:Thermodynamics}) and (\ref{sec:Quantum-mechanics}),
it is clear that thermodynamics and quantum mechanics are both successful
and well established physical theories developed and applied to different
aspects of natural phenomena. However, it is not a priori obvious
to understand how they connect. The research program coined as \textit{quantum
thermodynamics} (QT) is relatively young and has been emerging as
a meaningful scientific field during the last couple of decades. Its
name is self-explanatory\footnote{Interestingly, this is not the case for many things in science.}
since QT has the ambition to develop a consistent theory that both
extends and applies thermodynamics for general, and possibly nonequilibrium,
quantum systems and to describe the classical thermodynamic behaviour
from the underlying fundamental quantum dynamics. Both major goals
are clearly complementary and represent different valid perspectives
on the same subject, which partially explains the diverse community
forming around QT and the plethora of approaches and ideas coming
from distinct - but closely related - areas, especially from quantum
information theory and open quantum systems. For a general introduction to this topic, see. \cite{vinjanampathy2016quantum,millen2016perspective,anders2017focus,alicki2018introduction,auffeves2021short} A more detailed discussion on QT can be found in. \cite{gemmer2009quantum,binder2018thermodynamics,deffner2019quantum}

Although the recent impressive booming of the field, it is interesting
to notice that, historically, thermodynamics has always been assisting
quantum mechanics. In fact, it was present at the birth of quantum
theory \cite{planck1900theorie,einstein1905erzeugung} and, during the last century, supported some crucial
technological advances, in particular in the development of the lasers
and masers. \cite{einstein1916strahlungs,scovil1959three,geusic1959three,geusic1967quantum} However, despite being centenary theories, it
is no surprise, nor coincidence, that questions concerning the thermodynamics
of quantum systems have been rapidly coming to light during present
days: On the one hand, since the early 90s seminal results on nonequilibrium
thermodynamics \cite{evans1993probability,evans1994equilibrium,gallavotti1995dynamical,jarzynski1997nonequilibrium,crooks1998nonequilibrium}, the field of stochastic thermodynamics
flourished and has been successfully bridging the gap between the
understanding of macroscopic and microscopic thermodynamic processes
\cite{seifert2008stochastic,seifert2012stochastic}; On the other hand, the current state-of-the-art technology
allows the precise fabrication, control and measurement of truly quantum
objects in a wide variety of platforms, ranging from solid-state systems
to trapped ions and optical setups. In this context, questions concerning
whether or not it is possible to expand the laws of thermodynamics
to even smaller - and possibly quantum - systems is both a logical
step and a current technological urgency. In fact, along with quantum
computation and quantum information, QT is an important player in
the contemporary technological revolution, the so-called \textit{quantum
technologies 2.0}. Its primary purpose is to design quantum devices
capable of intentionally harvesting quantum properties, such as coherence
and entanglement, to outperform their classical counterparts and -
hopefully - execute functions that no classical machines could do,
even in principle. At the core of these technologies, it is necessary
to understand the underlying interplay between work, heat and information.
On this wise, QT provides the most natural framework to deal with
the design and manipulation of quantum heat engines \cite{scovil1959three,geusic1967quantum,Alicki19792,kosloff1984quantum,bender2000quantum,feldmann2004characteristics,quan2007quantum,zhang2007four,quan2009quantum,abah2012single,rahav2012heat,goswami2013thermodynamics,kosloff2017quantum,niedenzu2018quantum}, which is a fundamental
step toward the development of efficient and stable functioning quantum
machines. However, it is important to emphasize that, to this date,
it is still not clear to what extent such kinds of devices would be
feasible in practical scenarios. Physicists are optimistic, though. Currently, QT investigations
are already accessible and performed in the traditional experimental
setups for quantum mechanics studies, e.g., superconducting devices
\cite{martinez2014coherent,koski2015chip,solinas2016microwave,cottet2017observing,naghiloo2018information,zhang2018experimental,gardas2018quantum,buffoni2020thermodynamics,karimi2020reaching}, nitrogen-vacancy (NV)
centers \cite{klatzow2019experimental,hernandez2020experimental}, nuclear magnetic resonance (NMR) \cite{batalhao2014experimental,peterson2019experimental,de2019efficiency,mendoncca2020reservoir}, trapped
ions \cite{an2015experimental,rossnagel2016single,von2019spin,maslennikov2019quantum}, ultracold atoms \cite{brantut2013thermoelectric,cerisola2017using,bouton2021quantum} and others. In this sense, much experimental research in QT is driven by questions concerning cyclic processes, quantum heat engines and quantum refrigerators. \cite{rossnagel2016single,klatzow2019experimental,peterson2019experimental,de2019efficiency,von2019spin,maslennikov2019quantum,elouard2020interaction} For more details
and further references, see. \cite{myers2022quantum} Additionally, from a theoretical
perspective, QT has the potential to shed some light on fundamental
questions on the foundations of quantum mechanics and statistical
physics, especially those concerning the emergence of macroscopic
behaviour from first principles, e.g., the transition from quantum
to classical mechanics, irreversibility, thermalisation and the measurement
problem.

As briefly mentioned earlier, QT has been approached from many different
perspectives and attitudes, each one tackling particular problems
with its own set of tools. This diversity gave rise to smaller communities
and several research branches within QT. Along these lines, in addition
to studies with quantum heat engines, one can find investigations
on quantum thermometry \cite{correa2015individual,pasquale2018quantum,campbell2018precision,mehboudi2019thermometry,potts2019fundamental,rubio2021global,mok2021optimal}, quantum batteries \cite{allahverdyan2004maximal,hovhannisyan2013entanglement,alicki2013entanglement,campaioli2017enhancing,andolina2019extractable}, stochastic
quantum thermodynamics \cite{Campisi2011a,Elouard2016b,elouard2018efficient,manzano2018quantum,strasberg2019operational,strasberg2019repeated} and thermodynamics of quantum information
\cite{hilt2011landauer,deffner2013information,goold2015nonequilibrium,peterson2016experimental,cottet2017observing,campbell2017nonequilibrium}, just to name a few. All these subareas have been reporting
interesting and relevant theoretical progress, although not in a cohesive
manner yet. In particular, pertinent developments are being achieved
by resource theoretic notions imported from quantum information theory
and applied to QT. \cite{brandao2013resource,skrzypczyk2014work,horodecki2013fundamental,brandao2015second,gour2015resource,lostaglio2015description} Resource theories are robust axiomatic
and operational mathematical descriptions of the so-called \textit{resources}
and \textit{free operations}: while the former refers to desired -
and possibly scarce (just like any resource) - physical properties,
like entanglement and coherence, the latter refers to the set of feasible
and accessible operations. Interestingly, using this framework, one
can derive general results concerning possible state transformation
under a given set of constraints. At present, there is a plethora
of different resource theories, each one dealing with its particular
resources and set of operational restrictions. \cite{streltsov2017colloquium,chitambar2019quantum} In the thermodynamic
context, nonequilibrium states are seen as resources, while free operations
include the addition of thermal baths and global unitaries preserving
energy. \cite{ng2018resource,lostaglio2019introductory} It is interesting to appreciate how similar such
reasoning is compared with the usual phenomenological approach to
thermodynamics. 

Alternatively, instead of knowing whether or not a given transformation
is possible, one might be interested in the process itself \cite{kosloff2013quantum,kosloff2019quantum}, i.e.,
in all the dynamical aspects within a time-dependent transformation,
which certainly includes both energetic and entropic changes. In this
sense, describing the dynamics of any desired property in a given
protocol is of fundamental importance for a complete quantum thermodynamic
theory. Furthermore, such knowledge is clearly necessary for technological
applications where is expected a high degree of control during the
preparation and manipulation of sophisticated and fragile quantum
states. In fact, the most promising - and interesting - applications
will demand local descriptions of interacting quantum systems, especially
when dealing with quantum heat engines and contexts like quantum control and quantum sensing. For these reasons, the
framework of open quantum systems \cite{BRE02} is a natural - and unavoidable -
candidate to approach quantum thermodynamics and has been extensively
used in several contexts. Along these lines, both the understanding
of open quantum systems and the tools used to describe them are slowly
pushing the usual orthodox thermodynamic scenario to a wide range
of situations.

Given the relatively young age of QT, there is no surprise to notice
the presence of several open problems. In fact, considering its stage,
this is expected. However, despite all current advances and efforts,
the lack of consensus on central aspects of the theory is particularly
notorious. This situation can be explained, at least partially, by
the still unknown thermodynamic role of quantum properties, i.e.,
once considered interacting quantum systems, subtleties concerning
entanglement, coherence, and the interaction should be carefully scrutinized.
Of course, classical thermodynamics give us some expectations of how
things should occur. Nevertheless, such properties vanish at the appropriate
classical limit: while genuine quantum phenomena are absent in any
classical setting, the interaction is negligible for macroscopic systems.
In this sense, (remarkably) there is still no acceptable \textit{general}
definitions for the quantum counterparts of the most basic thermodynamic
quantities, which highlights the need for further investigations at
the foundational and conceptual levels.

Along these lines, the obtention of a general quantum thermodynamic
entropy remains elusive. Different alternatives can be found in the
literature \cite{Esposito_2010,Deffner2011,POLKOVNIKOV2011486,santos2017wigner,PhysRevA.99.010101}, but, unsurprisingly, most current approaches
are based on information-theoretic perspectives. \cite{RevModPhys.93.035008} Despite
its success for microscopic classical systems \cite{Seifert2005} and particular
- orthodox - quantum scenarios \cite{Alicki19792,santos2019role}, it does not satisfy the
expected properties for a proper generalization of thermodynamic entropy
and the second law. Unfortunately, this is also the case for internal
energy and the first law. There is no ambiguity in identifying the
internal energy of isolated quantum systems, i.e., this role is unambiguously
assigned to the expectation value of its Hamiltonian. However, it is
not clear how to proceed once considering arbitrary open quantum systems.
In such cases, the notion of physical local internal energy is clouded
by existing non-negligible interactions and correlations between the
system of interest and its surroundings. Naturally, any attempt description
of energy exchanges inherits such basic dubiety. In this sense, classical
thermodynamics states that energy flow is divided into two complementary
and fundamentally distinct categories, work and heat. Such splitting
for the quantum case also carries some intrinsic difficulties. As
mentioned earlier in Section (\ref{subsec:Classical-equilibrium-thermodyna}),
work and heat are defined along with trajectories, which establish
an extra conceptual barrier for directly translating them to the quantum
realm. \cite{Talkner2007} More importantly, it is uncertain how to account
for the energetic contributions of coherence and quantum correlations
into these elements. Still, most of the current strategies are grounded
on the classical - and pragmatical - reasoning that work is associated
with the energy transferred in a controllable and deterministic fashion
via the precise control of external parameters, and heat is linked
to randomly exchanged energy during a given process and entropic variation.
This perspective has been explored in different ways and with distinct
frameworks. On the one hand, many efforts were driven by the quest
for the quantum versions of FTs and measurement-based approaches. \cite{Campisi2011a,allahverdyan2014nonequilibrium,Elouard2016b}
On the other hand, a fully quantum dynamic description is also sought
from a less operational point of view. \cite{Alicki19792,kosloff2013quantum,kosloff2019quantum} Interestingly, it
should be emphasized that despite all the above discussion, even the
usual assumption of splitting the energy flow solely in terms of work
and heat can be debatable for the quantum case. \cite{Bernardo2020}

Additionally, another critical aspect concerns the enormous difficulty
of describing the dynamics of open quantum systems. As briefly mentioned
in the overview of Section (\ref{subsec:Open-systems-dynamics}),
the time-evolution of reduced states may be extremely convoluted and
rarely solvable exactly. Besides, unless we are dealing with simple
physical systems, this type of description and analysis is a formidable
task, even for numerical methods. Thus, most procedures assume several
approximations for microscopically deriving more tractable dynamical
equations, which inevitably restrains its validity for specific conditions
and scenarios. In particular, it is often considered weak-coupling
regimes and Markovian dynamics leading to the usual Lindblad-like
form of master equations.\footnote{In the literature, it is also commonly referred to as Gorini, Kossakowski,
Lindblad, and Sudarshan (GKLS) equation.} \cite{redfield1957theory,davies1974markovian,lindblad1976generators,gorini1976completely,Alicki19792,kosloff2019quantum,nathan2020universal} For more complex situations, when several interacting
subsystems are being described, alternative approaches also consider
the so-called local master equations by strategically neglecting some
interaction terms, which simplifies the analysis even further. \cite{de2018reconciliation,hewgill2021quantum}
Nevertheless, these approximations can lead to thermodynamic inconsistency
or unphysical situations when not taken into account carefully. \cite{anderloni2007redfield,wichterich2007modeling,argentieri2014violations,levy2014local,purkayastha2016out}
Alternatively, it was recently shown the possibility to derive thermodynamically
compatible master equations by employing additional selective hypotheses.
\cite{dann2021open} However, it is clear that relying on approximations and other restrictive assumptions also poses a critical limitation to the development of a general thermodynamic description of quantum systems, especially if one intends to characterize strongly-coupled systems and further arbitrary contexts. In this sense, there are efforts to extend the usual approach for broader scenarios. \cite{gelbwaser2015strongly,esposito2015quantum,esposito2015nature,katz2016quantum,miller2017entropy,Miller2018c,perarnau2018strong,dou2018universal,strasberg2019repeated,talkner2020colloquium,newman2020quantum,huang2020strong,Rivas2020,PhysRevLett.127.250601,Colla2021a}

In addition to the approximate treatment commonly employed in QT, a semi-classical description is also implicitly assumed, i.e., although not usually highlighted, the addition of a classical external agent is a fundamental part of the standard formalism, especially for closed quantum systems. Essentially, this agent is responsible for controlling the dynamics of the system of interest and inducing its Hamiltonian time-dependency, which yield the energy exchange directly interpreted as work. It is also the relevant party for measuring the system and, eventually, processing the accessible information. From a physical point of view, this control is mediated by the use of external fields interacting with the system in question in such a way that it maintains its quantum properties. However, they are not explicitly included in the physical description. Instead, they are regarded as classical fields whose interaction induces an effective time-dependency in the system's Hamiltonian. Despite its practical relevance, this perspective of the so-called "coherent control" \cite{warren1993coherent} does not contemplate autonomous quantum systems and, therefore, limits ourselves to the thermodynamic description of classically driven devices, where both the quantum nature of the fields is unimportant relative to the system of interest and the system itself does not affect the control's state. In contrast to a semi-classical approach, autonomous quantum systems are isolated and do not have Hamiltonian time-dependency, which means that all relevant parties - including the control agent - are quantized. Of course, it also implies that the semi-classical description is a limit regime. Along with this more general and fundamental picture, one is interested in understanding the thermodynamics \textit{within} isolated quantum systems, which would enable the design and characterization of autonomous quantum machines. In this sense, any attempt of developing a thermodynamic description of quantum systems that assumes a time-dependent Hamiltonian a priori is not fully quantum and, essentially, a phenomenological approach. Interestingly, this fundamental - and conceptual - issue is not mentioned very often in the literature of QT. For some discussions on autonomous quantum machines, see. \cite{tonner2005autonomous,youssef2010quantum,gelbwaser2013work,gelbwaser2014heat,brask2015autonomous,hofer2016autonomous,silva2016performance,erker2017autonomous,ghosh2018two,seah2018work,latune2019quantum,manzano2019autonomous,woods2019autonomous,niedenzu2019concepts,najera2020autonomous,hammam2021optimizing} More recently, \cite{dann2021quantum} presented a formal treatment (although restricted to the usual thermodynamic scenario) on this topic. It is worth mentioning that these questions are also at the heart of quantum (optimal) control theory. See \cite{lloyd2000coherent,dong2010quantum,shapiro2012quantum,koch2016controlling} for more discussions on this matter.

Finally, the recent global interest in developing a quantum thermodynamic
theory is growing fast each year. The contemporary version of the
industrial revolution is spearheaded by current progress in understanding
and developing genuine quantum technologies for information processing,
communication and sensing. In this context, a fully matured framework
of QT will certainly play a leading role in the design and operation
of functional quantum devices. Nevertheless, despite being a promising
field, considering its actual stage of achievements and open questions,
it is still in its infancy. More specifically, additional investigations
are needed at the conceptual level, as several fundamental aspects
are still under scrutiny. In short, QT is an interesting, stimulating
and young research field whose investigations and potential scientific
breakthroughs will, in one way or another, help to shape future technologies
and the understanding of quantum mechanics. Hopefully, future historians
of physics will view this chapter of science as we see past developments
today.

%% file: Chapters/Chapter3/Chapter3.tex
\chapter{Schmidt decomposition approach to quantum thermodynamics}
\label{part:Chapter-3--}

In Section (\ref{sec:Quantum-thermodynamics}), we briefly introduced
the current efforts and growing progress on the development of a self-consistent
thermodynamic theory of quantum systems. Despite clear advancements
toward this goal, the field of quantum thermodynamics (QT) still has
some fundamental issues to be addressed. In this sense, most theoretical
frameworks inherit some of the phenomenological spirit of the classical
theory and do not provide suitable tools to characterize and understand
thermodynamic processes within genuine autonomous quantum machines,
i.e., most modern approaches are built on the top of semi-classical
descriptions and approximative regimes, which clearly limit their
range of applicability.

In this chapter, we are interested to address and contribute to such
more foundational aspects of the theory. In the following pages, we
are going to present a novel approach to the thermodynamic analysis
of autonomous quantum systems. Our proposal is exact and based on
the well-known procedure of the Schmidt decomposition for bipartite
systems\footnote{Or single systems where there is coupling between diferent degrees of freedom, e.g., spin-orbit interaction.}. Interestingly, despite being simple and providing a powerful
statement, it is still not explicitly explored in the context of QT.
This framework will allow us to describe the dynamics and energetics
within generic interacting subsystems in a symmetrical fashion, i.e.,
regardless of their individual properties, details and dimension,
they will be treated on equal footing. In addition, it will not require
any complementary hypotheses and approximations, such as the commonly
used ones concerning the Hamiltonian structure, interaction regimes
and type of dynamics, i.e., strict energy conservation, weak-coupling,
markovianity, etc. Formally, we will introduce time-dependent local
effective Hamiltonians that naturally embrace both their respective
bare ones and the contributions of the interaction term. These elements
will be identified as the representative operators for characterizing
the subsystem's physical internal energies, in a way that will allow
us to extend the usual classical thermodynamic notion of energy additivity
to general interacting quantum systems\footnote{The following main discussions and results can be found in \cite{Malavazi}, submitted after this thesis defense.}.

The outline of this chapter is the following: Section (\ref{Section: Setup})
formally introduces and details the main setup of analysis, which
consists of an isolated pure bipartite quantum system. Also, it establishes
the mathematical notation used throughout the chapter. Then, Section
(\ref{SECTION: Schmidt-basis-dynamics}) presents the foundations
of our formalism. More specifically, it discusses and describes the
Schmidt basis dynamics and identifies their time-translation generators
as the so-called local effective Hamiltonians. The following Section
(\ref{SECTION: Local-states-dynamics}) briefly mentions how the local
dynamics are represented relative to these operators and the Schmidt
basis/coefficients. Then, in Section (\ref{subsec:Internal-energy-and}),
the notion of local internal energies is discussed in the context
of QT. Also, it is argued that the effective operators previously
defined are suitable candidates for characterizing the subsystem's
physical internal energy. Along these lines, Section (\ref{sec:Local-phase-gauge})
introduces and discusses the consequences of the intrinsic phase/frame
gauge degree of freedom underlying the Schmidt decomposition and,
therefore, inherited by our formalism. More importantly, it is presented
a procedure for fixing it. Section (\ref{Section: Thermodynamics})
focuses on the analysis of current approaches for defining quantum
versions of thermodynamic quantities. After that, Section (\ref{sec:Proof-of-principle})
shows a proof of principle of the proposed formalism. Then, Section
(\ref{Section: Generalization - mixed states}) generalizes the previous
results to mixed bipartite states. And, finally, Section (\ref{sec:Discussion-and-summary})
briefly discusses these results and summarizes the chapter.

\input{Chapters/Chapter3/Section1}
\input{Chapters/Chapter3/Section2}
\input{Chapters/Chapter3/Section3}
\input{Chapters/Chapter3/Section4}
\input{Chapters/Chapter3/Section5}

\input{Chapters/Chapter3/Section6}
\input{Chapters/Chapter3/Section7}

\input{Chapters/Chapter3/Section8}
\input{Chapters/Chapter3/Section9}

%% file: Chapters/Chapter3/Section1.tex
\section{The setting\label{Section: Setup}}

As mentioned earlier, we consider a finite, isolated and nondegenerate
pure quantum system composed of two smaller interacting subsystems.
Throughout this thesis, the whole system and its global quantities
will be labelled by $(0)$, while the parts and their relative local
properties will be identified by $(1)$ and $(2)$. Let $\mathcal{H}^{(k)}$,
with $k=0,1,2$, be their Hilbert spaces with dimensions $d^{(k)}:=dim(\mathcal{H}^{(k)})$,
such that $d^{(0)}=(d^{(1)}+d^{(2)})$ and - without any loss of generality
- $d^{(1)}\leq d^{(2)}$. Since we are interested in describing a
fully quantum autonomous object, the whole system Hamiltonian $\hat{H}^{(0)}$
generating its dynamics is time-independent and given by
\begin{equation}
\hat{H}^{(0)}:=\hat{H}^{(1)}\otimes\hat{1}^{(2)}+\hat{1}^{(1)}\otimes\hat{H}^{(2)}+\hat{H}_{int},\label{Total Hamiltonian}
\end{equation}
where $\hat{1}^{(1,2)}\in\mathcal{L}(\mathcal{H}^{(1,2)})$ are the
identity operators, $\hat{H}^{(1,2)}\in\mathcal{L}(\mathcal{H}^{(1,2)})$
are the local bare Hamiltonians of each subsystem and $\hat{H}_{int}\in\mathcal{L}(\mathcal{H}^{(0)})$
is the term that encompasses all the internal interactions between
them. It is important to emphasize that no additional hypothesis will
be considered, especially concerning dynamical features or the Hamiltonian/interaction
structure, e.g., markovianity, uncorrelated states, weak coupling, specific interaction Hamiltonian,
etc. The following description is general and exact.

At any time $t$ the whole pure system is described by a ket $|\Psi(t)\rangle\in\mathcal{H}^{(0)}=\mathcal{H}^{(1)}\otimes\mathcal{H}^{(2)}$.
Besides, since it is isolated, its dynamics is governed by the usual
Schrödinger equation $i\hbar\frac{d}{dt}|\Psi(t)\rangle=\hat{H}^{(0)}|\Psi(t)\rangle$;
thereby, for any initial state $|\Psi(t_{0})\rangle$ and $t\geq t_{0}$
we have
\begin{equation}
|\Psi(t)\rangle=\hat{\mathcal{U}}(t,t_{0})|\Psi(t_{0})\rangle,\label{Pure state time evolution}
\end{equation}
where $\hat{\mathcal{U}}(t,t_{0})=e^{-\frac{i}{\hbar}\hat{H}^{(0)}(t-t_{0})}\in\mathcal{L}(\mathcal{H}^{(0)})$
is the time-evolution operator of the whole. As usual, such bipartite
state could be written in any conceivable basis, e.g., $|\Psi(t)\rangle=\sum_{i=1}^{d^{(1)}}\sum_{j=1}^{d^{(2)}}\psi_{ij}(t)|b_{i}^{(1)},b_{j}^{(2)}\rangle$,
nevertheless the well known Schmidt decomposition guarantee the following
specific and convenient form (see (\ref{par:The-Schmidt-decomposition}))
\begin{equation}
|\Psi(t)\rangle=\sum_{j=1}^{d^{(1)}}\lambda_{j}(t)|\varphi_{j}^{(1)}(t)\rangle\otimes|\varphi_{j}^{(2)}(t)\rangle,\label{SchmidtDecomposition}
\end{equation}
for every instant $t$, where $\{\lambda_{j}(t)\geq0;j=1,...,d^{(1)}\}$
and $\{|\varphi_{j}^{(k)}(t)\rangle;j=1,...,d^{(1)}\}$\footnote{Unless it is necessary, from now on, any set will be represented simply
as $\{a_{j}\}_{j}$ where $j$ is the index counting its elements,
and the range is implicit.}$\in\mathcal{H}^{(k)}$ are the time-local Schmidt coefficients and
local Schmidt basis of subsystem $(k)$, respectively. The normalization
condition $\langle\Psi(t)|\Psi(t)\rangle=1$ implies that $\sum_{j=1}^{d^{(1)}}\lambda_{j}^{2}(t)=1$,
and the orthonormality of the local basis elements assure that $\langle\varphi_{m}^{(\alpha)}(t)|\varphi_{n}^{(\beta)}(t)\rangle=\delta_{\alpha\beta}\delta_{mn}$.
From now on, the Schmidt decomposition form showed in Eq. (\ref{SchmidtDecomposition})
will be the standard description of $|\Psi(t)\rangle$.

As already mentioned in chapter (\ref{part:Chapter-2--}), the representation above
is compelling and useful for a number of reasons: notice that despite
a potential huge discrepancy between $d^{(1)}$ and $d^{(2)}$ there
is a single sum bounded by the smallest dimension in question, by
hypothesis $d^{(1)}$\footnote{Even though the sum extends up to $d^{(1)}$ we may have null Schmidt
coefficients thus, in practice, the sum goes until the Schmidt rank,
defined as the number of non-zero coefficients. To keep track of these
dimensions, we will maintain $d^{(1)}$ in the sums.}; apart from that, it is symmetrical in the sense that for each ket
$|\varphi_{j}^{(1)}(t)\rangle$ from subsystem $(1)$ there is a related
ket $|\varphi_{j}^{(2)}(t)\rangle$ from $(2)$; additionally, as
we are going to present below, it turns out that it gives all necessary
information for representing the subsystem's local states; also, it
is guaranteed that the Schmidt coefficients are unambiguously defined,
while the Schmidt basis are unique up to eventual degenerate coefficients
and a phase degree of freedom, in the sense that Eq. (\ref{SchmidtDecomposition})
is invariant over simultaneous local phases changes (this point will
be discussed later in Section (\ref{sec:Local-phase-gauge})); finally,
it makes easy to verify whether the subsystems are entangled or not,
i.e., a product state will be obtained iff there is a single non-zero
Schmidt coefficient\footnote{Schmidt rank equal to one.} such that
$\lambda_{j}(t)=1$, $\lambda_{m}(t)=0$ for all $m\neq j$ and $|\Psi(t)\rangle=|\varphi_{j}^{(1)}(t)\rangle\otimes|\varphi_{j}^{(2)}(t)\rangle$.
As a last remark, it is important to highlight the fact that the set
$\{|\varphi_{j}^{(2)}(t)\rangle\}_{j}$ of Schmidt basis is not formally
complete since we only have $d^{(1)}$ elements, still we are always
allowed to find the remaining $(d^{(2)}-d^{(1)})$ orthonormal kets
to form a complete basis of $\mathcal{H}^{(2)}$.

Let us now turn our attention to the parts: the individual state description
of each subsystem is represented by the reduced density matrix of
the whole, whose own pure state density matrix is
\begin{equation}
\hat{\rho}^{(0)}(t)\equiv|\Psi(t)\rangle\langle\Psi(t)|.\label{Pure state whole}
\end{equation}
Formally, these local states are obtained by the usual procedure of
partial tracing Eq. (\ref{Pure state whole}) such that $\hat{\rho}^{(1,2)}(t)\equiv tr_{2,1}\{\hat{\rho}^{(0)}(t)\}$.
Thus, given Eq. (\ref{SchmidtDecomposition}) it is easy to see that,
for all $t$,
\begin{align}
\hat{\rho}^{(1)}(t) & =\sum_{j=1}^{d^{(1)}}\lambda_{j}^{2}(t)|\varphi_{j}^{(1)}(t)\rangle\langle\varphi_{j}^{(1)}(t)|,\label{local state 1}\\
\hat{\rho}^{(2)}(t) & =\sum_{j=1}^{d^{(1)}}\lambda_{j}^{2}(t)|\varphi_{j}^{(2)}(t)\rangle\langle\varphi_{j}^{(2)}(t)|.\label{local state 2}
\end{align}
Hopefully, the expressions above are sufficient to further elucidate
how convenient the Schmidt decomposition really is. As briefly mentioned
earlier, its provides all necessary information for inferring the
spectral decomposition of these states, i.e., their eigenvalues (also
referred as populations) and eigenvectors are given by the Schmidt
coefficients squared $\{\lambda_{j}^{2}(t)\}_{j}$ and Schmidt basis
$\{|\varphi_{j}^{(1,2)}(t)\rangle\}_{j}$, respectively. In particular,
notice that both local states are represented - in general - by mixed
density matrices\footnote{Iff the Schmidt rank is equal to one we have non-entangled systems
and pure local states. } and necessarily have the same spectrum whenever the whole bipartite
system is pure; more precisely, since $d^{(2)}\geq d^{(1)}$ the spectrum
of subsystem $(2)$ cointains the set $\{\lambda_{j}^{2}(t)\}_{j}$
plus $(d^{(2)}-d^{(1)})$ null eigenvalues, whose respective eigenvectors
form the nullspace of $\hat{\rho}^{(2)}(t)$\footnote{Obviously, if $d^{(2)}>d^{(1)}$ the density matrix $\hat{\rho}^{(2)}(t)$
is automatically singular.}. Additionally, this also implies that both subsystems will have the
same values for any local functional of their populations, most notably
purity and von-Neumann entropy.

%% file: Chapters/Chapter3/Section2.tex
\section{Schmidt basis dynamics and local effective Hamiltonians\label{SECTION: Schmidt-basis-dynamics}}

In this section, we are particularly interested in a \textit{local}
dynamical description of the Schmidt basis $\{|\varphi_{j}^{(1,2)}(t)\rangle\}_{j}$,
whereas by local we mean a description solely based in terms relative
to their respective Hilbert space $\mathcal{H}^{(1,2)}$. Firstly,
it is important to emphasize that both sets of Schmidt coefficients
and pair of basis are intrinsically connected to the whole system
state, in the sense that for every ket $|\Psi(t)\rangle$ there is
a single decomposition. Having that in mind, a pictorial representation
might by useful: at any time interval $[t_{0},t_{1}]$ the autonomous
time evolution performed by the whole can be visualized as a curve
$\mathcal{P}^{(0)}:|\Psi(t)\rangle,\:t\in[t_{0},t_{1}]$ in the total
Hilbert space $\mathcal{H}^{(0)}$; nevertheless, given Eq. (\ref{SchmidtDecomposition})
such path can be mapped into the simultaneous coupled trajectories
$\mathcal{P}_{j}^{(1,2)}:|\varphi_{j}^{(1,2)}(t)\rangle,\:t\in[t_{0},t_{1}]$
followed by the Schmidt basis in their own Hilbert spaces, and the
paths of the Schmidt coefficients, $\mathcal{\mathcal{P}}_{j}^{\lambda}:\lambda_{j}(t)\geq0,\:t\in[t_{0},t_{1}]$,
such that $\lambda_{j}^{2}(t)\in[0,1]$ for all $j$, and $\sum_{j=1}^{d^{(1)}}\lambda_{j}^{2}(t)=1$.
Obviously, once one has access to the initial state $|\Psi(t_{0})\rangle$,
$\mathcal{P}^{(0)}$ is fully characterized by the unitary operator
$\hat{\mathcal{U}}(t,t_{0})$ and Eq. (\ref{Pure state time evolution}),
while the correlated behaviours of $\{\mathcal{P}_{j}^{(1,2)}\}_{j}$
and $\{\mathcal{P}_{j}^{\lambda}\}_{j}$ are direct byproducts of
the former. For instance, at any time $t$ we have the following expression
for the coefficients
\begin{equation}
\lambda_{j}(t)=\sum_{k=1}^{d^{(1)}}\lambda_{k}(t_{0})\langle\varphi_{j}^{(1)}(t),\varphi_{j}^{(2)}(t)|\hat{\mathcal{U}}(t,t_{0})|\varphi_{k}^{(1)}(t_{0}),\varphi_{k}^{(2)}(t_{0})\rangle,
\end{equation}
where it clearly depends on the initial state of the whole, its time-evolution
operator and the instantaneous Schmidt basis. However, such global
knowledge is rarely accessible in most realistic scenarios, thus what
we are really interested in is to inferring and describing the individual
effective dynamics portrayed by $\mathcal{P}_{j}^{(1,2)}$ such that defining \textit{local} thermodynamic quantities is meaningful and universal, in the sense that the theoretical machinery can be applied for both parts without any further adjustments. Such a procedure will prove very useful
in describing the subsystems dynamics and their energetic flux.

Let us now put it more formally: initially, we define the local dynamical
map $\tilde{\mathcal{U}}^{(k)}:\mathcal{H}^{(k)}\rightarrow\mathcal{H}^{(k)}$
($k=1,2$) that serves as a time-evolution operator and reproduces
the paths $\{\mathcal{P}_{j}^{(k)}\}_{j}$ in a way that every Schmidt
basis ket continuously follows
\begin{equation}
|\varphi_{j}^{(k)}(t)\rangle=\tilde{\mathcal{U}}^{(k)}(t,t_{0})|\varphi_{j}^{(k)}(t_{0})\rangle,
\end{equation}
for any $t\geq t_{0}$, with $\underset{t\rightarrow t_{0}}{lim}|\varphi_{j}^{(k)}(t)\rangle=|\varphi_{j}^{(k)}(t_{0})\rangle$
or $\underset{t\rightarrow t_{0}}{lim}\,\tilde{\mathcal{U}}^{(k)}(t,t_{0})=\hat{1}^{(k)}$.
Since the Schmidt basis at distinct times corresponds to a different
orthonormal basis for the same Hilbert space, the relationship above
is trivially guaranteed\footnote{In fact, we already know it should be unitary.}.
Additionally, it is required both that
\begin{equation}
\langle\varphi_{l}^{(k)}(t)|\varphi_{j}^{(k)}(t)\rangle=\langle\varphi_{l}^{(k)}(t_{0})|\varphi_{j}^{(k)}(t_{0})\rangle=\delta_{lj},
\end{equation}
and
\begin{equation}
\tilde{\mathcal{U}}^{(k)}(t_{2},t_{0})=\tilde{\mathcal{U}}^{(k)}(t_{2},t_{1})\tilde{\mathcal{U}}^{(k)}(t_{1},t_{0}),
\end{equation}
for $t_{2}>t_{1}>t_{0}$, where the former condition guarantees the
orthonormality during the entire dynamics, and the latter demands
the composition of the maps for intermediate times. It is worth mentioning
that, in general, these local maps are not directly related to the
whole time-evolution operator, such that $\hat{\mathcal{U}}(t,t_{0})\neq\tilde{\mathcal{U}}^{(1)}(t,t_{0})\otimes\tilde{\mathcal{U}}^{(2)}(t,t_{0})$.
The previous conditions are automatically fulfilled if the time evolution
operator $\tilde{\mathcal{U}}^{(k)}(t,t_{0})$ is unitary, i.e., $\tilde{\mathcal{U}}^{(k)\dagger}(t,t_{0})\tilde{\mathcal{U}}^{(k)}(t,t_{0})=\hat{1}^{(k)}$,
and have the form presented below for infinitesimal temporal displacements
$dt$,
\begin{equation}
\tilde{\mathcal{U}}^{(k)}(t+dt,t)=\hat{1}^{(k)}-\frac{i}{\hbar}\tilde{H}^{(k)}(t)dt,
\end{equation}
 where $\tilde{H}^{(k)}(t)=\tilde{H}^{(k)\dagger}(t)\in\mathcal{L}(\mathcal{H}^{(k)})$
is a hermitian and possibly time-dependent operator. Thus, one can
show that 
\begin{equation}
i\hbar\frac{d}{dt}\tilde{\mathcal{U}}^{(k)}(t,t_{0})=\tilde{H}^{(k)}(t)\tilde{\mathcal{U}}^{(k)}(t,t_{0}),\label{Time evolution - Schmidt Basis}
\end{equation}
and, therefore,
\begin{equation}
i\hbar\frac{d}{dt}|\varphi_{j}^{(k)}(t)\rangle=\tilde{H}^{(k)}(t)|\varphi_{j}^{(k)}(t)\rangle,\label{Schmidt basis dynamics}
\end{equation}
for all $j$. The Eq. (\ref{Schmidt basis dynamics}) describes exactly
what we wanted, where the new operator $\tilde{H}^{(k)}(t)$ introduced
above plays the role of the time-translation generator of the Schmidt
basis $\{|\varphi_{j}^{(k)}(t)\rangle\}_{j}$ along with $\{\mathcal{P}_{j}^{(k)}\}_{j}$,
and, from now on, will be referred to as the \textit{local effective
Hamiltonian} of subsystem $(k)$.

Now, the most natural question is \textquotedbl What exactly is the
form of $\tilde{H}^{(k)}(t)$?\textquotedbl . It certainly should
depend on the context, in the sense that different initial conditions
will give rise to distinct Schmidt basis trajectories and, therefore,
a new effective Hamiltonian. In fact, this is clear once noticed that
the equality above can be reversed such that 
\begin{equation}
\tilde{H}^{(k)}(t)\equiv i\hbar\sum_{j=1}^{d^{(k)}}\frac{d}{dt}|\varphi_{j}^{(k)}(t)\rangle\langle\varphi_{j}^{(k)}(t)|,\label{Effective Hamiltonian}
\end{equation}
in other words, knowing the Schmidt basis and its dynamics one could
obtain its respective local effective Hamiltonian. Interestingly,
despite this convoluted nature with the whole pure quantum system
state, such operator is locally accessible in principle, since the
Schmidt basis are exactly the eigenbasis of the local state $\hat{\rho}^{(k)}(t)$
in question. Nevertheless, we can go one step further and show that
Eq. (\ref{Effective Hamiltonian}) can be directly related to the
local bare Hamiltonian $\hat{H}^{(k)}$. To do so, we first explicitly
write down the latter in its spectral decomposition
\begin{equation}
\hat{H}^{(k)}\equiv\sum_{j=1}^{d^{(k)}}b_{j}^{(k)}|b_{j}^{(k)}\rangle\langle b_{j}^{(k)}|,\label{local bare hamiltonian}
\end{equation}
where $\{b_{j}^{(k)}\}_{j}$ and $\{|b_{j}^{(k)}\rangle\}_{j}$ are
its respective bare eigenenergies and eigenbasis. Then we define the
projection $\langle b_{j}^{(k)}|\varphi_{l}^{(k)}(t)\rangle:=r_{jl}^{(k)}(t)e^{-\frac{i}{\hbar}b_{j}^{(k)}t}$,
in such a way that
\begin{equation}
\langle b_{j}^{(k)}|\frac{d}{dt}|\varphi_{l}^{(k)}(t)\rangle=\left(\frac{d}{dt}r_{jl}^{(k)}(t)\right)e^{-\frac{i}{\hbar}b_{j}^{(k)}t}-\frac{i}{\hbar}b_{j}^{(k)}r_{jl}^{(k)}(t)e^{-\frac{i}{\hbar}b_{j}^{(k)}t}.
\end{equation}
Besides, given the basis orthonormality $\langle b_{\alpha}^{(k)}|b_{\beta}^{(k)}\rangle=\delta_{\alpha\beta}$,
we also have
\begin{equation}
\sum_{l=1}^{d^{(k)}}r_{\alpha l}^{(k)}(t)\left(r_{\beta l}^{(k)}(t)\right)^{*}e^{\frac{i}{\hbar}\left(b_{\beta}^{(k)}-b_{\alpha}^{(k)}\right)t}=\delta_{\alpha\beta}.
\end{equation}
Finally, by casting $\tilde{H}^{(k)}(t)$ in the bare eigenbasis representation
and using the previous relations, one can rewrite the local effective
Hamiltonian as follows
\begin{equation}
\tilde{H}^{(k)}(t)=\hat{H}^{(k)}+\hat{H}_{LS}^{(k)}(t)+\hat{H}_{X}^{(k)}(t),\label{Effective Hamiltonian 2}
\end{equation}
where
\begin{align}
\hat{H}_{LS}^{(k)}(t) & :=i\hbar\sum_{j=1}^{d^{(k)}}\left(\sum_{l=1}^{d^{(k)}}\left(\frac{d}{dt}r_{jl}^{(k)}(t)\right)r_{jl}^{(k)*}(t)\right)|b_{j}^{(k)}\rangle\langle b_{j}^{(k)}|,\\
\hat{H}_{X}^{(k)}(t):= & i\hbar\sum_{j=1}^{d^{(k)}}\sum_{m\neq j}^{d^{(k)}}\left(\sum_{l=1}^{d^{(k)}}\frac{d}{dt}r_{jl}^{(k)}(t)r_{ml}^{(k)*}(t)\right)e^{\frac{i}{\hbar}\left(b_{m}^{(k)}-b_{j}^{(k)}\right)t}|b_{j}^{(k)}\rangle\langle b_{m}^{(k)}|.
\end{align}
Thus, the local effective Hamiltonian can be split into the sum of
three distinct elements, including the bare Hamiltonian. Note that
the additional operators are responsible for the time-dependency of
$\tilde{H}^{(k)}(t)$, where $\hat{H}_{LS}^{(k)}(t)$ is a general
Lamb-shift like term, in the sense that it is diagonal in the bare
eigenbasis, i.e., $[\hat{H}_{LS}^{(k)}(t),\hat{H}^{(k)}]=0$ for all
$t$, and $\hat{H}_{X}^{(k)}(t)$ contains only non-diagonal elements.
Intuitively, such extra quantities are expected to be somehow related
to the interaction term. Notice that in its absence, each subsystem
would unitarily evolve in time according to their individual bare
Hamiltonians, since $\hat{\mathcal{U}}(t,t_{0})=e^{-\frac{i}{\hbar}\hat{H}^{(0)}(t-t_{0})}=e^{-\frac{i}{\hbar}\hat{H}^{(1)}(t-t_{0})}e^{-\frac{i}{\hbar}\hat{H}^{(2)}(t-t_{0})}$
for this scenario. Thus, for a initial state $|\Psi(t_{0})\rangle$
we would have
\begin{equation}
|\Psi(t)\rangle_{bare}=\sum_{j=1}^{d^{(1)}}\lambda_{j}(t_{0})|\varphi_{j}^{(1)}(t)\rangle_{bare}\otimes|\varphi_{j}^{(2)}(t)\rangle_{bare},
\end{equation}
where $|\varphi_{j}^{(k)}(t)\rangle_{bare}=e^{-\frac{i}{\hbar}\hat{H}^{(k)}(t-t_{0})}|\varphi_{j}^{(k)}(t_{0})\rangle$
represents the free evolution of the initial Schmidt basis, for $t\geq t_{0}$.
Hence, in this case, the Schmidt coefficients $\{\lambda_{j}(t_{0})\}_{j}$
are constant and the local effective Hamiltonians are simply identified
as the bare ones, i.e, $\tilde{H}^{(k)}(t)=\hat{H}^{(k)}$. This,
however, can be seen from the usual time-independent theory approach.
Let us now add a dimensionless parameter $\varepsilon\in[0,1]$ and
make $\hat{H}_{int}\rightarrow\varepsilon\hat{H}_{int}$, such that
we can write the following formal perturbative series
\begin{align}
\lambda_{j}(t) & =\lambda_{j}(t_{0})+\varepsilon\lambda_{j,1}(t)+\mathcal{O}_{\lambda}(\varepsilon^{2}),\\
|\varphi_{j}^{(k)}(t)\rangle & =|\varphi_{j}^{(k)}(t)\rangle_{bare}+\varepsilon|\varphi_{j}^{(k)}(t)\rangle_{1}+\mathcal{O}^{(k)}(\varepsilon^{2}),
\end{align}
where $\lambda_{j,1}(t)$ and $|\varphi_{j}^{(k)}(t)\rangle_{1}$
are the respective first order perturbation elements for the Schmidt
coefficients and basis, and $\mathcal{O}_{\lambda}(\varepsilon^{2})$
and $\mathcal{O}^{(k)}(\varepsilon^{2})$ are the terms including
higher orders corrections. Thus, for sufficiently small $\varepsilon$,
the first order approximation for the whole bipartite state is simply
\begin{equation}
|\Psi(t)\rangle\approx|\Psi(t)\rangle_{bare}+\varepsilon|\Psi(t)\rangle_{1},
\end{equation}
and, given Eq. (\ref{Effective Hamiltonian}),
\begin{equation}
\tilde{H}^{(k)}(t)\approx\hat{H}^{(k)}+\varepsilon\hat{H}_{1}^{(k)}(t),
\end{equation}
where $|\Psi(t)\rangle_{1}$ and $\hat{H}_{1}^{(k)}$ are their first
order components. Hence, any deviation from the bare Hamiltonian is
due to the interaction between the subsystems and, as expected, if
$\varepsilon\rightarrow0$ it is clear that $\tilde{H}^{(k)}(t)\rightarrow\hat{H}^{(k)}$.
In short, both additional quantities are the local effective by-products
of the interaction term $\hat{H}_{int}$. In fact, later will be shown
the functional relationship between their expectation values.

%% file: Chapters/Chapter3/Section3.tex
\section{Local states dynamics\label{SECTION: Local-states-dynamics}}

In Section (\ref{Section: Setup}) we obtained the exact form of the
reduced density matrices of our pure bipartite system. We also emphasize
how remarkably symmetrical their representation are despite any eventual
dimensional difference: we might be dealing with two interacting qubits
or a single qubit interacting with a highly complex reservoir, in
both scenarios, we would obtain density matrices with equal spectrum,
identified by the Schmidt coefficients. Now, we are interested in
describing their time-evolution, i.e., to write down the dynamical
equation $\frac{d}{dt}\hat{\rho}^{(1,2)}(t)$ for both subsystems.

As a starting point, recall that the whole isolated bipartite system
dynamics is unitary and described by the Schrödinger equation. The
equivalent description for its density matrix is expressed by the
Liouville-von Neumann equation
\begin{equation}
i\hbar\frac{d}{dt}\hat{\rho}^{(0)}(t)=[\hat{H}^{(0)},\hat{\rho}^{(0)}(t)].\label{LiouvilleVNeumann}
\end{equation}
Of course, when dealing with the subsystems dynamics such behaviour
is not expected, and additional terms should be taken into account
to properly describe non-unitary features commonly observed in open
quantum systems, e.g., dissipation and decoherence. The first obvious
approach for our goal is partial tracing the previous equation and
using the total Hamiltonian expression (\ref{Total Hamiltonian}),
such that
\begin{equation}
i\hbar\frac{d}{dt}\hat{\rho}^{(k)}(t)=[\hat{H}^{(k)},\hat{\rho}^{(k)}(t)]+tr_{\bar{k}}\{[\hat{H}_{int},\hat{\rho}^{(0)}(t)]\},\label{Master Eq. 1}
\end{equation}
where $k=1,2$, and $\bar{k}$ is its complement (if $k=1$ we have $\bar{k}=2$ and vice versa). As we can see, the local time-evolutions are clearly
separated into a unitary part, guided by the bare Hamiltonians, and
a non-unitary part, represented by the partial trace. Also note that
the latter explicitly depends on global properties, represented by
the commutation relation between the interaction term and the state
of the whole quantum system. It is well known that in some particular
scenarios and under specific hypotheses the previous expression can
be cast in more relatable forms, usually into time-local master equations
with the canonical Lindblad form. Until very recently, a similar simplification
for broad dynamics, constraints and initial conditions was elusive,
however, S. Alipour \textit{et. al.} \cite{Alipour2020} showed that the general exact expression
for the dynamics of the reduced states, presented in Eq. (\ref{Master Eq. 1}),
can, in fact, be cast in a universal Lindblad-like form.

Nevertheless, from Eq. (\ref{Schmidt basis dynamics}) and the direct
derivative of Eqs. (\ref{local state 1}), (\ref{local state 2})
we can write an alternative dynamical expression for both subsystems
in terms of their local effective Hamiltonians, and the Schmidt basis
and coefficients:
\begin{equation}
i\hbar\frac{d}{dt}\hat{\rho}^{(k)}(t)=[\tilde{H}^{(k)}(t),\hat{\rho}^{(k)}(t)]+i\hbar\sum_{j=1}^{d^{(1)}}\frac{d}{dt}\lambda_{j}^{2}(t)|\varphi_{j}^{(k)}(t)\rangle\langle\varphi_{j}^{(k)}(t)|.\label{Master Eq. 2}
\end{equation}
 Interestingly, by defining the operators $\hat{G}_{\alpha\beta}^{(k)}(t)\equiv|\varphi_{\alpha}^{(k)}(t)\rangle\langle\varphi_{\beta}^{(k)}(t)|$
and rates $\gamma_{\alpha\beta}(t)\equiv\frac{1}{d^{(1)}\lambda_{\beta}^{2}(t)}\frac{d}{dt}\lambda_{\alpha}^{2}(t)$
we can also put it into a non-linear (the rates $\gamma_{\alpha\beta}(t)$ do depend on the state of the whole) Lindblad-like form
\begin{equation}
i\hbar\frac{d}{dt}\hat{\rho}^{(k)}(t)=[\tilde{H}^{(k)}(t),\hat{\rho}^{(k)}(t)]+\mathfrak{\hat{D}}_{t}^{(k)}\hat{\rho}^{(k)}(t),\label{lindblad form}
\end{equation}
where 
\begin{equation}
\mathfrak{\hat{D}}_{t}^{(k)}\hat{\rho}^{(k)}(t):=i\hbar\sum_{\alpha,\beta=1}^{d^{(1)}}\gamma_{\alpha\beta}(t)\left(\hat{G}_{\alpha\beta}^{(k)}(t)\hat{\rho}^{(k)}(t)\hat{G}_{\alpha\beta}^{(k)\dagger}(t)-\frac{1}{2}\left\{ \hat{G}_{\alpha\beta}^{(k)\dagger}(t)\hat{G}_{\alpha\beta}^{(k)}(t),\hat{\rho}^{(k)}(t)\right\} \right).
\end{equation}
In these expressions, the unitary part is governed by the local effective
Hamiltonian instead of the bare one, while the non-unitary part explicitly
depends on the population time dependency and, therefore, is directly
related to the entanglement change during the time evolution. Eq.
(\ref{Master Eq. 2}) can be seen as a parametric expression for the
curve $\mathcal{C}^{(k)}:\hat{\rho}^{(k)}(t),\:t\in[t_{0},t_{1}]$
followed by the subsystem $(k)$ in its respective density operator
space $\mathscr{D}(\mathcal{H}^{(k)})$, such that the unitary contribution
is due to the generator of the Schmidt basis paths $\{\mathcal{P}_{j}^{(k)}\}_{j}$,
and the non-unitary factor is given by the population's trajectories
$\{\mathcal{P}_{j}^{(\lambda)}\}_{j}$. Thus, it is important to highlight
that $\mathcal{C}^{(k)}$ is a byproduct of the whole system dynamics,
in the sense that different trajectories $\mathcal{P}^{(0)}$ of $|\Psi(t)\rangle$
originated from distinct initial states, will result in distinct
density matrices curves and parametric descriptions.

Finally, since both Eqs. (\ref{Master Eq. 1}) and (\ref{Master Eq. 2})
are exact expressions for dealing with the same dynamics they can
be directly associated. In fact, given Eq. (\ref{Effective Hamiltonian 2}),
both unitary contributions satisfy
\begin{equation}
[\tilde{H}^{(k)}(t),\hat{\rho}^{(k)}(t)]=[\hat{H}^{(k)},\hat{\rho}^{(k)}(t)]+[\hat{H}_{LS}^{(k)}(t),\hat{\rho}^{(k)}(t)]+[\hat{H}_{X}^{(k)}(t),\hat{\rho}^{(k)}(t)],
\end{equation}
which leads to the following equality for all time $t$
\begin{equation}
tr_{\bar{k}}\{[\hat{H}_{int},\hat{\rho}^{(0)}(t)]\}=i\hbar\sum_{j=1}^{d^{(1)}}\frac{d}{dt}\lambda_{j}^{2}(t)|\varphi_{j}^{(k)}(t)\rangle\langle\varphi_{j}^{(k)}(t)|+[\hat{H}_{LS}^{(k)}(t)+\hat{H}_{X}^{(k)}(t),\hat{\rho}^{(k)}(t)].
\end{equation}

\subsection{Unitary dynamics}

The previous equations are exact, nevertheless, note that whenever
the second term of Eq. (\ref{Master Eq. 2}) is negligible compared
with the first one, the state dynamics is approximately unitary. Hence,
it is valuable to identify - at least qualitatively - what conditions
are necessary for having such behaviour. Suppose an arbitrary initial
state $|\Psi(t_{0})\rangle=\sum_{k=1}^{d^{(1)}}\lambda_{k}(t_{0})|\varphi_{k}^{(1)}(t_{0})\rangle\otimes|\varphi_{k}^{(2)}(t_{0})\rangle$,
the Schmidt coefficients at any time $t$ are simply $\lambda_{j}(t)=\langle\varphi_{j}^{(1)}(t),\varphi_{j}^{(2)}(t)|\Psi(t)\rangle$
and
\begin{align}
i\hbar\frac{d}{dt}\lambda_{j}(t) & =\lambda_{j}(t)\langle\varphi_{j}^{(1)}(t),\varphi_{j}^{(2)}(t)|\left(\hat{H}_{int}-\hat{H}_{LSX}^{(1)}(t)-\hat{H}_{LSX}^{(2)}(t)\right)|\varphi_{j}^{(1)}(t),\varphi_{j}^{(2)}(t)\rangle\nonumber \\
 & +\sum_{k\neq j}^{d^{(1)}}\lambda_{k}(t)\langle\varphi_{j}^{(1)}(t),\varphi_{j}^{(2)}(t)|\hat{H}_{int}|\varphi_{k}^{(1)}(t),\varphi_{k}^{(2)}(t)\rangle,
\end{align}
where $\hat{H}_{LSX}^{(k)}(t):=\hat{H}_{LS}^{(k)}(t)+\hat{H}_{X}^{(k)}(t)$
and $\langle\varphi_{j}^{(1)}(t),\varphi_{j}^{(2)}(t)|\hat{H}_{int}|\varphi_{k}^{(1)}(t),\varphi_{k}^{(2)}(t)\rangle$
with $k\neq j$ are the factors responsible for coupling distinct
$\lambda$'s. If these terms have a minor contribution to the coefficients
dynamics, we can write the following approximative solution
\begin{equation}
\lambda_{j}(t)\approx\lambda_{j}(t_{0})\overleftarrow{\mathcal{T}}e^{-\frac{i}{\hbar}\int_{t_{0}}^{t}ds\,\langle\varphi_{j}^{(1)}(s),\varphi_{j}^{(2)}(s)|\left(\hat{H}_{int}-\hat{H}_{LSX}^{(1)}(s)-\hat{H}_{LSX}^{(2)}(s)\right)|\varphi_{j}^{(1)}(s),\varphi_{j}^{(2)}(s)\rangle},
\end{equation}
where $\overleftarrow{\mathcal{T}}$ is the usual chronological time-ordering operator. However, since $\{\lambda_{j}(t)\}_{j}$ are real numbers we must
also have
\begin{equation}
\langle\varphi_{j}^{(1)}(t),\varphi_{j}^{(2)}(t)|\left(\hat{H}_{int}-\hat{H}_{LSX}^{(1)}(t)-\hat{H}_{LSX}^{(2)}(t)\right)|\varphi_{j}^{(1)}(t),\varphi_{j}^{(2)}(t)\rangle=0
\end{equation}
 for all $j$ and $t$, therefore,
\begin{equation}
\lambda_{j}(t)\approx\lambda_{j}(t_{0}),
\end{equation}
and
\begin{equation}
|\Psi(t)\rangle\approx\tilde{\mathcal{U}}^{(1)}(t,t_{0})\tilde{\mathcal{U}}^{(2)}(t,t_{0})|\Psi(t_{0})\rangle\quad\Rightarrow\quad i\hbar\frac{d}{dt}\hat{\rho}^{(k)}(t)\approx[\tilde{H}^{(k)}(t),\hat{\rho}^{(k)}(t)].
\end{equation}
Thus, as long as $\langle\varphi_{j}^{(1)}(t),\varphi_{j}^{(2)}(t)|\hat{H}_{int}|\varphi_{k}^{(1)}(t),\varphi_{k}^{(2)}(t)\rangle\approx0$
is satisfied for all $j\neq k$ both subsystems evolves approximately
unitarily and the degree of entanglement remains conserved. Interestingly,
observe that this is true despite the other matrix elements of $\hat{H}_{int}$.

\subsubsection{Semi-classical external drive}

Under the previous approximation, if the initial Schmidt rank is equal to one, such that $|\Psi(t_{0})\rangle=|\varphi_{\eta}^{(1)}(t_{0})\rangle\otimes|\varphi_{\eta}^{(2)}(t_{0})\rangle$, we would guarantee uncorrelated local pure states for all $t$, i.e.,
\begin{equation}
|\Psi(t)\rangle\approx|\varphi_{\eta}^{(1)}(t)\rangle\otimes|\varphi_{\eta}^{(2)}(t)\rangle,
\end{equation}
where $|\varphi_{\eta}^{(k)}(t)\rangle=\langle\varphi_{\eta}^{(\bar{k})}(t)|\Psi(t)\rangle$ and
\begin{equation}
i\hbar\frac{d}{dt}|\varphi_{\eta}^{(k)}(t)\rangle\approx(\hat{H}^{(k)}+\langle\varphi_{\eta}^{(\bar{k})}(t)|\hat{H}_{int}|\varphi_{\eta}^{(\bar{k})}(t)\rangle-\langle\varphi_{\eta}^{(\bar{k})}(t)|\hat{H}_{LSX}^{(\bar{k})}(t)|\varphi_{\eta}^{(\bar{k})}(t)\rangle)|\varphi_{\eta}^{(k)}(t)\rangle.
\end{equation}
Thus, from Eq. (\ref{Schmidt basis dynamics}) it is clear that, in
such cases, the local effective Hamiltonian simply becomes
\begin{equation}
\tilde{H}^{(k)}(t)\approx\hat{H}^{(k)}+tr_{\bar{k}}\{(\hat{H}_{int}-\hat{H}_{LSX}^{(\bar{k})}(t))\hat{\rho}^{(\bar{k})}(t)\},
\end{equation}
where $\hat{H}_{LSX}^{(k)}(t)\approx tr_{\bar{k}}\{(\hat{H}_{int}-\hat{H}_{LSX}^{(\bar{k})}(t))\hat{\rho}^{(\bar{k})}(t)\}$,
and $\langle\varphi_{\eta}^{(\bar{k})}(t)|\hat{X}|\varphi_{\eta}^{(\bar{k})}(t)\rangle=tr_{\bar{k}}\{\hat{X}\hat{\rho}^{(\bar{k})}(t)\}$
for any operator $\hat{X}$. Note that these approximative equations
are still symmetrical for both subsystems, nevertheless, it is easier
to see how asymmetrical physical systems might lead to distinct effective
behaviours. For instance, if subsystem
$(2)$ is sufficiently large to be regarded as a macroscopic system,
it is expected that both the interaction and subsystem $(1)$ dynamics
would have a negligible effect on $(2)$, in such a way that $tr_{1}\{(\hat{H}_{int}-\hat{H}_{LSX}^{(1)}(t))\hat{\rho}^{(1)}(t)\}\approx\hat{H}_{LSX}^{(2)}(t)\approx0$,
and its local effective Hamiltonian is indistinguishable from the
bare one:
\begin{align}
\tilde{H}^{(2)}(t) & \approx\hat{H}^{(2)}.
\end{align}
This, however, still not true for the subsystem $(1)$, since
\begin{equation}
\tilde{H}^{(1)}(t)\approx\hat{H}^{(1)}+tr_{2}\{\hat{H}_{int}\hat{\rho}^{(2)}(t)\},
\end{equation}
i.e., its local effective Hamiltonian is determined by the macroscopic
state of $(2)$ and the interaction term. In particular, if the former
is somehow controllable by a set of time-dependent parameters $\{\boldsymbol{R}_{t}\}$,
such that $\hat{\rho}^{(2)}(t)=\hat{\rho}^{(2)}(\boldsymbol{R}_{t})$,
we obtain an approximative description of a quantum system whose dynamics
is driven by an external semi-classical agent, where
\begin{equation}
i\hbar\frac{d}{dt}\hat{\rho}^{(1)}(t)\approx[\tilde{H}^{(1)}(\boldsymbol{R}_{t}),\hat{\rho}^{(1)}(t)]
\end{equation}
and
\begin{equation}
\tilde{H}^{(1)}(\boldsymbol{R}_{t})\approx\hat{H}^{(1)}+tr_{2}\{\hat{H}_{int}\hat{\rho}^{(2)}(\boldsymbol{R}_{t})\}.
\end{equation}
Hence, as expected, from a fully autonomous quantum description one
might obtain an - approximative - asymmetric effective behaviour,
under the right conditions. This is exactly the case of a single spin
weakly interacting with a magnetic field, for instance, whose controllable
parameter is the field intensity. In such cases, the full quantization
is possible and desirable, yet, this description level corresponds
to a highly complex task for many realistic scenarios, and the expressions
above correspond to valid approximative characterization of the local
dynamics. Nevertheless, from a quantum thermodynamic point of view,
it is important to highlight that outside this specific scope the
neglected energetic contributions will result in incomplete thermodynamic
descriptions.

%% file: Chapters/Chapter3/Section4.tex
\section{Internal energy and additivity\label{subsec:Internal-energy-and}}

As mentioned earlier, there is no ambiguity in identifying the internal
energy of isolated quantum systems: this role is naturally assigned
to the expectation value of the Hamiltonian generating its dynamics.
However, it is not entirely clear how to obtain a consistent and meaningful
analogous to arbitrary open quantum systems. In such cases, the notion
of local internal energy is blurred by non-negligible interactions
and correlations that might exist within the whole. Thus, any general
and coherent definition should somehow account for these elements.
Additionally, it is important to emphasize that a clear understanding
of internal energy is the most obvious first step toward proper definitions
of other fundamental thermodynamic quantities in the quantum regime,
especially quantum heat and work. In this section, we argue that the
local effective Hamiltonians are the representative physical operators
for characterizing the subsystems internal energies. Furthermore,
given this identification, we show that the thermodynamic notion of
energy additivity is naturally recovered.

\subsection{The whole and the parts internal energies}

By hypothesis, the whole system is closed and autonomous, which means
that no energy flows inward or outward. Thus, we immediately identify
the total internal energy $U^{(0)}$ as the expectation value of the
total Hamiltonian $\hat{H}^{(0)}$, i.e.,
\begin{equation}
U^{(0)}\equiv\langle\hat{H}^{(0)}\rangle=\langle\Psi(t)|\hat{H}^{(0)}|\Psi(t)\rangle,\label{Total Internal Energy}
\end{equation}
where $|\Psi(t)\rangle=\hat{\mathcal{U}}(t,t_{0})|\Psi(t_{0})\rangle$
and $\hat{\mathcal{U}}(t,t_{0})=e^{-\frac{i}{\hbar}\hat{H}^{(0)}(t-t_{0})}$.
Moreover, it is easy to see that this quantity is indeed conserved\footnote{The unitary evolution and constant Hamiltonian guarantee the following
equality for all $t$: $\langle\Psi(t)|\hat{H}^{(0)}|\Psi(t)\rangle=\langle\Psi(t_{0})|\hat{H}^{(0)}|\Psi(t_{0})\rangle$.},
\begin{equation}
\frac{d}{dt}U^{(0)}=0.
\end{equation}
However, since we are dealing with a bipartite system, from Eq. (\ref{Total Hamiltonian})
we are able to rewrite $U^{(0)}$ as
\begin{equation}
U^{(0)}=\langle\hat{H}^{(1)}\rangle(t)+\langle\hat{H}^{(2)}\rangle(t)+\langle\hat{H}_{int}\rangle(t)\label{Energy balance1}
\end{equation}
where $\langle.\rangle\equiv\langle\Psi(t)|(.)|\Psi(t)\rangle$ and,
therefore, 
\begin{equation}
\langle\hat{H}^{(k)}\rangle(t)=tr_{k}\{\hat{H}^{(k)}\hat{\rho}^{(k)}(t)\}=\sum_{j=1}^{d^{(1)}}\lambda_{j}^{2}(t)\langle\varphi_{j}^{(k)}(t)|\hat{H}^{(k)}|\varphi_{j}^{(k)}(t)\rangle.
\end{equation}
Notice that the total internal energy is the sum of the expectation values
of the bare Hamiltonians plus the interaction between the subsystems.
It is worth mentioning that, even though these operators are constant,
the time-dependency of their expectation values is due to the state dynamics,
in such a way that any change in $\langle\hat{H}^{(1)}\rangle(t)+\langle\hat{H}^{(2)}\rangle(t)$
induces the negative variation in $\langle\hat{H}_{int}\rangle(t)$,
i.e.,
\begin{equation}
\frac{d}{dt}\langle\hat{H}^{(1)}\rangle(t)+\frac{d}{dt}\langle\hat{H}^{(2)}\rangle(t)=-\frac{d}{dt}\langle\hat{H}_{int}\rangle(t).
\end{equation}
From the previous equations, it is not clear how to properly assign
internal energies for each subsystem. Given that the interaction term
actively influences their local dynamics, it is reasonable to assume
that its contribution should be somehow shared between them. Besides,
it is also desirable two relevant properties for the local energies:
\textbf{(i)} be obtained by local measurements, i.e., associated with the expectation value of local operators; \textbf{(ii)} be
an additive quantity (extensive property). While the first condition
guarantees a local description and accessibility, the second also
allows the intuitive picture of energy flowing from one system to
another without including energetic sinks or sources, i.e., the sum
of the local internal energies is a conserved quantity. These features,
of course, are not trivial, especially because the interaction term
acts on the whole Hilbert space, which means that it is a global property
per se. This fact, however, suggests that an effective approach for
describing local internal energy provides the most promising route.
Otherwise, a global picture would be necessary for fully characterizing
the energetic flux, which is impractical for most realistic scenarios.

Different approaches for how accounting the interaction input can
be found in the literature. \cite{Weimer2008,Hossein-Nejad2015,Alipour2016a,Miller2018c,Rivas2020,Colla2021a} Nevertheless, the most common
route in quantum thermodynamics is to directly identify local internal
energies as the expectation values of the bare Hamiltonians. Consequently,
the sum of the parts is not equal to the whole, and the total internal
energy is not additive, in general (in contrast with classical scenarios).
To circumvent this issue, it is also usually necessary to assume additional
hypotheses\footnote{Although sometimes not explicitly.} on the form
and/or strength of the interaction operator. After all, if the interaction
could be ignored the recognition of local internal energies would
be immediate, and the two desired properties would be automatically
satisfied. In this sense, the so-called weak-coupling approximation
is the most frequent assumption when dealing with open quantum systems
dynamics. It explicitly assumes that the interaction term is small
enough to be treated as a perturbation. The formal procedure follows
the usual perturbative recipe: which consists in scaling the interaction
term, such that $\hat{H}_{int}\rightarrow\alpha\hat{H}_{int}$, then
expanding the local states in a series of $\alpha$ and, finally,
discarding high order terms, i.e.,
\begin{equation}
\hat{\rho}^{(k)}(t)=e^{\frac{i}{\hbar}\hat{H}^{(k)}(t-t_{0})}\hat{\rho}^{(k)}(t_{0})e^{-\frac{i}{\hbar}\hat{H}^{(k)}(t-t_{0})}+\alpha\hat{\rho}_{1}^{(k)}(t)+\alpha^{2}\hat{\rho}_{2}^{(k)}(t)+\mathcal{O}(\alpha^{3}),\label{state interaction perturbation}
\end{equation}
where $\hat{\rho}_{n}^{(k)}(t)$ is the $n$-th order correction.
In such expansion, the zeroth-order component is simply the initial
state time-evolved under the local bare Hamiltonian. Thus non-trivial
behaviour is only achieved if considered at least first-order contributions,
especially for obtaining non-unitary dynamics. This reasoning alone,
nevertheless, is not enough to justify the previous local internal
energy identification, given that the interaction term itself is also
in the first order and, consequently, it is still relevant in the
energetic balance, i.e., $U^{(0)}=\langle\hat{H}^{(1)}\rangle(t)+\langle\hat{H}^{(2)}\rangle(t)+\alpha\langle\hat{H}_{int}\rangle(t)$.
Interestingly, in the classical macroscopic thermodynamic setting
- \textit{a priori} - we would also have $U^{(0)}=U^{(1)}+U^{(2)}+U_{int}$,
however, the interaction input is several orders of magnitude smaller
than the other two individual elements, which supports its prompt
negligence\footnote{In the paradigmatic example of two ideal gases separated by a partition,
the interaction is intermediated by the former, and its energy is
proportional to its surface. This energy, however, is negligible compared
with the ones stored in each gas.}. It is worth mentioning that the usual derivation of master equations
in the Lindblad-like form is based on the second-order expansions
and also relies on other restrictive approximations, like Markov and
secular ones. \cite{BRE02} Still, even in such cases, there is no reason for not
considering the interaction. As an alternative, another common approach
is to assign the role of local internal energy for the operator arising
in the unitary part of the dynamical equation, which may automatically
contain the local bare Hamiltonian plus a correction due to the interaction.
\cite{Valente2018} In this context, the generator usually has the following
superoperator structure $\hat{\mathfrak{L}}_{t}=\mathfrak{\hat{H}}_{t}+\mathfrak{\hat{D}}_{t}$,
where $\mathfrak{H}_{t}\hat{\rho}^{(k)}(t)=[\hat{h}(t),\hat{\rho}^{(k)}(t)]$
is the desired unitary element and $\mathfrak{\hat{D}}_{t}$ is the
dissipator. Such procedure, however, should be carefully considered
since $\mathfrak{\hat{H}}_{t}$ may change over transformations that
keep the generator invariant. Thus complementary hypothesis may be
required for unambiguously fixing $\hat{h}(t)$. For a recent proposal,
see. \cite{Colla2021a}

Instead of focusing on the interaction strength, one might assume
specific Hamiltonian structures. For instance, given the expectation
value $\langle\hat{H}_{int}\rangle(t)=\langle\Psi(t)|\hat{H}_{int}|\Psi(t)\rangle$
it is easy to see that its time-evolution obeys the following equation
\begin{equation}
i\hbar\frac{d}{dt}\langle\hat{H}_{int}\rangle(t)=\langle[\hat{H}_{int},(\hat{H}^{(1)}\otimes\hat{1}^{(2)}+\hat{1}^{(1)}\otimes\hat{H}^{(2)})]\rangle.\label{SEC condition}
\end{equation}
The \textit{strict energy conservation} (SEC) condition is the assumption
that the commutator above is null, i.e., $[\hat{H}_{int},(\hat{H}^{(1)}\otimes\hat{1}^{(2)}+\hat{1}^{(1)}\otimes\hat{H}^{(2)})]=0$
\cite{dann2021open}, which leads to a constant expectation value of $\langle\hat{H}_{int}\rangle(t)=cte$
and, therefore,
\begin{equation}
SEC:\qquad\frac{d}{dt}\langle\hat{H}^{(1)}\rangle(t)=-\frac{d}{dt}\langle\hat{H}^{(2)}\rangle(t).
\end{equation}
Hence, if we accept the bare Hamiltonians as the representative operators,
despite the local internal energies still not being additive, the
energy flowing from one subsystem is necessarily obtained by the other,
i.e., the interaction neither captures nor releases any additional
energy. This fact justifies neglecting the interaction term into the
dynamical energetic analysis within this quantum system. Besides,
the SEC condition also implies that $\hat{\mathcal{U}}(t,t_{0})$
is a so-called \textit{energy-preserving unitary} (EPU)\footnote{It also implies that simultaneous local Gibbs states are fixed points
of the dynamics, since $\hat{\mathcal{U}}(t,t_{0})(e^{-\beta\hat{H}^{(1)}}\otimes e^{-\beta\hat{H}^{(2)}})\hat{\mathcal{U}}^{\dagger}(t,t_{0})=e^{-\beta\hat{H}^{(1)}}\otimes e^{-\beta\hat{H}^{(2)}}$.
In fact, any state that is a function of its bare Hamiltonian would
be.}, i.e.,
\begin{equation}
SEC\Rightarrow[\hat{\mathcal{U}}(t,t_{0}),(\hat{H}^{(1)}\otimes\hat{1}^{(2)}+\hat{1}^{(1)}\otimes\hat{H}^{(2)})]=0,
\end{equation}
which constitute a free operation in the context of resource theory
of thermal operations \cite{Lostaglio_2019}. Even though both SEC and EPU represents
useful scenarios, they also serve as very restrictive conditions in
the form of $\hat{H}_{int}$.

Despite its versatility and relevance, it is clear that approximative
procedures and particular hypotheses are only suitable for especific
regimes, and a general approach is necessary for developing a fully
quantum thermodynamic description of arbitrary systems.

\subsubsection{Local effective internal energy\label{subsec:Local-effective-internal}}

In Section (\ref{SECTION: Schmidt-basis-dynamics}), we obtained the
local effective Hamiltonian $\tilde{H}^{(k)}(t)$ as the generator
of the local Schmidt basis dynamics of subsystem $(k)$ (Eq. (\ref{Schmidt basis dynamics})),
and showed that it can be directly related to the bare Hamiltonian
through Eq. (\ref{Effective Hamiltonian 2}). In Section (\ref{SECTION: Local-states-dynamics}),
we also showed that the unitary part of the local state dynamics $i\hbar\frac{d}{dt}\hat{\rho}^{(k)}(t)$
is parametrized by $\tilde{H}^{(k)}(t)$. Let us now argue that such
Hamiltonians can be seen as the representative local operators for
characterizing the physical internal energy. 

First, by definition, both $\tilde{H}^{(1)}(t)$ and $\tilde{H}^{(2)}(t)$
are local objects, which means that they can be accessible by local
measurements, such that their expectation values are simply $\langle\tilde{H}^{(k)}(t)\rangle=\langle\Psi(t)|\tilde{H}^{(k)}(t)|\Psi(t)\rangle=tr_{k}\{\tilde{H}^{(k)}(t)\hat{\rho}^{(k)}(t)\}$.
Now, we are interested to investigate their relationship with the
whole internal energy $U^{(0)}$. Given Eq. (\ref{SchmidtDecomposition})
and Eq. (\ref{Schmidt basis dynamics}) for the Schmidt decomposition
and the Schmidt basis dynamics, we have the following equation
\begin{equation}
i\hbar\frac{d}{dt}|\Psi(t)\rangle=\sum_{j=1}^{d^{(1)}}\left(i\hbar\frac{d}{dt}\lambda_{j}(t)\right)|\varphi_{j}^{(1)}(t)\rangle\otimes|\varphi_{j}^{(2)}(t)\rangle+\left(\tilde{H}^{(1)}(t)+\tilde{H}^{(2)}(t)\right)|\Psi(t)\rangle.
\end{equation}
Thus, since $i\hbar\frac{d}{dt}|\Psi(t)\rangle=\hat{H}^{(0)}|\Psi(t)\rangle$,
it is easy to see that
\begin{align}
\langle\Psi(t)|\hat{H}^{(0)}|\Psi(t)\rangle & =\langle\Psi(t)|\sum_{j=1}^{d^{(1)}}\left(i\hbar\frac{d}{dt}\lambda_{j}(t)\right)|\varphi_{j}^{(1)}(t)\rangle\otimes|\varphi_{j}^{(2)}(t)\rangle\nonumber \\
 & +\langle\Psi(t)|\tilde{H}^{(1)}(t)|\Psi(t)\rangle+\langle\Psi(t)|\tilde{H}^{(2)}(t)|\Psi(t)\rangle.
\end{align}
However, notice that due to normalization of $|\Psi(t)\rangle$, the
first contribution is necessarily null,
\begin{equation}
\langle\Psi(t)|\sum_{j=1}^{d^{(1)}}\left(i\hbar\frac{d}{dt}\lambda_{j}(t)\right)|\varphi_{j}^{(1)}(t)\rangle\otimes|\varphi_{j}^{(2)}(t)\rangle=i\hbar\sum_{j=1}^{d^{(1)}}\lambda_{j}(t)\frac{d}{dt}\lambda_{j}(t)=0,
\end{equation}
since $\sum_{j=1}^{d^{(1)}}\lambda_{j}^{2}(t)=1$, and $\sum_{j=1}^{d^{(1)}}\lambda_{j}(t)\frac{d}{dt}\lambda_{j}(t)=\frac{1}{2}\frac{d}{dt}\sum_{j=1}^{d^{(1)}}\lambda_{j}^{2}(t)$.
Hence, surprisingly, the expectation value of the whole Hamiltonian $\hat{H}^{(0)}$
is exactly equal to the sum of the expectation values of the local effective
ones, i.e.,
\begin{equation}
\langle\hat{H}^{(0)}\rangle=\langle\tilde{H}^{(1)}(t)\rangle+\langle\tilde{H}^{(2)}(t)\rangle=U^{(0)}.\label{Energy balance2}
\end{equation}
If we identify $\langle\tilde{H}^{(1)}(t)\rangle$ and $\langle\tilde{H}^{(2)}(t)\rangle$
as the physical \textit{local (effective) internal energies} along
the respective paths $\{\mathcal{P}_{j}^{(1)}\}_{j}$ and $\{\mathcal{P}_{j}^{(2)}\}_{j}$,
such that
\begin{equation}
U^{(k)}(t):=\langle\tilde{H}^{(k)}(t)\rangle=i\hbar\sum_{j=1}^{d^{(1)}}\lambda_{j}^{2}(t)\langle\varphi_{j}^{(k)}(t)|\frac{d}{dt}|\varphi_{j}^{(k)}(t)\rangle,\label{eq:valor esperado}
\end{equation}
we automatically account for both the bare and interaction contributions,
\begin{equation}
U^{(k)}=\langle\hat{H}^{(k)}\rangle(t)+\langle\hat{H}_{LS}^{(k)}(t)\rangle+\langle\hat{H}_{X}^{(k)}(t)\rangle,\label{Local Internal energy}
\end{equation}
in such a way that we directly guarantee the additivity of energy
(extensive property),
\begin{equation}
U^{(0)}=U^{(1)}(t)+U^{(2)}(t),
\end{equation}
and, consequently, that the energy flowing from subsystem $(1)$ is
fully captured by subsystem $(2)$ and vice versa, i.e., 
\begin{equation}
\frac{d}{dt}U^{(1)}(t)=-\frac{d}{dt}U^{(2)}(t).
\end{equation}
It is important to emphasize that such equalities are exact, and no
additional hypotheses were required\footnote{While in classical thermodynamics energy additivity is an approximative
idealization (justified by the negligible interaction), in this context,
this is an exact statement.}.

Furthermore, it was mentioned earlier that, intuitively, both operators
$\hat{H}_{LS}^{(k)}(t)$ and $\hat{H}_{X}^{(k)}(t)$ are byproducts
of the interaction term. In fact, it is possible to show that their
expectation values are directly related. From Eq. (\ref{Energy balance1})
and Eq.(\ref{Energy balance2}), we have the following equality
\begin{equation}
\langle\hat{H}^{(1)}\rangle(t)+\langle\hat{H}^{(2)}\rangle(t)+\langle\hat{H}_{int}\rangle(t)=\langle\tilde{H}^{(1)}(t)\rangle+\langle\tilde{H}^{(2)}(t)\rangle,
\end{equation}
and, therefore, since Eq. (\ref{Local Internal energy}),
\begin{equation}
\langle\hat{H}_{int}\rangle(t)=\langle\hat{H}_{LS}^{(1)}(t)\rangle+\langle\hat{H}_{X}^{(1)}(t)\rangle+\langle\hat{H}_{LS}^{(2)}(t)\rangle+\langle\hat{H}_{X}^{(2)}(t)\rangle.\label{Interaction balance}
\end{equation}
The previous equation states how the energetic contribution coming
from the interaction term is symmetrically shared between the subsystems\footnote{Of course, this does not imply that their modulus cannot be extremely
different.}, and how its change affects the local internal energies. Interestingly,
notice that the SEC condition, Eq. (\ref{SEC condition}), is analogous
to supposing that
\begin{equation}
SEC:\qquad\frac{d}{dt}\langle\hat{H}_{LS}^{(1)}(t)+\hat{H}_{X}^{(1)}(t)\rangle=-\frac{d}{dt}\langle\hat{H}_{LS}^{(2)}(t)+\hat{H}_{X}^{(2)}(t)\rangle,
\end{equation}
i.e., the change of the subsystem $(1)$ local internal energy due
to $\hat{H}_{LS}^{(1)}(t)+\hat{H}_{X}^{(1)}(t)$ dynamics is perfectly
balanced by $\hat{H}_{LS}^{(2)}(t)+\hat{H}_{X}^{(2)}(t)$, in a way
that their net change is null.

In summary, the recognition of the local effective Hamiltonians as
the representative operators for describing the physical internal
energies allow us to consistently refer to these local quantities
without explicitly mentioning global properties. Most importantly,
this procedure is exact and general and, thus, applicable to \textit{any}
setting and regime. We consider this as one of our main results.

%% file: Chapters/Chapter3/Section5.tex
\section{Local phase gauge\label{sec:Local-phase-gauge}}

It was mentioned earlier that even though the Schmidt coefficients
$\{\lambda_{j}(t)\}_{j}$ are unambiguously defined by the Schmidt
decomposition, its basis $\{|\varphi_{j}^{(k)}(t)\rangle\}_{j,k}$
are unique up to degeneracy and a phase component. In this section,
we will present and investigate how the latter ambiguity influences
our local effective description. Then we will discuss and emphasize
its consequences at the energetic level and argue how to, possibly,
fix such a freedom. 

Generally speaking, phases are intrinsic to the mathematical formalism
of quantum mechanics. In fact, it is a direct consequence of representing
physical quantum states in Hilbert spaces. It is often common to introduce
the concept of global and relative phases. While the former is usually
treated as simply artefacts, the latter are viewed as sources of fundamentally
quantum behaviour, e.g., coherence. However, it is worth mentioning
that despite being superfluous for any physical description and measurement,
global phases are deeply connected with the underlying geometry of
these abstract structures and far from being unimportant\footnote{The notion of geometric phase, for instance, arises in such context
and, since initial developments by Berry \cite{M.V.Berry03081984}, it became clear
its importance for the complete understanding of quantum mechanics.}. More specifically, given a Hilbert space $\mathcal{H}$, physical
states are not uniquely related to kets from $\mathcal{H}$, i.e.,
both $|\Psi\rangle\in\mathcal{H}$ and $|\Psi^{\prime}\rangle=e^{i\theta}|\Psi\rangle\in\mathcal{H}$,
simply differing by $e^{i\theta}$, represents the same physical system
for any $\theta$ real. Given the measurement postulate, note that
all possible extractable information from $|\Psi\rangle$ is also
equally encoded by $|\Psi^{\prime}\rangle$. Such phases invariance
illustrates a fundamental gauge transformation inbuilt in the core
of the theory. That said, let us now consider an arbitrary Schmidt
decomposition, presented by Eq. (\ref{SD}). It is clear that the
simultaneous addition of local phases $\{\theta_{\eta}\}_{\eta}$,
such that 
\begin{align}
|\varphi_{\eta}^{(1)}\rangle & \rightarrow|\varphi_{\eta}^{\prime(1)}\rangle=e^{i\theta_{\eta}}|\varphi_{\eta}^{(1)}\rangle,\\
|\varphi_{\eta}^{(2)}\rangle & \rightarrow|\varphi_{\eta}^{\prime(2)}\rangle=e^{-i\theta_{\eta}}|\varphi_{\eta}^{(2)}\rangle,
\end{align}
maintains the whole quantum state structure unchanged since the phases
cancel out and, therefore,
\begin{equation}
|\Psi\rangle=\sum_{\eta=1}^{n}\lambda_{\eta}|\varphi_{\eta}^{(1)}\rangle\otimes|\varphi_{\eta}^{(2)}\rangle=\sum_{\eta=1}^{n}\lambda_{\eta}|\varphi_{\eta}^{\prime(1)}\rangle\otimes|\varphi_{\eta}^{\prime(2)}\rangle=|\Psi^{\prime}\rangle.\label{Ambiguity}
\end{equation}
Interestingly, from a local point of view, any phase gauge transformation
is valid in the sense that it keeps describing the same physical state.
Nevertheless, from a global perspective, the same phases (in modulus)
should be included in the remaining Schmidt basis to guarantee consistency.
Otherwise, we would be adding relative phases and changing the whole
system state $|\Psi\rangle$. Such \textquotedbl flexibility\textquotedbl{}
corresponds to an internal freedom within the Schmidt decomposition
itself\footnote{More fundamentally, it comes from the singular value decomposition.}
that, naturally, will be inherited by our description. Along these
lines, considering the set of simultaneous coupled trajectories $\mathcal{P}_{j}^{(1,2)}$
followed by the Schmidt basis $|\varphi_{j}^{(1,2)}(t)\rangle$ in
$\mathcal{H}^{(1,2)}$, we might define a set of real functions $\{\theta_{j}(t)\}_{j}$
for $t\in[t_{0},t_{1}]$ that transform the curves $\{\mathcal{P}_{j}^{(1,2)}\}_{j}$
into a new phase gauge $\{\mathcal{P}_{j}^{\prime(1,2)}\}_{j}$ that
still represent the whole state trajectory $\mathcal{P}^{(0)}$ of
$|\Psi(t)\rangle$ in the total Hilbert space $\mathcal{H}^{(0)}$,
i.e.,
\begin{align}
\mathcal{P}_{j}^{(1)} & \rightarrow\mathcal{P}_{j}^{\prime(1)}:|\varphi_{j}^{\prime(1)}(t)\rangle=e^{i\theta_{j}(t)}|\varphi_{j}^{(1)}(t)\rangle,\:t\in[t_{0},t_{1}],\label{Phase gauge transf 1}\\
\mathcal{P}_{j}^{(2)} & \rightarrow\mathcal{P}_{j}^{\prime(2)}:|\varphi_{j}^{\prime(2)}(t)\rangle=e^{-i\theta_{j}(t)}|\varphi_{j}^{(2)}(t)\rangle,\:t\in[t_{0},t_{1}],\label{Phase gauge transf 2}
\end{align}
such that $\mathcal{P}^{(0)}\rightarrow\mathcal{P}^{\prime(0)}=\mathcal{P}^{(0)}$
and $\langle\varphi_{\alpha}^{\prime(k)}(t)|\varphi_{\beta}^{\prime(k)}(t)\rangle=e^{(-1)^{k-1}i(\theta_{\beta}(t)-\theta_{\alpha}(t))}\delta_{\alpha\beta}$.
Of course, given that the physical kets are invariant under such transformations,
it is clear that both local density operators $\hat{\rho}^{(1,2)}(t)$
(Eqs. (\ref{local state 1}, \ref{local state 2})) and their trajectories
$\mathcal{C}^{(1,2)}:\hat{\rho}^{(1,2)}(t),\:t\in[t_{0},t_{1}]$ along
the state space $\mathscr{D}(\mathcal{H}^{(k)})$ are not sensible
to phase changes, $\hat{\rho}^{\prime(1,2)}(t)=\hat{\rho}^{(1,2)}(t)$.
Nevertheless, since the local effective Hamiltonians $\tilde{H}^{(1,2)}(t)$,
given by Eq. (\ref{Effective Hamiltonian}), are functionals of the
Schmidt basis, it is straightforward to see that they intrinsically
depend on the chosen gauge. If we perform the transformations $\mathcal{P}_{j}^{(1,2)}\rightarrow\mathcal{P}_{j}^{\prime(1,2)}$
above and use the expression $i\hbar\frac{d}{dt}|\varphi_{j}^{(k)}(t)\rangle=\tilde{H}^{(k)}(t)|\varphi_{j}^{(k)}(t)\rangle$
we obtain
\begin{align}
i\hbar\frac{d}{dt}|\varphi_{j}^{\prime(1)}(t)\rangle & =\left(\tilde{H}^{(1)}(t)-\hbar\frac{d\theta_{j}(t)}{dt}\right)|\varphi_{j}^{\prime(1)}(t)\rangle,\\
i\hbar\frac{d}{dt}|\varphi_{j}^{\prime(2)}(t)\rangle & =\left(\tilde{H}^{(2)}(t)+\hbar\frac{d\theta_{j}(t)}{dt}\right)|\varphi_{j}^{\prime(2)}(t)\rangle,
\end{align}
for the Schmidt basis dynamics. Then, it is clear that the local effective
Hamiltonians in the new gauge, $\tilde{H}^{\prime(1,2)}(t)$, might
be directly related to those from the old ones, such that
\begin{align}
\tilde{H}^{\prime(1)}(t) & \equiv i\hbar\sum_{j=1}^{d^{(1)}}\frac{d}{dt}|\varphi_{j}^{\prime(1)}(t)\rangle\langle\varphi_{j}^{\prime(1)}(t)|=\tilde{H}^{(1)}(t)-\hbar\sum_{j=1}^{d^{(1)}}\left(\frac{d\theta_{j}(t)}{dt}\right)|\varphi_{j}^{(1)}(t)\rangle\langle\varphi_{j}^{(1)}(t)|,\label{New Frame Effective Hamiltonian 1}\\
\tilde{H}^{\prime(2)}(t) & \equiv i\hbar\sum_{j=1}^{d^{(2)}}\frac{d}{dt}|\varphi_{j}^{\prime(2)}(t)\rangle\langle\varphi_{j}^{\prime(2)}(t)|=\tilde{H}^{(2)}(t)+\hbar\sum_{j=1}^{d^{(2)}}\left(\frac{d\theta_{j}(t)}{dt}\right)|\varphi_{j}^{(2)}(t)\rangle\langle\varphi_{j}^{(2)}(t)|.\label{New Frame Effective Hamiltonian 2}
\end{align}
Observe that a gauge change adds an extra term, $\hbar\sum_{j=1}^{d^{(1,2)}}\left(\frac{d\theta_{j}(t)}{dt}\right)|\varphi_{j}^{(1,2)}(t)\rangle\langle\varphi_{j}^{(1,2)}(t)|$,
that only depends on the time derivative of the phases and are diagonal
on their respective Schmidt basis. In general, these additional quantities
will change the operator structure in a way that both their eigenbasis
and eigenvalues will be affected. This implies that, for most cases,
the spectral gaps will not maintain fixed\footnote{If $\{\epsilon_{j}^{(k)}(t)\}_{j}$ and $\{\epsilon_{j}^{\prime(k)}(t)\}_{j}$
are the eigenvalues of $\tilde{H}^{(k)}(t)$ and $\tilde{H}^{\prime(k)}(t)$,
respectively. Then, the gaps are simply defined by the changes $\hbar\omega{}_{\alpha\beta}^{(k)}(t)=\epsilon_{\beta}^{(k)}(t)-\epsilon{}_{\alpha}^{(k)}(t)$.
And, in general, for two different gauges, we have $\hbar\omega{}_{\alpha\beta}^{\prime(k)}(t)=\epsilon'{}_{\beta}^{(k)}(t)-\epsilon'{}_{\alpha}^{(k)}(t)\neq\epsilon{}_{\beta}^{(k)}(t)-\epsilon{}_{\alpha}^{(k)}(t)=\hbar\omega{}_{\alpha\beta}^{(k)}(t)$.} and $[\tilde{H}^{\prime(k)}(t),\tilde{H}^{(k)}(t)]\neq0$. Besides,
the extra term form means that $[\sum_{j=1}^{d^{(1,2)}}\left(\frac{d\theta_{j}(t)}{dt}\right)|\varphi_{j}^{(1,2)}(t)\rangle\langle\varphi_{j}^{(1,2)}(t)|,\hat{\rho}^{(1,2)}(t)]=0$
for all $t$, which also guarantees the invariance of the local state
dynamics written in Eq. (\ref{Master Eq. 2}). Concerning the expectation
values of $\tilde{H}^{\prime(1,2)}(t)$, by directly computing $\langle\tilde{H}^{\prime(k)}(t)\rangle=\langle\Psi^{\prime}(t)|\tilde{H}^{\prime(k)}(t)|\Psi^{\prime}(t)\rangle$
we see that a phase gauge transformation just perform a shift in the
mean value obtained from the previous gauge, i.e.,
\begin{align}
\langle\tilde{H}^{\prime(1)}(t)\rangle & =\langle\tilde{H}^{(1)}(t)\rangle-\hbar\sum_{j=1}^{d^{(1)}}\lambda_{j}^{2}(t)\left(\frac{d\theta_{j}(t)}{dt}\right),\label{New gauge effective Hamiltonians expected value 1}\\
\langle\tilde{H}^{\prime(2)}(t)\rangle & =\langle\tilde{H}^{(2)}(t)\rangle+\hbar\sum_{j=1}^{d^{(1)}}\lambda_{j}^{2}(t)\left(\frac{d\theta_{j}(t)}{dt}\right).\label{New gauge effective Hamiltonians expected value 2}
\end{align}
Interestingly, the shift accumulated by subsystem $(1)$, $\hbar\sum_{j=1}^{d^{(1)}}\lambda_{j}^{2}(t)\left(\frac{d\theta_{j}(t)}{dt}\right)$,
is compensated by the one acquired by subsystem $(2)$\footnote{Since there are only $d^{(1)}$ Schmidt coefficients, we have $\lambda_{\eta}(t)=0$ for $d^{(1)}<\eta\leq d^{(2)}$, and the following equality $\sum_{j=1}^{d^{(2)}}\lambda_{j}^{2}(t)\left(\frac{d\theta_{j}(t)}{dt}\right)=\sum_{j=1}^{d^{(1)}}\lambda_{j}^{2}(t)\left(\frac{d\theta_{j}(t)}{dt}\right)$ is satisfied.}. Hence, as expected, such a phase gauge automatically ensures that the additivity
property obtained in Eq. (\ref{Energy balance2}) is still satisfied
for any transformation:
\begin{equation}
U^{(0)}=\langle\hat{H}^{(0)}\rangle=\langle\tilde{H}^{(1)}(t)\rangle+\langle\tilde{H}^{(2)}(t)\rangle=\langle\tilde{H}^{\prime(1)}(t)\rangle+\langle\tilde{H}^{\prime(2)}(t)\rangle,
\end{equation}
and, therefore,
\begin{equation}
\frac{d}{dt}\langle\tilde{H}^{\prime(1)}(t)\rangle=-\frac{d}{dt}\langle\tilde{H}^{\prime(2)}(t)\rangle.
\end{equation}

Finally, it is clear that for every phase gauge, there is a pair of
local effective Hamiltonians $\tilde{H}^{\prime(1,2)}(t)$ attained
to it. However, this poses a fundamental obstacle for interpreting their expectation values as the \textit{physical} local effective internal energies: after all, if all gauges correspond to the same physical state, which one is the representative one for characterizing the internal energies?

Of course, different choices may lead to very distinct conclusions.
Yet, not all possibilities preserve some of the desired properties.
For instance, since any set $\{\theta_{j}(t)\}_{j}$ represents a
valid gauge (as long as it is a real function), let us strategically
choose
\begin{equation}
\hbar\frac{d}{dt}\theta_{j}(t)=\langle\varphi{}_{j}^{(1)}(t)|\tilde{H}^{(1)}(t)|\varphi_{j}^{(1)}(t)\rangle,\label{eq:parallel transport}
\end{equation}
for all $j$. Thus, Eq. (\ref{New Frame Effective Hamiltonian 1})
becomes
\begin{equation}
\tilde{H}^{\prime(1)}(t)=\tilde{H}^{(1)}(t)-\sum_{j=1}^{d^{(1)}}\langle\varphi{}_{j}^{(1)}(t)|\tilde{H}^{(1)}(t)|\varphi_{j}^{(1)}(t)\rangle|\varphi_{j}^{(1)}(t)\rangle\langle\varphi_{j}^{(1)}(t)|,
\end{equation}
which, essentially, subtracts from $\tilde{H}^{(1)}(t)$ its diagonal
elements in the instantaneous Schmidt basis representation, $\langle\varphi{}_{j}^{(1)}(t)|\tilde{H}^{(1)}(t)|\varphi_{j}^{(1)}(t)\rangle$.
Also, notice it implies the mean value from Eq. (\ref{New gauge effective Hamiltonians expected value 1})
is null for all $t$, 
\begin{equation}
\langle\tilde{H}^{\prime(1)}(t)\rangle=0
\end{equation}
since $\hbar\sum_{j=1}^{d^{(1)}}\lambda_{j}^{2}(t)\left(\frac{d\theta_{j}(t)}{dt}\right)=\langle\tilde{H}^{(1)}(t)\rangle$.
Nevertheless, this difference is compensated by subsystem $(2)$,
such that
\begin{equation}
\langle\tilde{H}^{\prime(2)}(t)\rangle=\langle\tilde{H}^{(2)}(t)\rangle+\langle\tilde{H}^{(1)}(t)\rangle,
\end{equation}
and $U^{(0)}=\langle\tilde{H}^{\prime(2)}(t)\rangle=\langle\tilde{H}^{(1)}(t)\rangle+\langle\tilde{H}^{(2)}(t)\rangle$
remains invariant. As we can see, according to this phase gauge choice,
if we identify $U^{(k)}(t):=\langle\tilde{H}^{\prime(k)}(t)\rangle$,
there is no energetic flow since $\frac{d}{dt}\langle\tilde{H}^{\prime(1,2)}(t)\rangle=0$,
and all the internal energy of the whole bipartite system is gathered
exclusively by the subsystem $(2)$, which eliminates our symmetrical
perspective. Moreover, the opposite conclusion is obtained if considered
$\hbar\frac{d}{dt}\theta_{j}(t)=-\langle\varphi{}_{j}^{(2)}(t)|\tilde{H}^{(2)}(t)|\varphi_{j}^{(2)}(t)\rangle$
instead. Interestingly, given Eq. (\ref{New Frame Effective Hamiltonian 1}),
the assumption from Eq. (\ref{eq:parallel transport}) translates
to $\langle\varphi_{j}^{\prime(1)}(t)|\frac{d}{dt}|\varphi_{j}^{\prime(1)}(t)\rangle=0$
for all $j$, i.e., we are dealing with the gauge where all kets $\{|\varphi_{j}^{\prime(1)}(t)\rangle\}_{j}$
are parallel transported. \cite{tong2004kinematic}

Alternatively, we may also assume the following gauge
\begin{equation}
\hbar\frac{d\theta_{j}(t)}{dt}=\langle\varphi_{j}^{(1)}(t)|\hat{H}_{LS}^{(1)}(t)|\varphi_{j}^{(1)}(t)\rangle+\langle\varphi_{j}^{(1)}(t)|\hat{H}_{X}^{(1)}(t)|\varphi_{j}^{(1)}(t)\rangle.
\end{equation}
 In this case, we have $\langle\varphi_{j}^{\prime(1)}(t)|\tilde{H}^{\prime(1)}(t)|\varphi_{j}^{\prime(1)}(t)\rangle=\langle\varphi_{j}^{(1)}(t)|\hat{H}^{(1)}|\varphi_{j}^{(1)}(t)\rangle$
for all $j$ and, therefore,
\begin{equation}
\hbar\sum_{j=1}^{d^{(1)}}\lambda_{j}^{2}(t)\left(\frac{d\theta_{j}(t)}{dt}\right)=\langle\hat{H}_{LS}^{(1)}(t)\rangle+\langle\hat{H}_{X}^{(1)}(t)\rangle,
\end{equation}
which also implies that all energetic contributions for subsystem
$(1)$ except coming from its bare Hamiltonian are eliminated, such
that 
\begin{equation}
\langle\tilde{H}^{\prime(1)}(t)\rangle=\langle\hat{H}^{(1)}\rangle(t),
\end{equation}
and, consequently, $\langle\tilde{H}^{\prime(2)}(t)\rangle=\langle\tilde{H}^{(2)}(t)\rangle+\langle\hat{H}_{LS}^{(1)}(t)\rangle+\langle\hat{H}_{X}^{(1)}(t)\rangle.$
However, given Eq. (\ref{Interaction balance}), we obtain that
\begin{equation}
\langle\tilde{H}^{\prime(2)}(t)\rangle=\langle\hat{H}^{(2)}\rangle(t)+\langle\hat{H}_{int}\rangle(t),
\end{equation}
 i.e., if we identify $U^{(k)}(t):=\langle\tilde{H}^{\prime(k)}(t)\rangle$,
it would be concluded that the local internal energy of subsystem
$(1)$ is fully characterized by its bare Hamiltonian, $\hat{H}^{(1)}$,
and all the interaction contributions belong entirely to subsystem
$(2)$. Again, this perspective also breaks our desired symmetrical
description, and, clearly, by correctly changing the gauge, the opposite
roles can be easily obtained. Moreover, it is interesting to notice
that, essentially, this gauge corresponds to the usual identification
of the local internal energies.

\subsection{Frame change}

Before attempting to answer the earlier questioning, let us now recast
the previous discussion in the language of frame changes. In general,
frame changes are represented by unitary operators $\hat{\Theta}(t)$
that map - in a convenient way - a given physical state from one representation
to another, i.e., $|\psi\rangle\rightarrow|\psi'\rangle=\hat{\Theta}(t)|\psi\rangle$\footnote{Essentially, the well known Heisenberg and interaction pictures are
specific cases of these kinds of transformations. The usual \textquotedbl rotating
frame\textquotedbl{} is also an example.}. Such unitaries keep observation outcomes from different frames invariant,
i.e., the observables also change in a way that maintains their spectra
unaffected. This behaviour, however, is not observed for the whole
Hamiltonian in particular, since the Schrödinger equation must be
covariant under these frame transformations. This result is analogous
to changing the frame of reference and, consequently, the potential
energy in classical mechanics: the internal energy computed by different
external observers will depend on their particular frames, and despite
observing distinct dynamics on their subjective perspective, they
will agree with the mean values. Additionally, this also implies there
is no fundamental privileged frame since all possibilities are equally
valid for describing the system of interest. Thus, by defining the
following phase operators
\begin{align}
\hat{\Theta}^{(1)}(t) & =\sum_{j=1}^{d^{(1)}}e^{i\theta_{j}(t)}|\varphi_{j}^{(1)}(t)\rangle\langle\varphi_{j}^{(1)}(t)|,\\
\hat{\Theta}^{(2)}(t) & =\sum_{j=1}^{d^{(2)}}e^{-i\theta_{j}(t)}|\varphi_{j}^{(2)}(t)\rangle\langle\varphi_{j}^{(2)}(t)|,
\end{align}
and
\begin{equation}
\hat{\Theta}^{(0)}(t)=\hat{\Theta}^{(1)}(t)\otimes\hat{\Theta}^{(2)}(t),
\end{equation}
such that $\hat{\Theta}^{(k)}(t)\hat{\Theta}^{(k)\dagger}(t)=\hat{1}^{(k)}$
for $k\in[0,2]$, we can write the whole state transformation as
\begin{equation}
|\Psi'(t)\rangle=\sum_{j=1}^{d^{(1)}}\lambda_{j}(t)|\varphi_{j}^{\prime(1)}(t)\rangle\otimes|\varphi_{j}^{\prime(2)}(t)\rangle=\hat{\Theta}^{(0)}(t)|\Psi(t)\rangle,
\end{equation}
while the Schmidt basis phase gauge transformation that are shown
in Eqs. (\ref{Phase gauge transf 1}, \ref{Phase gauge transf 2})
become
\begin{align}
|\varphi_{j}^{\prime(1)}(t)\rangle & =\hat{\Theta}^{(1)}(t)|\varphi_{j}^{(1)}(t)\rangle,\\
|\varphi_{j}^{\prime(2)}(t)\rangle & =\hat{\Theta}^{(2)}(t)|\varphi_{j}^{(2)}(t)\rangle.
\end{align}
Thus, given the basis time-evolution in the old frame $|\varphi_{j}^{(k)}(t)\rangle=\tilde{\mathcal{U}}^{(k)}(t,t_{0})|\varphi_{j}^{(k)}(t_{0})\rangle$,
the time evolution operator in the new frame is simply
\begin{equation}
|\varphi_{j}^{\prime(k)}(t)\rangle=\hat{\Theta}^{(k)}(t)\tilde{\mathcal{U}}^{(k)}(t,t_{0})\hat{\Theta}^{(k)\dagger}(t_{0})|\varphi_{j}^{\prime(k)}(t_{0})\rangle,
\end{equation}
where $\tilde{\mathcal{U}}^{\prime(k)}(t,t_{0})=\hat{\Theta}^{(k)}(t)\tilde{\mathcal{U}}^{(k)}(t,t_{0})\hat{\Theta}^{(k)\dagger}(t_{0})$
and the inverse transformation $|\varphi_{j}^{(k)}(t_{0})\rangle=\hat{\Theta}^{(k)\dagger}(t_{0})|\varphi_{j}^{\prime(k)}(t_{0})\rangle$
is automatically guaranteed by the unitarity of the frame change operator.
Besides, given the Schrödinger equation $i\hbar\frac{d}{dt}|\Psi(t)\rangle=\hat{H}^{(0)}|\Psi(t)\rangle$
and the basis dynamical equations Eq. (\ref{Schmidt basis dynamics}),
it is straightforward to show that their forms are covariant over
these frame transformations, in a way that
\begin{equation}
i\hbar\frac{d}{dt}|\Psi'(t)\rangle=\hat{H}^{\prime(0)}(t)|\Psi'(t)\rangle
\end{equation}
and
\begin{equation}
i\hbar\frac{d}{dt}|\varphi_{j}^{\prime(k)}(t)\rangle=\tilde{H}^{\prime(k)}(t)|\varphi_{j}^{\prime(k)}(t)\rangle,
\end{equation}
where
\begin{align}
\hat{H}^{\prime(0)}(t) & \equiv\hat{\Theta}^{(0)}(t)\hat{H}^{(0)}\hat{\Theta}^{(0)\dagger}(t)+i\hbar\left(\frac{d}{dt}\hat{\Theta}^{(0)}(t)\right)\hat{\Theta}^{(0)\dagger}(t),\\
\tilde{H}^{\prime(1,2)}(t) & \equiv\hat{\Theta}^{(1,2)}(t)\tilde{H}^{(1,2)}(t)\hat{\Theta}^{(1,2)\dagger}(t)+i\hbar\left(\frac{d}{dt}\hat{\Theta}^{(1,2)}(t)\right)\hat{\Theta}^{(1,2)\dagger}(t)\label{Frame transformation effective Hamiltonians}
\end{align}
are the global and local effective Hamiltonians represented in the
new frame, respectively. Interestingly, notice that the - once time-independent
- global Hamiltonian explicitly depends on time in the transformed
frame, in a way that $\langle\Psi^{\prime}(t)|\left(\frac{d}{dt}\hat{\Theta}^{(0)}(t)\right)\hat{\Theta}^{(0)\dagger}(t)|\Psi^{\prime}(t)\rangle=0$
and, therefore,
\begin{equation}
\langle\hat{H}^{\prime(0)}(t)\rangle=\langle\hat{H}^{(0)}\rangle=U^{(0)},
\end{equation}
i.e., despite $\hat{H}^{\prime(0)}(t)$ being time-dependent, its
mean value is constant and the conservation of the whole internal
energy is still satisfied, as expected. Furthermore, for the mean
values of the local effective Hamiltonians in the new frame we have
\begin{align}
\langle\tilde{H}^{\prime(1,2)}(t)\rangle & =\langle\tilde{H}^{(1,2)}(t)\rangle+i\hbar\left\langle \left(\frac{d}{dt}\hat{\Theta}^{(1,2)}(t)\right)\hat{\Theta}^{(1,2)\dagger}(t)\right\rangle .\label{Frame transformation effective Hamiltonians expected value}
\end{align}
It is easy to check that the frame change expressions above, Eqs.
(\ref{Frame transformation effective Hamiltonians}, \ref{Frame transformation effective Hamiltonians expected value}),
are the same as the ones presented previously in Eqs. (\ref{New Frame Effective Hamiltonian 1},
\ref{New Frame Effective Hamiltonian 2}, \ref{New gauge effective Hamiltonians expected value 1},
\ref{New gauge effective Hamiltonians expected value 2})\footnote{It is straightforward to see that
\begin{equation*}
\langle\Psi^{\prime}(t)|\left(\frac{d}{dt}\hat{\Theta}^{(k)}(t)\right)\hat{\Theta}^{(k)\dagger}(t)|\Psi^{\prime}(t)\rangle=(-1)^{k-1}i\sum_{j=1}^{d^{(k)}}\left(\frac{d\theta_{j}(t)}{dt}\right)\lambda_{j}^{2}(t).
\end{equation*}}.

Thus, the internal phase gauge freedom corresponds to a
family of frames that consistently describe the whole and the local
dynamics. In a way that each possible frame characterizes different
Schmidt basis dynamics and, therefore, distinct local effective Hamiltonians,
i.e., they are not frame-invariant. Nevertheless, this set of allowed
frame transformations guarantees that the whole internal energy remains
the same. Hence, the question previously posed becomes: which frame
(or frames) is (are) the relevant ones for characterizing the local internal
energies?

\subsection{Recap and gauge fixing proposal\label{Recap and gauge}}

In retrospect, during Section (\ref{subsec:Local-effective-internal}),
we argued that the local effective Hamiltonians $\tilde{H}^{(1,2)}(t)$
are interesting candidates for being the representative operators
for characterizing the physical internal energies in an exact and
complete general way. More specifically, it was shown that these operators
are hermitian, local - by construction - and also satisfy the usual
notion of energy additivity (or extensivity). Then, in the current
Section, we just identified that for a unique bipartite physical system
there is an intrinsic phase gauge freedom within the Schmidt
decomposition structure (see Eq. (\ref{Ambiguity})). Such ambiguity
translates into a degeneracy for defining those local operators and
might be interpreted as the result of the existence of a set of possible
frames that consistently describes the subsystems dynamics and the
global energetics. In short, for the same physical system and behaviour,
we can identify a large family of frames, each one with a particular
pair of coupled local effective Hamiltonians $\{\tilde{H}^{\prime(1)}(t),\tilde{H}^{\prime(2)}(t)\}$
satisfying $\frac{d}{dt}\langle\tilde{H}^{\prime(1)}(t)\rangle=-\frac{d}{dt}\langle\tilde{H}^{\prime(2)}(t)\rangle$,
in a way that it is not clear which one should be considered for quantifying
the internal energies $U^{(1,2)}(t)$.

Interestingly, this kind of ambiguity is not exclusive to our discussion
since it also happens in the classical mechanics context, so let us
digress a little bit to the classical realm. In a very general sense,
once established the usual Lagrangian formulation of mechanics one
may restructure the theory to a Hamiltonian picture simply by performing
the following Legendre transformation 
\begin{equation}
H(\boldsymbol{q},\boldsymbol{p},t)=\dot{\boldsymbol{q}}.\boldsymbol{p}-L(\boldsymbol{q},\dot{\boldsymbol{q}},t),\label{Legendre transform}
\end{equation}
where $\dot{\boldsymbol{q}}=\frac{d\boldsymbol{q}}{dt}$, and $\boldsymbol{q}=\{q_{j}\}_{j}^{N}$
and $\boldsymbol{p}=\{p_{j}=\frac{\partial L}{\partial\dot{q}_{j}}\}_{j}^{N}$
are the respective $N$-dimensional set of generalized coordinates
and conjugate momenta, while $L(\boldsymbol{q},\dot{\boldsymbol{q}},t)$
and $H(\boldsymbol{q},\boldsymbol{p},t)$ are the Lagrangian describing
the system and its Hamiltonian, respectively. Thus, to obtain the
Hamiltonian of a given problem, it is just required to follow a specific
set of mathematical steps. Although straightforward, at least in theory,
this might be a very complex procedure. Interestingly, for particular
circumstances, there is a formal and justifiable shortcut to this
recipe: if the generalized coordinates $\boldsymbol{q}$ do not explicitly
depend on time and there are only conservative potentials, the Hamiltonian
is necessarily equal to the system's total energy $U$, i.e., 
\begin{equation}
H=K+V=U,\label{Hamiltonian equal total energy}
\end{equation}
where $K$ and $V$ are the kinetic and potential energies, respectively.
Therefore, if one of these conditions is unfulfilled the Hamiltonian
is not automatically equal to the internal energy. This, nevertheless,
represents the most generic situation, with Eq. (\ref{Hamiltonian equal total energy})
being the particular case. Also, in contrast to the Lagrangian, it
is clear that the Hamiltonian description is intrinsically bound to
the generalized coordinates, i.e., $L(\boldsymbol{q},\dot{\boldsymbol{q}},t)$
may functionally depend on this choice but its numerical value (magnitude)
maintains fixed for generalized coordinate changes, while, as emphasized
by H. Goldstein in \cite{goldstein:mechanics}, \textquotedbl\textit{the Hamiltonian
is dependent both in magnitude and in functional form upon the initial
choice of generalized coordinates}\textquotedbl . Hence, for the
same physical system, one may construct distinct Hamiltonians by employing
different sets of generalized coordinates in the definition presented
by Eq. (\ref{Legendre transform}). In particular, a conserved Hamiltonian
described by one set of coordinates might be time-dependent in another,
which illustrates the fact that the conditions required for the Hamiltonian
be the total energy are not the same for being a conserved quantity.
For a detailed discussion and examples on these matters, see Chapter
8 of. \cite{goldstein:mechanics}

So, from the discussion above, it is now clear that the relationship
between Hamiltonian and total energy is not always straightforward,
even in classical mechanics. In such a context, the existence of different
possible generalized coordinates for representing a given system produces
an ambiguity that might generate Hamiltonians of different forms,
time-dependency and magnitude. Of course, despite not being necessarily
equal to the system's total energy, any choice is suitable for consistently
describing the dynamics. That said, this situation is analogous to
what we obtained before: in both cases, there is a set of valid Hamiltonians
and a source of ambiguity, where the frames/phase gauge plays a similar role played by the classical coordinates. However, in the classical scenario,
the identification of the Hamiltonian that correctly describes the
internal energy is easily checked, i.e., Eq. (\ref{Hamiltonian equal total energy})
provides an independent prescription for calculating this quantity
and comparing the results. Interestingly, even though there is no comparable straightforward manner to inspecting the relationship between the local effective Hamiltonians and energy, we can still identify the set of physically consistent phases and fix the relevant gauges.

First, let us emphasize that the addition of local phases is a consequence of the mathematical freedom within the Schmidt decomposition and does not depend on the Hamiltonian structure of the whole system, i.e., the phases are arbitrary and are independent of the local bare Hamiltonians $\hat{H}^{(1,2)}$ and, more importantly, of the interaction term $\hat{H}_{int}$. Then, as shown earlier in Section (\ref{SECTION: Schmidt-basis-dynamics}), given the following form for the local effective Hamiltonian $\tilde{H}^{(k)}(t)=\hat{H}^{(k)}+\hat{H}_{LS}^{(k)}(t)+\hat{H}_{X}^{(k)}(t)$, it is clear that the additional terms $\hat{H}_{LS}^{(k)}(t)$ and $\hat{H}_{X}^{(k)}(t)$ are by-products of the existing interaction between the subsystems, i.e., in the absence of $\hat{H}_{int}$ both subsystems would behave independently as isolated objects and their local effective Hamiltonians would be simply identified as their bare ones, $\tilde{H}^{(k)}(t)=\hat{H}^{(k)}$. Note that a similar conclusion should be true \textit{regardless} of the chosen gauge. To see how to guarantee this, let us rewrite Eqs. (\ref{New Frame Effective Hamiltonian 1}, \ref{New Frame Effective Hamiltonian 2}) as follows:
\begin{align}
\tilde{H}^{\prime(1)}(t) & =\hat{H}^{(1)}+\hat{H}_{LS}^{(1)}(t)+\hat{H}_{X}^{(1)}(t)-\hbar\sum_{j=1}^{d^{(1)}}\left(\frac{d\theta_{j}(t)}{dt}\right)|\varphi_{j}^{(1)}(t)\rangle\langle\varphi_{j}^{(1)}(t)|,\\
\tilde{H}^{\prime(2)}(t) & =\hat{H}^{(2)}+\hat{H}_{LS}^{(2)}(t)+\hat{H}_{X}^{(2)}(t)+\hbar\sum_{j=1}^{d^{(2)}}\left(\frac{d\theta_{j}(t)}{dt}\right)|\varphi_{j}^{(2)}(t)\rangle\langle\varphi_{j}^{(2)}(t)|.
\end{align}
If we add a dimensionless parameter $\varepsilon\in[0,1]$, such that $\hat{H}_{int}\rightarrow\varepsilon\hat{H}_{int}$ and make $\varepsilon\rightarrow0$, for the new gauge we would obtain
\begin{align}
\tilde{H}^{\prime(1)}(t) & =\hat{H}^{(1)}-\hbar\sum_{j=1}^{d^{(1)}}\left(\frac{d\theta_{j}(t)}{dt}\right)|\varphi_{j}^{(1)}(t)\rangle_{bare}\langle\varphi_{j}^{(1)}(t)|_{bare},\\
\tilde{H}^{\prime(2)}(t) & =\hat{H}^{(2)}+\hbar\sum_{j=1}^{d^{(2)}}\left(\frac{d\theta_{j}(t)}{dt}\right)|\varphi_{j}^{(2)}(t)\rangle_{bare}\langle\varphi_{j}^{(2)}(t)|_{bare}.
\end{align}
where $|\varphi_{j}^{(k)}(t)\rangle_{bare}=e^{-\frac{i}{\hbar}\hat{H}^{(k)}(t-t_{0})}|\varphi_{j}^{(k)}(t_{0})\rangle$ is the free evolution of the initial Schmidt basis for $t\geq t_{0}$\footnote{As shown in Section (\ref{SECTION: Schmidt-basis-dynamics}), the perturbative series of the Schmidt basis elements are given by $|\varphi_{j}^{(k)}(t)\rangle=|\varphi_{j}^{(k)}(t)\rangle_{bare}+\varepsilon|\varphi_{j}^{(k)}(t)\rangle_{1}+\mathcal{O}^{(k)}(\varepsilon^{2})$.}. Observe that the phases are still relevant, but, despite being mathematically allowed, not all sets of phases are necessarily physically consistent with the expected behaviour in the absence of the interaction. Of course, if we constrain $\frac{d}{dt}\theta_{j}(t)=0$ for all $j$, we automatically obtain $\tilde{H}^{\prime(k)}(t)=\hat{H}^{(k)}$. However, we can be more general: instead, if we require $\frac{d}{dt}\theta_{j}(t)=\alpha\in\mathbb{R}$ for all $j$, the limiting expressions above would provide equivalent conclusions, in the sense that they would still consistently describe the same local energy measurement differences since
\begin{align}
\tilde{H}^{\prime(1)}(t) & =\hat{H}^{(1)}-\hbar\alpha\hat{1}^{(1)},\\
\tilde{H}^{\prime(2)}(t) & =\hat{H}^{(2)}+\hbar\alpha\hat{1}^{(2)},
\end{align}
and the additive constant $\hbar\alpha\hat{1}^{(k)}$ just equally shifts the energy spectrum, i.e., as long the phases are linear functions of time, such that $\theta_{j}(t)=\alpha t+c_{0}$ with $c_{0}\in\mathbb{R}$ being an arbitrary constant for all $j$, the expected compatibility with the limiting behaviour is guaranteed.

Hence, in order to assure physical consistency, we must only consider gauges such that $\left\{ \frac{d}{dt}\theta_{j}(t)=\alpha\right\} _{j}$ with $\alpha\in\mathbb{R}$. In this scenario, Eqs. (\ref{New Frame Effective Hamiltonian 1}, \ref{New Frame Effective Hamiltonian 2},
\ref{New gauge effective Hamiltonians expected value 1}, \ref{New gauge effective Hamiltonians expected value 2}) simplify to
\begin{align}
\tilde{H}^{\prime(1)}(t) & =\tilde{H}^{(1)}(t)-\hbar\alpha\hat{1}^{(1)},\\
\tilde{H}^{\prime(2)}(t) & =\tilde{H}^{(2)}(t)+\hbar\alpha\hat{1}^{(2)},
\end{align}
and
\begin{align}
\langle\tilde{H}^{\prime(1)}(t)\rangle & =\langle\tilde{H}^{(1)}(t)\rangle-\hbar\alpha,\\
\langle\tilde{H}^{\prime(2)}(t)\rangle & =\langle\tilde{H}^{(2)}(t)\rangle+\hbar\alpha,
\end{align}
respectively. Interestingly, notice that, within this set of gauges, all the local effective Hamiltonians possess the same gap structure, while their expectation values simply differ from one another by an additive constant. Under these circumstances, if we identify the local physical internal energies as $\langle\tilde{H}^{(k)}(t)\rangle=U^{(k)}(t)$, even though different gauges would provide distinct absolute energy values, we guarantee identical energy measurement differences. Along these lines, the remaining freedom $\alpha$ just shifts the energy by $\pm\hbar\alpha$ and is analogous to the classical thermodynamic freedom in the definition of internal energy. \cite{Prigogine}

In summary, despite the broad mathematical freedom, we were able to identify and fix the set of physically consistent phases, $\left\{ \frac{d}{dt}\theta_{j}(t)=\alpha\in\mathbb{R}\right\} _{j}$, that recovers the expected limiting behaviour. In this sense, it is worth mentioning that such a procedure and reasoning explicitly demanded knowledge about the interaction $\hat{H}_{int}$ to obtain the correct physical phase gauge. Thus, in order to construct a local and consistent energy description for the subsystems, one cannot rely solely on local features (see \cite{rodrigo} for a recent discussion). Finally, as long it is chosen a gauge that belongs to this relevant set, one can identify physical local internal energies up to an additive constant, such that the physical local effective Hamiltonians are guaranteed to possess invariant (unambiguous) spectral gaps.

%% file: Chapters/Chapter3/Section6.tex
\section{Thermodynamics\label{Section: Thermodynamics}}

In Section (\ref{sec:Quantum-thermodynamics}), we briefly introduced
the general context and main motivations underlying the formulation
of a thermodynamic theory for non-equilibrium quantum systems. We
also stressed that its development is, currently, a work in progress
and that several fundamental issues are still under scrutiny. In fact,
the lack of consensus and understanding on some key aspects emphasize
a real challenge to establishing the foundations of the theory and
highlights the urgency of fostering discussion at the fundamental
level. Along these lines, the most critical barrier lies in the proper
identification of general quantum versions of the most basic classical
thermodynamic quantities, such as internal energy, work, heat and
entropy. On the one hand, the definitions of work and heat require
both the previous recognition of physical internal energy and the
understanding of the essential features that characterize these quantities;
on the other, the concept of entropy and irreversibility, despite
being a central concept in modern physics and several branches of
science, remains elusive for general scenarios. Hence, it is unclear
how to state the well-known laws of thermodynamics for situations
whose both non-equilibrium processes and quantum features play essential
roles. While the former is consistently contemplated in the formalism
of stochastic thermodynamics, the latter is still obscure. One can
find in the literature several proposals for accounting for these
questions. However, most current approaches still rely on thermal
states and baths, semi-classical asymmetrical descriptions and approximative
regimes, which makes them not suitable for describing the thermodynamics
within fully quantum (autonomous) systems.

Let us now briefly present and discuss part of the current efforts
concerning the generalization of the usual equilibrium thermodynamic
concepts and results to arbitrary non-equilibrium quantum systems.
Also, when possible, we will consider our pure bipartite setup and
mention how our approach fits into this context.

\subsection{Remarks on thermal states}

In thermodynamics, both classical and quantum, we are most of the
time interested in describing and characterizing systems at equilibrium.
These states are dynamical fixed points attained asymptotically in
time and constrained by conserved physical properties. More frequently,
however, it is focused on systems whose internal energy remains fixed.
In those cases, the usual Gibbs canonical ensemble, given by 
\begin{equation}
\hat{\rho}_{th}=\frac{e^{-\beta\hat{H}}}{Z},\label{Thermal state}
\end{equation}
is the appropriate steady-state, where $\hat{H}$ is the Hamiltonian,
$\beta=\frac{1}{k_{B}T}$ is the inverse of temperature $T$ and $Z\equiv tr\{e^{-\beta\hat{H}}\}$
is the partition function\footnote{If more conserved quantities should be considered, then different
steady-states are reached, and a generalized Gibbs ensemble (GGE)
takes place.}. Interestingly, by the usual Lagrange multipliers method, one can
easily show that Eq. (\ref{Thermal state}) is exactly the state that
maximizes the von Neumann entropy $S_{vN}$\footnote{This quantity will be properly defined below.},
once assumed that the expectation value of the Hamiltonian, $\langle\hat{H}\rangle=tr\left\{ \hat{\rho}\hat{H}\right\} $,
is a constant quantity. From a more fundamental perspective, nevertheless,
such states might be derived and justified by the usual statistical
physics recipe of assuming that the system of interest is weakly interacting
with a - much larger - heat bath and invoking the equal \textit{a
priori} probability postulate, which basically assumes that the whole
bipartition is statistically described by the microcanonical ensemble.
Essentially, assuming the total Hamiltonian given by Eq. (\ref{Total Hamiltonian}),
if the state of the whole system is maximally mixed, $\hat{\rho}^{(0)}=\frac{\hat{1}^{(0)}}{d^{(0)}}$,
and $d^{(2)}\gg d^{(1)}$ with $\hat{H}_{int}$ negligible, one can
show that
\begin{equation}
\hat{\rho}^{(1)}=tr_{2}\{\hat{\rho}^{(0)}\}=\frac{e^{-\beta\hat{H}^{(1)}}}{Z^{(1)}}.\label{equal a priori}
\end{equation}
Of course, once considered pure states for the whole system, it is not obvious if Eq. (\ref{equal a priori}) should also be true or not. Along these lines, arguments on canonical typicality show that, in fact, the postulate above might be dismissed, i.e., if the whole system is restricted by a given arbitrary condition $R$ such that $|\Psi\rangle\in\mathcal{H}^{(R)}\subseteq\mathcal{H}^{(0)}=\mathcal{H}^{(1)}\otimes\mathcal{H}^{(2)}$, it was shown that for almost every pure state $|\Psi\rangle$, the local state $\hat{\rho}^{(1)}=tr_{2}\{|\Psi\rangle\langle\Psi|\}$ of a sufficiently small subsystem ($d^{(1)}\ll d^{(2)}$) is approximately equal to $\hat{\rho}^{(1)}\approx tr_{2}\left\{ \frac{\hat{1}^{(R)}}{d^{(R)}}\right\} $, where $\frac{\hat{1}^{(R)}}{d^{(R)}}$ is the maximally mixed state of the whole system, once considered the states consistent with the restriction. This result is completely general but can be directly connected to thermal states. One can show that as long $R$ translates into the total internal energy being close to a fixed value, the subsystems are weakly-coupled, and the density of states of subsystem $(2)$ increases approximately exponentially with energy, the local state $\hat{\rho}^{(1)}$ is approximately equal to the equilibrium state \cite{Popescu2006,PhysRevLett.96.050403}, i.e., 
\begin{equation}
\hat{\rho}^{(1)}=tr_{2}\{|\Psi\rangle\langle\Psi|\}\approx\frac{e^{-\beta\hat{H}^{(1)}}}{Z^{(1)}}.
\end{equation}
Despite powerful statements, these results are still restrictive to a very asymmetrical treatment and regime, and far from being applicable to many relevant scenarios within QT\footnote{For instance, when considering systems of similar dimensionality $d^{(2)}\approx d^{(1)}$.}. Clearly, thermal states are extremely useful and important, and one might still assume previous thermal states preparations regardless of the system's nature. However, the thermodynamic analysis of finite and quantum systems, in general, will have to deal with non-equilibrium states. Strictly speaking, the classical setting of two interacting subsystems in individual local thermal states are not allowed for pure bipartite systems: given $|\Psi\rangle$, the Schmidt decomposition,
Eq. (\ref{SchmidtDecomposition}), guarantee that the local states
- independently of the subsystems - must have the same spectrum $\{\lambda_{j}\}_{j}$,
which means that the number of non-zero populations are bounded by
the smallest dimension in question, i.e., if $\hat{H}^{(1,2)}\equiv\sum_{j=1}^{d^{(1,2)}}b_{j}^{(1,2)}|b_{j}^{(1,2)}\rangle\langle b_{j}^{(1,2)}|$
are the local bare Hamiltonians, our framework clearly shows that there is no pure state $|\Psi\rangle\langle\Psi|$
such that
\begin{equation}
\hat{\rho}^{(1)}=tr_{2}\{|\Psi\rangle\langle\Psi|\}=\sum_{j}^{d^{(1)}}\frac{e^{-\beta b_{j}^{(1)}}}{Z_{1}}|b_{j}^{(1)}\rangle\langle b_{j}^{(1)}|,
\end{equation}
and
\begin{equation}
\hat{\rho}^{(2)}=tr_{1}\{|\Psi\rangle\langle\Psi|\}=\sum_{j}^{d^{(2)}}\frac{e^{-\beta b_{j}^{(2)}}}{Z_{2}}|b_{j}^{(2)}\rangle\langle b_{j}^{(2)}|,
\end{equation}
simultaneously.

On the other hand, even if the whole system $\hat{\rho}^{(0)}$ is
assumed (or previously prepared) to be mixed and thermal with a temperature
$T$, once interaction within the bipartition becomes appreciable
local divergences from Eq. (\ref{Thermal state}) are expected to
appear. Along these lines, the understanding of the thermodynamic
behaviour at the (ultra)strong coupling regime has also recently been
investigated. The current approach is based on earlier work by \cite{doi:10.1063/1.1749657}
and consists of the definition of a local effective Hamiltonian for
a subsystem interacting with a large reservoir, the so-called Hamiltonian
of mean force (HMF), that guarantee the usual canonical Gibbs state
form (see \cite{Gelin2009,Miller2018c} and \cite{PhysRevLett.127.250601} for further references), i.e.,
assuming $\hat{\rho}^{(0)}=\hat{\rho}_{th}^{(0)}$ one can show that
the local state $\hat{\rho}^{(1)}$ can be cast as
\begin{equation}
\hat{\rho}^{(1)}=tr_{2}\{\hat{\rho}_{th}^{(0)}\}\equiv\frac{e^{-\beta\hat{H}_{eff}^{(1)}}}{Z_{eff}^{(1)}}\label{HMF gibbs state}
\end{equation}
where $Z_{eff}^{(1)}\equiv tr_{1}\{e^{-\beta\hat{H}_{eff}^{(1)}}\}$
is the new partition function and
\begin{equation}
\hat{H}_{eff}^{(1)}:=-k_{B}T\,ln(tr_{2}\{e^{-\beta\hat{H}^{(0)}}\}/tr_{2}\{e^{-\beta\hat{H}^{(2)}}\}),
\end{equation}
is the HMF, that clearly depends on the temperature $T$ and the interaction
$\hat{H}_{int}$. Since the functional thermal structure in Eq. (\ref{HMF gibbs state})
is preserved, this formalism allows a straightforward connection with
the usual equilibrium thermodynamic expressions and mathematical machinery,
in particular the definition of effective versions of the thermodynamic
potentials, such as the Helmholtz free energy, given by $F_{eff}^{(1)}\equiv-\beta^{-1}ln(Z_{eff}^{(1)})$.

In short, despite omnipresence and undeniable importance in equilibrium
thermodynamics, the emergence and use of thermal states represent
particular cases of more broad scenarios of quantum dynamic processes.
In this sense, there are many open questions and it is imperative
to develop a quantum thermodynamic formalism able to deal with arbitrary
states, systems, non-thermal baths and general coupling regimes.

\subsection{Remarks on quantum thermodynamic entropy}

Along with the concept of energy, entropy has also reached the status
of one of the most fundamental quantities in modern physics. Interestingly,
despite being initially introduced in thermodynamics, its use transcended
the scope of its initial conception, and now it is being used across
several disciplines. Despite this universality, its understanding
remains elusive, especially when considering questions regarding the
second law and its extension to describe non-equilibrium quantum systems.
Let us now briefly discuss the current status and approaches of entropy
in the context of QT. For more discussions, see. \cite{RevModPhys.93.035008}

\subsubsection{General context}

In Section (\ref{sec:Thermodynamics}), we briefly introduced the
second law of thermodynamics, the notion of thermodynamic entropy
and its intrinsic relationship with irreversibility. As we saw, essentially,
it states that the entropy production of an isolated system, $\Sigma=\Delta S_{th}$,
along any path, should increase or remain the same, i.e., $\Sigma\geq0$,
where the equality is satisfied if and only if reversible processes
have taken place. From a pragmatical point of view, such a powerful
statement imposes fundamental and universal constraints on any possible
physical transformation. Further progress on the understanding of
entropy, and thermodynamics in general, came along with contemporary
developments in the study of non-equilibrium systems. In this sense,
stochastic thermodynamics both pushed the boundaries of thermodynamics
to once uncharted regimes and provided novel insights into the foundations
of the theory. As mentioned earlier, in this context, entropy - and
other relevant thermodynamic quantities - are fluctuating quantities
defined and characterized at the individual phase space trajectory
level. \cite{Seifert2005,seifert2008stochastic,seifert2012stochastic} If $p(q,t)$ is the probability of finding a given
physical system in state $q$ at time $t$, the stochastic entropy
is simply defined by the following expression \cite{Seifert2005} 
\begin{equation}
s(t)\equiv-k_{B}ln(p(q,t)),
\end{equation}
Interestingly, such a
formulation naturally implies that there will be some trajectories
with negative entropy production, which is also implicit in the Fluctuation
Theorems, in particular $\langle e^{-\frac{\Sigma}{k_B}}\rangle=1$. Nevertheless,
the usual second law statement is satisfied once considered an ensemble
analysis and mean values, such that 
\begin{equation}
S_{Gibbs}(t)\equiv-k_{\beta}\int dq\,p(q,t)ln(p(q,t))=\langle s(t)\rangle,\label{Gibbs entropy}
\end{equation}
where $S_{Gibbs}(t)$ is the well known Gibbs entropy (or Shannon
entropy if $k_{B}$ is not considered): this expression is well defined
for any probability distribution $p(q,t)$ and it is commonly regarded
as a proper choice for nonequilibrium extension of the thermodynamic
entropy, $S_{th}$; in fact, if the system is at thermal equilibrium
with inverse of temperature $\beta$, the probabilities
are given by the equilibrium Boltzmann distribution $p_{eq}(q)=Z^{-1}(T)e^{-\beta E(q)}$,
where $E(q)$ is the energy for the $q$th state, and the Gibbs expression
above becomes equal to the usual thermodynamic entropy relation
\begin{equation}
S_{Gibbs}=\frac{1}{T}\left(\langle E\rangle-F(T)\right)=S_{th},
\end{equation}
where $\langle E\rangle$ is the system's internal energy, $F(T)\equiv-\beta^{-1}\,ln(Z(T))$
is the Helmholtz free energy and $Z(T)\equiv\int dq\,e^{-\beta E(q)}$
is the partition function; in addition, considering a proper identification
of heat $Q$, one can show that 
\begin{equation}
dS_{Gibbs}=\Sigma+d\Phi,
\end{equation}
for arbitrary changes, where $\Sigma\geq0$ is the positive entropy
production and $d\Phi=\frac{dQ}{T}$ is the entropy flux due to heat
exchange with a heat bath with temperature $T$. In short, once identified
the Eq. (\ref{Gibbs entropy}) as the general quantifier of entropy,
one recovers the usual thermodynamic expressions and the second law
behaviour for both equilibrium and non-equilibrium scenarios.

\paragraph{Fully informational perspective\label{par:Fully-informational-perspective}}

From now on, let us explicitly assume that $k_{B}=1$\footnote{At the end of the day, it does not influence our discussion.}.
The quantum counterpart of the Gibbs/Shannon entropy is given by the
von Neumann entropy
\begin{equation}
S_{vN}[\hat{\rho}]\equiv-tr\left\{ \hat{\rho}ln\left(\hat{\rho}\right)\right\} ,\label{ENTROPIA DE VON NEUMANN}
\end{equation}
where $\hat{\rho}$ is the system's density matrix. Before discussing
further, first, let us recall some of the basic properties of the
Eq. (\ref{ENTROPIA DE VON NEUMANN}) \cite{nielsen00}:
\begin{enumerate}
\item \textbf{Positivity:} $S_{vN}[\hat{\rho}]\geq0$ for all $\hat{\rho}$.
The equality is obtained if and only if the state is pure, i.e., $\hat{\rho}=|\psi\rangle\langle\psi|$;
\item \textbf{Unitary invariance:} $S_{vN}[\hat{\rho}]=S_{vN}[\hat{U}\hat{\rho}\hat{U}^{\dagger}]$
for any unitary $\hat{U}$;
\item \textbf{Subadditivity:} For bipartite systems described by $\hat{\rho}^{(0)}$
and reduced states $\hat{\rho}^{(1,2)}=tr_{2,1}\{\hat{\rho}^{(0)}\}$
we have the following inequality, $S_{vN}[\hat{\rho}^{(0)}]\leq S_{vN}[\hat{\rho}^{(1)}]+S_{vN}[\hat{\rho}^{(2)}]$.
Equality is guaranteed only for uncorrelated systems, such that $\hat{\rho}^{(0)}=\hat{\rho}^{(1)}\otimes\hat{\rho}^{(2)}$.
\end{enumerate}
In analogy with the classical definition of mutual information, from
property 3. one might define its quantum counterpart as the following
difference
\begin{equation}
I_{12}\equiv S[\hat{\rho}^{(0)}||\hat{\rho}^{(1)}\otimes\hat{\rho}^{(2)}]=S_{vN}[\hat{\rho}^{(1)}]+S_{vN}[\hat{\rho}^{(2)}]-S_{vN}[\hat{\rho}^{(0)}]\geq0,\label{Mutual information}
\end{equation}
where $S[\hat{\rho}||\hat{\sigma}]=tr\{\hat{\rho}[ln(\hat{\rho})-ln(\hat{\sigma})]\}$
is the quantum relative entropy. The expression above is clearly positive
(due to the subadditivity) and quantifies the total correlations -
classical \textit{and} quantum - within the whole quantum system $\hat{\rho}^{(0)}$.

In the context of QT, Eq. (\ref{ENTROPIA DE VON NEUMANN}) is the
most commonly chosen candidate for quantifying quantum entropy. Its
popularity is partially inherited by the success of Eq (\ref{Gibbs entropy})
in the classical domain, but also because it behaves properly for
some key scenarios: again, if we consider thermal states, such the
ones given by Eq. (\ref{Thermal state}), it is easy to see that
the Eq (\ref{ENTROPIA DE VON NEUMANN}) automatically satisfy the
thermodynamic entropy relation below
\begin{equation}
S_{vN}[\hat{\rho}_{th}]=-tr\left\{ \hat{\rho}_{th}ln\left(\hat{\rho}_{th}\right)\right\} =\frac{1}{T}(U-F)=S_{th},
\end{equation}
where $U\equiv tr\left\{ \hat{\rho}_{th}\hat{H}\right\} =\langle\hat{H}\rangle$
is identified as the internal energy and $F\equiv-\beta^{-1}\,ln(Z)$
is the Helmholtz free energy; also, if we consider a quantum system
initially described by $\hat{\rho}(t_{0})$ and weakly coupled to
a thermal bath with temperature $T$, such that $\hat{\rho}_{th}$
is the system's asymptotic state, one can easily check the expression
$S[\hat{\rho}(t)||\hat{\rho}_{th}]=-S_{vN}[\hat{\rho}(t)]+\frac{1}{T}\left(\langle\hat{H}\rangle(t)-F\right)$,
where $\langle\hat{H}\rangle(t)=tr\left\{ \hat{\rho}(t)\hat{H}\right\} $.
Then, if $\hat{\rho}(t)$ is the state at instant $t$, by identifying
entropy production and heat as $\Sigma=-(S[\hat{\rho}(t)||\hat{\rho}_{th}]-S[\hat{\rho}(t_{0})||\hat{\rho}_{th}])$
and $Q=tr\left\{ (\hat{\rho}(t)-\hat{\rho}(t_{0}))\hat{H}\right\} $,
respectively, we obtain
\begin{equation}
\Delta S_{vN}=\Sigma+\frac{1}{T}Q.
\end{equation}
The positivity of $\Sigma$ is assured by the fact that $S[\Lambda[\hat{\rho}]||\Lambda[\hat{\sigma}]]\leq S[\hat{\rho}||\hat{\sigma}]$
for CPTP maps $\Lambda[(.)]$: if $\hat{\rho}(t)=\Lambda_{t,t_{0}}[\hat{\rho}(t_{0})]$
and since $\hat{\rho}_{th}$ is a fixed point for this dynamics, i.e.,
$\Lambda_{t,t_{0}}[\hat{\rho}_{th}]=\hat{\rho}_{th}$, we guarantee
that $S[\Lambda_{t,t_{0}}[\hat{\rho}(t_{0})]||\Lambda_{t,t_{0}}[\hat{\rho}_{th}]]=S[\hat{\rho}(t)||\hat{\rho}_{th}]\leq S[\hat{\rho}(t_{0})||\hat{\rho}_{th}]$
and, therefore,
\begin{equation}
\Sigma=-(S[\hat{\rho}(t)||\hat{\rho}_{th}]-S[\hat{\rho}(t_{0})||\hat{\rho}_{th}])\geq0.
\end{equation}
Similar statements for entropy production can be made even if assuming
slightly broader cases, such as considering explicitly Hamiltonian
time-dependency. \cite{Alicki19792,Deffner2011}

Nevertheless, the use of the von Neumann entropy often comes along
with a fully information-theoretic perspective of entropy production,
in which the thermodynamic relevant scenarios mentioned above are
seen as particular cases. Along these lines, irreversibility and,
therefore, $\Sigma$, only appears once information is omitted (or
becomes inaccessible) from a local point-of-view. Such reasoning was
put forward by reference \cite{Esposito_2010}, but see \cite{RevModPhys.93.035008} for further
discussions. More specifically, if at $t=t_{0}$ a system of interest
$(1)$ is put into contact with another arbitrary system $(2)$, such
that $\hat{\rho}^{(0)}(t_{0})=\hat{\rho}^{(1)}(t_{0})\otimes\hat{\rho}^{(2)}(t_{0})$,
the local von Neumann entropy change $\Delta S_{vN}[\hat{\rho}^{(1)}]=S_{vN}[\hat{\rho}^{(1)}(t)]-S_{vN}[\hat{\rho}^{(1)}(t_{0})]$
can be separated into
\begin{equation}
\Delta S_{vN}[\hat{\rho}^{(1)}]=S[\hat{\rho}^{(0)}(t)||\hat{\rho}^{(1)}(t)\otimes\hat{\rho}^{(2)}(t_{0})]+tr_{2}\{(\hat{\rho}^{(2)}(t)-\hat{\rho}^{(2)}(t_{0}))ln(\hat{\rho}^{(2)}(t_{0}))\},
\end{equation}
where $\hat{\rho}^{(1,2)}(t)=tr_{2,1}\{\hat{\rho}^{(0)}(t)\}$ and
$\hat{\rho}^{(0)}(t)$ evolves unitarily. While the second term from
the right-hand side is identified as entropy flux $\Phi(t)$, entropy
production is defined as
\begin{equation}
\Sigma(t)=S[\hat{\rho}^{(0)}(t)||\hat{\rho}^{(1)}(t)\otimes\hat{\rho}^{(2)}(t_{0})],\label{Info-theoretic entropy production EQ 1}
\end{equation}
such that the usual thermodynamic form $\Delta S_{vN}[\hat{\rho}^{(1)}]=\Sigma(t)+\Phi(t)$
is recovered. By construction, this quantity is non-negative $\Sigma(t)\geq0$
for all $t$, with equality being satisfied if and only if the whole
time-evolved state remains uncorrelated and subsystem $(2)$ keeps
unchanged throughout the dynamics, i.e., $\hat{\rho}^{(0)}(t)=\hat{\rho}^{(1)}(t)\otimes\hat{\rho}^{(2)}(t_{0})$.
Thus, essentially, $\Sigma(t)$ measures how far the actual whole
system's state $\hat{\rho}^{(0)}(t)$ is from $\hat{\rho}^{(1)}(t)\otimes\hat{\rho}^{(2)}(t_{0})$.
Nevertheless, an information-theoretic interpretation of entropy production
becomes more explicit if Eq. (\ref{Info-theoretic entropy production EQ 1})
above is cast as follows
\begin{equation}
\Sigma(t)=I_{12}(t)+S[\hat{\rho}^{(2)}(t)||\hat{\rho}^{(2)}(t_{0})].
\end{equation}
Along these lines, $\Sigma(t)$ emerges from the lost information
encoded both by the correlations within the whole system and the time
evolution of the inaccessible subsystem $(2)$, represented by $I_{12}(t)$
and $S[\hat{\rho}^{(2)}(t)||\hat{\rho}^{(2)}(t_{0})]$, respectively.
Notice that, ultimately, this perspective is very different from the
classical second law notion: instead of focusing on quantifying the
entropy change of the whole system, one concentrate on the analysis
of the entropic dynamics - measured by the von Neumann entropy - of
a local state. Having that in mind, despite inheriting some of the
desired properties for a possible candidate of quantum thermodynamic
entropy (at least for some paradigmatic scenarios), the use of the
von Neumann entropy has some fundamental and challenging issues.

\subsubsection{Issues of the von Neumann entropy\label{Issues of the von Neumann entropy}}

As mentioned earlier, in the classic context, the thermodynamic entropy
of the whole system (commonly referred to as the universe), or the
total entropy production, increases or remains the same for arbitrary
dynamical processes. Thus, it is expected that any consistent quantum
thermodynamic entropy definition, $S_{Qth}$, both generalizes the
classical notions and capture this very general statement, i.e., $\Delta S_{Qth}\geq0$.
Interestingly, despite the prevailing use of the von Neumann entropy, it does not work as desired for arbitrary scenarios.

The unitary invariance of $S_{vN}[\hat{\rho}]$ (property \textbf{2.})
implies that for any isolated system, whose dynamics is fully characterized
by the Schrödinger equation, the von Neumann entropy remains fixed,
i.e., 
\begin{equation}
\hat{\rho}(t)=\hat{U}(t,t_{0})\hat{\rho}(t_{0})\hat{U}^{\dagger}(t,t_{0})\;\Rightarrow\;\Delta S_{vN}[\hat{\rho}]=S_{vN}[\hat{\rho}(t)]-S_{vN}[\hat{\rho}(t_{0})]=0,
\end{equation}
for all $t$. This strict equality is consistent with the second law, but there is no room for global entropy increases. Thus, in order to observe any entropic variation, it is
required non-unitarity induced by some interaction with another system
(at least in the first order of the interaction term, Eq. (\ref{state interaction perturbation})).
In fact, as we saw earlier, the local von Neumann entropies of subsystems
within a bigger composite one might change in time. Along these lines,
the general subadditivity of $S_{vN}[\hat{\rho}]$ (property \textbf{3.})
also represents a fundamental difference compared with the additivity
observed in the thermodynamic entropy, i.e., the sum of the local
entropies is equal to the whole's. For bipartite systems, for instance,
along with Eq. (\ref{Mutual information}), we see that the sum of
local variations in their von Neumann entropies is equal to the change
in the mutual information, such that
\begin{equation}
\Delta S_{vN}[\hat{\rho}^{(1)}]+\Delta S_{vN}[\hat{\rho}^{(2)}]=\Delta I_{12},\label{mutual information change}
\end{equation}
where $\Delta I_{12}=I_{12}(t)-I_{12}(t_{0})$ - a priori - could
assume positive or negative values. Since $I_{12}(t)$ is necessarily
positive for any $t$, in the special case of assuming initial uncorrelated
systems ($\hat{\rho}^{(0)}(t_{0})=\hat{\rho}^{(1)}(t_{0})\otimes\hat{\rho}^{(2)}(t_{0})$
and $I_{12}(t_{0})=0$), the expression above becomes $\Delta S_{vN}[\hat{\rho}^{(1)}]+\Delta S_{vN}[\hat{\rho}^{(2)}]=I_{12}(t)\geq0,$which
is similar to the thermodynamic statement but not general enough.

Let us now consider the setup described in Section (\ref{Section: Setup}):
a generic pure bipartite system. In this context, the von Neumann
entropy is commonly referred to as \textit{entanglement entropy} since
it directly measures the degree of entanglement between the partitions.
First, from property \textbf{1.}, we know that the von Neumann entropy
of pure states is null, thus for the whole system $S_{vN}[\hat{\rho}^{(0)}(t)]=0$
throughout any - unitary - dynamics; then, given the symmetric description
elucidated by the Schmidt decomposition (Eq. (\ref{SchmidtDecomposition})),
we have the local mixed states $\hat{\rho}^{(1,2)}(t)$ showed on
Eqs. (\ref{local state 1}, \ref{local state 2}), whose populations
are given by the Schmidt coefficients squared $\{\lambda_{j}^{2}(t)\}_{j}$ and,
therefore,
\begin{equation}
S_{vN}[\hat{\rho}^{(1)}(t)]=-\sum_{i=1}^{d^{(1)}}\lambda_{i}^{2}(t)ln\left(\lambda_{i}^{2}(t)\right)=S_{vN}[\hat{\rho}^{(2)}(t)]
\end{equation}
for all $t$\footnote{Of course, as mentioned earlier, any local functionals purely dependent
on the Schmidt coefficients will be equal for both subsystems.}. It is clear that if the Schmidt rank is equal to one, the whole
system is separable ($|\Psi(t)\rangle=|\varphi^{(1)}(t)\rangle\otimes|\varphi^{(2)}(t)\rangle$)
and $S_{vN}[\hat{\rho}^{(1,2)}(t)]=0$. In fact, given Eq. (\ref{Mutual information}),
for arbitrary bipartite systems, the local entropies are proportional
to the mutual information quantifying the total correlation
\begin{equation}
S_{vN}[\hat{\rho}^{(1)}(t)]=S_{vN}[\hat{\rho}^{(2)}(t)]=\frac{1}{2}I_{12}(t)\geq0,
\end{equation}
and 
\begin{equation}
\Delta S_{vN}[\hat{\rho}^{(1)}]=\Delta S_{vN}[\hat{\rho}^{(2)}]=\frac{1}{2}\Delta I_{12},
\end{equation}
which also clearly contrasts with the second law statement applied
to a classical bipartition
\begin{equation}
\Delta S_{th}^{(1)}\geq-\Delta S_{th}^{(2)}.
\end{equation}

In addition, the thermodynamic entropy is also directly linked with
the energetics within a given process and, more specifically, the
heat exchanged. Except for the cases already mentioned, namely when
the systems are functionals of the Hamiltonians, and even more particular,
for Gibbs states, it is not clear how to generalize a relationship
between von Neumann entropy variation with energy flux. In fact, as
we are going to discuss below, it also points-out a difficulty in identifying
a candidate for quantum heat, since in classical thermodynamics heat
is often defined as the energy exchange that also is accompanied by
some entropic flux.

In short, despite the extensive use of the von Neumann entropy in
QT and relative success for particular scenarios, it does not satisfy
the expected properties for a proper microscopic generalization of
the thermodynamic entropy, and it is not clear how to proceed. This,
however, is already stressed by some authors. Along these lines, alternative
proposals for a quantum thermodynamic entropy is represented by the
observational \cite{PhysRevA.99.010101,Strasberg2021a} and diagonal \cite{POLKOVNIKOV2011486} ones, while the former
relies on a coarse-graining process, the latter uses the von Neumann
entropy form calculated considering only the instantaneous diagonal
elements of the density matrix in the energy basis. In the end, such
a lack of agreement and understanding on this very basic quantity
highlights the necessity of further investigations at the fundamental
level. Otherwise, the thermodynamic role played by genuine quantum
features, such as coherence and entanglement, will remain elusive,
and developments of future quantum technologies will be affected.

\subsection{Remarks on quantum work and heat\label{subsec:quantum work and heat}}

Once noticed that even the most basic concept of internal energy is
still under scrutiny in the QT community, it should not be surprising
the fact that the quantum versions of energy-based thermodynamic quantities,
such as work and heat, are also elusive. Let us now briefly review
work and heat in the general context of QT and, subsequently, consider
them according to our effective internal energy description.

\subsubsection{General context}

For pedagogical purposes, let us divide the current approaches for
defining quantum work and heat into two major distinct categories:

\paragraph{Operational approach}

Among all thermodynamic quantities, quantum work is - by far - the
most discussed one. On the one hand, given recent technological progress
in the fabrication and manipulation of quantum systems, it is no surprise
to find the concept of work in the spotlight of QT. Its complete understanding
and control are one of the pinnacles of the field and is expected
to fuel all sorts of technological applications, just like its classical
counterpart did in the past. On the other hand, work has also been
a central issue in former - although still contemporary - fundamental
discussions concerning the generalization of thermodynamics to classical
microscopic settings and the genesis of stochastic thermodynamics.
\cite{seifert2008stochastic,seifert2012stochastic}

Unsurprisingly, this context served as an important stage for a considerable
amount of efforts into the search for quantum work, especially for
closed quantum systems. The usual setting consists of a quantum object,
depicted by a state $\hat{\rho}(t)$, submitted to an externally controlled
protocol, represented by a time-dependent Hamiltonian, such that $\hat{H}(t_{0})\rightarrow\hat{H}(t_{1})$
with $t_{1}\geq t_{0}$. The closed time-evolution guarantees no other
interactions and unitary dynamics, which commonly justify the interpretation
of any energetic change as work $W(t)$ from/to the external agent,
$\langle\hat{H}(t_{1})\rangle-\langle\hat{H}(t_{0})\rangle=W(t_{1})$.
In particular, the quest for quantum versions of FTs was a major
driving force that helped to shed some light on very important aspects.
For instance, it was argued that work $W$ is not a fundamentally
time-local entity, instead, it is characterized by processes and trajectories.
Such reasoning is an extension of the classical thermodynamic conclusion
that work is a path-dependent quantity and not a state function. This
is often stated in the literature by the phrase \textquotedbl work
is not an observable\textquotedbl{} \cite{Talkner2007}, which means that work
should not be simply understood and represented by a Hermitian operator
$\hat{\Omega}$ such that $W:=tr\{\hat{\Omega}\hat{\rho}\}$ and whose
eigenvalues encode all the possible work measurements. Nevertheless,
it is worth mentioning that it is not entirely clear if whether or
not the concept of \textquotedbl work operator\textquotedbl{} is
well-founded, which corresponds to an active debate in the field. \cite{PhysRevE.71.066102,Beyer2020,Silva2021}
Furthermore, this route also encouraged discussions about possible
meaningful definitions of quantum work fluctuations and the role of
the measurement back-action: while the former demands a consistent
stochastic description, i.e., a set $\{w_{k}\}_{k}$ of possible work
outcomes and its distribution $P(w)$; the latter, highlight the invasive
nature of measurements and its potential thermodynamic cost. Of course,
both questions have their subtleties concerning foundational aspects
of quantum mechanics. On the one hand, the definition of quantum analogous
of classical stochastic trajectories is challenging and far from being
trivial\footnote{For instance, see \cite{PhysRevA.97.012131} for a Bohmian perspective approach.};
on the other hand, discussions and criticisms concerning the underlying
nature of the measurement postulate of quantum mechanics are as old
as the first developments of the theory.

Along these lines, the orthodox approach of quantum stochastic trajectories
in QT is based on the sequential projective measurements of the externally
driven quantum system. Thus, stochasticity naturally emerges due to
the probabilistic nature of the measurement process, and the set of
outcomes establishes the dynamical path that the measured system follows
during the intercalation of unitary time-evolution and measurements.
While the drive plays the role of the classical external parameters
changes, the measurement is analogous to random displacements due
to noise. From this operational perspective, it became clear the possibility
of defining the fluctuating work of performing a given protocol $\hat{H}(t_{0})\rightarrow\hat{H}(t_{1})$,
as the difference of two projective energy measurements outcomes:
if $E_{k}(t)$ is the kth eigenenergy at time $t$, hence fluctuating
work is simply $w_{ji}(t_{1})=E_{j}(t_{1})-E_{i}(t_{0})$. Therefore,
the average work $\langle W(t_{1})\rangle$ is directly obtained from
an ensemble of protocol realizations. Interestingly, it was shown
that the statistics $P_{TPM}(w)$ associated with such procedure,
commonly known as the Two Projective Measurement protocol (TPM), both
satisfy the classical FTs form \cite{Tasaki2000,Campisi2011a} and corresponds to the classical
work distribution under the semi-classical limit, $P_{TPM}(w)\rightarrow P_{class}(w)$.
\cite{jarzynski2015quantum,zhu2016quantum} Despite being the most popular approach and experimentally
verifiable \cite{batalhao2014experimental,an2015experimental,cerisola2017using,zhang2018experimental,hernandez2020experimental}, the TPM scheme has some very important issues
once considered initial coherent states (in the energy basis) and
the unavoidable destructive effect of obtaining these informations.
In such cases, the statistics of measurements (unsurprisingly) fail
to capture the expected internal energy change of the unperturbed
close quantum time-evolution, i.e., the first measurement eliminates
all initial coherence and, therefore, influences the future state
dynamics such that $\sum_{i,j}P_{TPM}(w_{ij})w_{ij}(t_{1})\neq\langle\hat{H}(t_{1})\rangle-\langle\hat{H}(t_{0})\rangle$\footnote{Of course, for non-coherent initial states, there is no such problem.}.
Additionally, the identification of energy measurements differences
as representative values of work is disputable once considered open
quantum systems. Such questions were partially addressed by a no-go
theorem presented in \cite{Perarnau-Llobet2017}, which states that there is no fluctuating
work definition $w$ and $P(w)$ that simultaneously satisfy the TPM
statistics for non-coherent initial states ($P(w)=P_{TPM}(w)$), and
$\sum_{w}P(w)w(t_{1})=\langle\hat{H}(t_{1})\rangle-\langle\hat{H}(t_{0})\rangle$
for any initial state\footnote{In \cite{Baumer2018} there is a refinement and one more condition is considered,
namely, the linearity of the distributions associated with different
measurement protocols.}. The former condition explicitly assumes that the success of the
TPM protocol in the FTs context is enough evidence for considering
it as the correct distribution, while the latter is motivated by the
assumption that the system is closed and, therefore, any energy exchange
is due to work. Along these lines, other operational approaches are
proposed in the literature, considering weak measurements, POVMS,
etc. \cite{Baumer2018}

Also, in this context, the concept of \textit{quantum heat} was introduced
to account for the energetic price of performing a measurement. \cite{Elouard2016b}
In general, if a given state $|\psi\rangle$ is measured, it is induced
an irreversible transformation $|\psi\rangle\rightarrow|\psi^{\prime}\rangle$,
such that $\langle\psi|\hat{H}|\psi\rangle\neq\langle\psi^{\prime}|\hat{H}|\psi^{\prime}\rangle$.
This sudden change have no classical analogous and intrinsically depends
on the eventual coherence in the chosen basis: of course, if the system
is already in a given eigenstate of the measured observable, nothing
will change. In this sense, the measurement apparatus is treated as
the source of stochasticity for the time-evolution and plays a similar
role played by thermal baths for the classical stochastic trajectories.
From a thermodynamic point of view, heat is usually associated with
irreversibility and entropic changes, which commonly justifies the
identification of this energetic difference as a fully quantum analogous
of heat, $q(t):=\langle\psi^{\prime}|\hat{H}(t)|\psi^{\prime}\rangle-\langle\psi|\hat{H}(t)|\psi\rangle$.

Finally, notice that operational perspectives of quantum thermodynamic
quantities fundamentally depends on the assumption of external classical
agents capable of performing certain protocols and measuring states.
While the former is responsible for inducing a deterministic Hamiltonian
time dependency, the latter introduces irreversible random outcomes.
Consequently, and more importantly, such an approach prevents any
further discussions concerning work and heat within fully isolated
interacting quantum subsystems and, therefore, is not suitable for
describing autonomous quantum machines (at least in any straightforward
manner).

\paragraph{Dynamical approach\label{Dynamical approach}}

Instead of focusing on stochastic trajectories and fluctuating variables
in an operational sense, one might be interested in a quantum dynamical
description of these thermodynamic entities. Along these lines, work
and heat are treated from an ensemble perspective and directly defined
by changes in the internal energy, commonly identified by the expectation value of a given Hamiltonian $\langle\hat{H}(t)\rangle:=tr\{\hat{H}(t)\hat{\rho}(t)\}$.
Thus, given the dynamical equation for both $\hat{\rho}(t)$ and $\hat{H}(t)$,
in principle, one might be able to compute the thermodynamic quantities
related to the followed dynamics. In this sense, the open quantum
systems formalism provides a suitable mathematical framework for dealing
both with the dynamics and energetics of general scenarios of interacting
quantum systems.

Nevertheless, as mentioned earlier, it is not clear how to unambiguously
identify quantum counterparts of work and heat. In fact, in the literature,
there are several proposals and strategies for approaching these definitions
from a dynamical point of view. However, most ideas are quantum versions
of the following reasoning: consider a classical system with discrete
states, indexed by $j$, the internal energy is simply given by the
ensemble average $U_{cla}\equiv\sum_{j}E_{j}P_{j}$, where $E_{j}$
is the energy of the $j$th state and $P_{j}$ is its occupation probability,
with $\sum_{j}P_{j}=1$ \cite{REI65,VANDENBROECK20156}; it is also assumed that the energy
states depends on the state of an external time-dependent control
parameter $\eta(t)$, such that $E_{j}=E_{j}(\eta)$; thus, the internal
energy change rate can be separated into two distinct categories
\begin{equation}
\frac{d}{dt}U_{cla}=\sum_{j}\frac{dE_{j}}{dt}P_{j}+\sum_{j}E_{j}\frac{dP_{j}}{dt},\label{Classical energy separation}
\end{equation}
i.e., a change due to the state's energy alteration and a change coming
from the occupation probability adjustment. The first contribution
is interpreted as work rate ($\frac{d}{dt}W_{cla}(t)$) since it is
the controlled energy transfering from/to the external agent, while
the remaining change is identified as heat flow ($\frac{d}{dt}Q_{cla}(t)$)
due to its relationship with transitioning states. Hence,
\begin{align}
W_{cla}(t) & :=\sum_{j}\int_{t_{0}}^{t}ds\,\left(\frac{dE_{j}}{ds}\right)P_{j}=\sum_{j}\int_{t_{0}}^{t}ds\,\left(\frac{dE_{j}}{d\eta}\right)\left(\frac{d\eta}{ds}\right)P_{j},\label{Classical work}\\
Q_{cla}(t) & :=\sum_{j}\int_{t_{0}}^{t}ds\,E_{j}\left(\frac{dP_{j}}{ds}\right),\label{Classical heat}
\end{align}
and the first law of thermodynamics form is obtained $\frac{d}{dt}U_{cla}(t)=\frac{d}{dt}W_{cla}(t)+\frac{d}{dt}Q_{cla}(t)$.
As just mentioned, such an approach provides an interesting route
for defining the desired quantities. Nevertheless, purely quantum
features do not allow a direct and unique analogy.

Along these lines, in \cite{Alicki19792} Alicki proposed a similar classification
of quantum work and heat for open quantum systems weakly coupled to
- possibly - $N$ thermal reservoirs. It was also assumed a slowly
driven Hamiltonian $\hat{H}(t)=\hat{H}_{0}+\hat{h}(t)$, containing
both the bare one and a time-dependent contribution representing the
externally controlled parameters, given by $\hat{H}_{0}$ and $\hat{h}(t)$
respectively. Under these conditions, the state $\hat{\rho}(t)$ dynamics
is described by a Markovian master equation in the usual Lindblad
form, such that
\begin{equation}
i\hbar\frac{d}{dt}\hat{\rho}(t)=\left[\hat{H}(t),\hat{\rho}(t)\right]+\sum_{k}^{N}\mathfrak{\hat{D}}_{t}^{(k)}\hat{\rho}(t),\label{Alicki Dynamics}
\end{equation}
and $\mathfrak{\hat{D}}_{t}^{(k)}\hat{\rho}_{\beta_{k}}=0$ for all
$k$, where $\mathfrak{\hat{D}}_{t}^{(k)}$ is the non-unitary superoperator
(also refered as dissipator) relative to the interaction with $k$th
reservoir with inverse of temperature $\beta_{k}$ and $\hat{\rho}_{\beta_{k}}=Z^{-1}e^{-\beta_{k}\hat{H}_{0}}$
is the usual Gibbs (thermal) state\footnote{The equality $\mathfrak{\hat{D}}_{t}^{(k)}\hat{\rho}_{\beta_{k}}=0$
implies that $\hat{\rho}_{\beta_{k}}$ is a fixed-point relative to
the $k$th reservoir. Essentially, it means that each reservoir alone
would thermalize the system to its temperature.}. The internal energy is recognized by the expectation value of $\hat{H}(t)$,
i.e.,
\begin{equation}
U(t):=tr\{\hat{H}(t)\hat{\rho}(t)\},\label{Energia interna}
\end{equation}
whose change rate is simply
\begin{equation}
\frac{d}{dt}U(t)=tr\left\{ \frac{d}{dt}\hat{H}(t)\hat{\rho}(t)\right\} +tr\left\{ \hat{H}(t)\frac{d}{dt}\hat{\rho}(t)\right\} .\label{Energy Split}
\end{equation}
In complete analogy with the classical case depicted by Eqs. (\ref{Classical energy separation}-\ref{Classical heat}),
work is associated with the controlled energetic change, depicted
by the Hamiltonian switch $\frac{d}{dt}\hat{H}(t)$, while heat is
linked with the probabilities change encoded by the density operator
dynamics $\frac{d}{dt}\hat{\rho}(t)$, i.e.,
\begin{align}
W_{1}(t) & :=\int_{t_{0}}^{t}ds\,tr\left\{ \frac{d}{ds}\hat{H}(s)\hat{\rho}(s)\right\} ,\label{Alicki work}\\
Q_{1}(t) & :=\int_{t_{0}}^{t}ds\,tr\left\{ \hat{H}(s)\frac{d}{ds}\hat{\rho}(s)\right\} .\label{Alicki heat}
\end{align}
It is clear that such identifications automatically fulfil a quantum
dynamical version of the first law, stated by $U(t)-U(t_{0})=W_{1}(t)+Q_{1}(t)$\footnote{The subscripts will be necessary to differentiate distinct work and
heat definitions.}. Besides, notice that heat $Q_{1}(t)$ does not depend on the unitary
component of Eq. (\ref{Alicki Dynamics})\footnote{Given the trace ciclic property, it is easy to see that $tr\left\{ \hat{H}(t)\left[\hat{H}(t),\hat{\rho}(t)\right]\right\} =0$.},
in such a way that $Q_{1}(t)=\sum_{k}^{N}\int_{t_{0}}^{t}ds\,tr\{\hat{H}(s)\mathfrak{\hat{D}}_{t}^{(k)}\hat{\rho}(s)\}=\sum_{k}^{N}Q_{1}^{(k)}(t)$,
where $Q_{1}^{(k)}(t)\equiv tr\{\hat{H}(s)\mathfrak{\hat{D}}_{t}^{(k)}\hat{\rho}(s)\}$
is understood as the energy supplied by the $k$th reservoir. Thus,
as expected, heat is the energy transferred due to the coupling with
other systems, which also implies that $Q_{1}^{(k)}(t)=0$ for all
$k$ iff the dynamics is unitary/closed. Work, on the other hand,
explicitly relies on a semi-classical description for accounting for
the external dynamical control and Hamiltonian time-dependency, which
is assimilated by $\hat{h}(t)$. Thus, work is simply $W_{1}(t)\equiv\int_{t_{0}}^{t}ds\,tr\{\frac{d}{ds}\hat{h}(s)\hat{\rho}(s)\}$
and, therefore, vanishes iff the quantum system is isolated. As we
can see, those definitions are consistent with the intuition behind
their classical counterparts, however, it is not clear if they are
still compatible with general dynamics, and - more importantly - it
also prevents further discussions concerning the thermodynamics of
autonomous quantum machines. Despite these issues, Alicki's definition
proposal is widely accepted and used in the literature since their entropic predictions are consistent with the second law when considering weak-coupling and Markov approximations. Also, it is
worth mentioning that alternative approaches of quantum work and heat
also adopt the Eqs. (\ref{Alicki work}, \ref{Alicki heat}) forms,
but instead of $\hat{H}(t)$ it is assumed different sorts of effective
Hamiltonians. \cite{Weimer2008,PhysRevE.81.021118,Hossein-Nejad2015,Alipour2016a,Valente2018,Celeri2021,Colla2021a}

Interestingly, note that the energy change rate splitting into two
distinct terms is not unique, i.e., one might perform the trace in
Eq. (\ref{Energia interna}) on any conceivable basis, and separate
its time derivative into two arbitrary components. For instance, if $\hat{H}(t)=\sum_{j}\epsilon_{j}(t)|\epsilon_{j}(t)\rangle\langle\epsilon_{j}(t)|$
is the instantaneous Hamiltonian spectral decomposition, we can rewrite
Eq. (\ref{Energia interna}) in the following way
\begin{equation}
U(t)\equiv\sum_{j}\epsilon_{j}(t)p_{j}(t),
\end{equation}
where $p_{j}(t):=\langle\epsilon_{j}(t)|\hat{\rho}(t)|\epsilon_{j}(t)\rangle$
is the $j$th diagonal element of $\hat{\rho}(t)$ in the instantaneous
Hamiltonian eigenbasis representation and quantify the probability
of the system being in the state $|\epsilon_{j}(t)\rangle$. Then,
in analogy with Eqs. (\ref{Classical energy separation}-\ref{Classical heat}),
work and heat might be defined after the internal energy change rate
\begin{equation}
\frac{d}{dt}U(t)=\sum_{j}\frac{d\epsilon_{j}(t)}{dt}p_{j}(t)+\sum_{j}\epsilon_{j}(t)\frac{dp_{j}(t)}{dt},
\end{equation}
such that,
\begin{align}
W_{2}(t) & :=\sum_{j}\int_{t_{0}}^{t}ds\,\frac{d\epsilon_{j}(s)}{ds}p_{j}(s),\label{Segunda forma de trabalho}\\
Q_{2}(t) & :=\sum_{j}\int_{t_{0}}^{t}ds\,\epsilon_{j}(s)\frac{dp_{j}(s)}{ds},\label{Segunda forma de calor}
\end{align}
in a way that the first law equation form $U(t)-U(t_{0})=W_{2}(t)+Q_{2}(t)$
is still satisfied. In general, the expressions above are different
from the ones presented in Eqs. (\ref{Alicki work}, \ref{Alicki heat}).
In particular, instead of being associated with the whole Hamiltonian
change, work is only related to modifications in the energy spectrum
$\frac{d\epsilon_{j}(t)}{dt}$ while heat depends on the dynamics
of both the whole state $\hat{\rho}(t)$ and the energy eigenstates
$\{|\epsilon_{j}(t)\rangle\}_{j}$, encoded by the populations $\{p_{j}(t)\}_{j}$.
These quantities are directly related to Alicki's proposal in the
following way
\begin{align}
W_{2}(t) & =W_{1}(t)-\sum_{j}\int_{t_{0}}^{t}ds\,\epsilon_{j}(s)\left(\frac{d}{ds}\left(\langle\epsilon_{j}(s)|\right)\hat{\rho}(s)|\epsilon_{j}(s)\rangle+\langle\epsilon_{j}(s)|\hat{\rho}(s)\frac{d}{ds}\left(|\epsilon_{j}(s)\rangle\right)\right),\label{Diferença work}\\
Q_{2}(t) & =Q_{1}(t)+\sum_{j}\int_{t_{0}}^{t}ds\,\epsilon_{j}(s)\left(\frac{d}{ds}\left(\langle\epsilon_{j}(s)|\right)\hat{\rho}(s)|\epsilon_{j}(s)\rangle+\langle\epsilon_{j}(s)|\hat{\rho}(s)\frac{d}{ds}\left(|\epsilon_{j}(s)\rangle\right)\right),\label{Diferença heat}
\end{align}
where it is clear that the only difference between both definitions
is where the Hamiltonian basis change contribution is considered,
and $U(t)-U(t_{0})\equiv W_{1}(t)+Q_{1}(t)\equiv W_{2}(t)+Q_{2}(t)$.
Conceptually, however, such approaches are very distinct and might
predict conflicting scenarios. For instance, while for unitary processes
$Q_{1}(t)=0$ for all $t$, which implies that there is no heat involved,
$Q_{2}(t)$ does not necessarily vanish for closed systems dynamics.
In this sense, contrasting with the previous definition, the energy
transferred by the external agent is divided into heat and work. Still,
both approaches fundamentally rely on the classical picture that work
is the energy externally provided in a controlled fashion.

Alternatively, instead of focusing on the energy basis with a prior
role for work, more recently in \cite{Ahmadi2019a,Alipour2019b}, it was suggested to use
the instantaneous basis of the density matrix $\{|\phi_{j}(t)\rangle\}_{j}$
and concentrate on the identification of heat. Along these lines,
if $\hat{\rho}(t)=\sum_{j}\varrho_{j}(t)|\phi_{j}(t)\rangle\langle\phi_{j}(t)|$
is the time-local spectral decomposition of $\hat{\rho}(t)$, Eq.
(\ref{Energia interna}) can also be written as
\begin{equation}
U(t)\equiv\sum_{j}\varepsilon_{j}(t)\varrho_{j}(t),
\end{equation}
where $\{\varrho_{j}(t)\}$ are the instantaneous populations of $\hat{\rho}(t)$
and $\varepsilon_{j}(t):=\langle\phi_{j}(t)|\hat{H}(t)|\phi_{j}(t)\rangle$
is interpreted as the energy relative to the $j$th pure state $|\phi_{j}(t)\rangle$.
Then, the energy change rate can be divided into 
\begin{equation}
\frac{d}{dt}U(t)=\sum_{j}\frac{d\varepsilon_{j}(t)}{dt}\varrho_{j}(t)+\sum_{j}\varepsilon_{j}(t)\frac{d\varrho_{j}(t)}{dt}
\end{equation}
 and, again in analogy with Eqs. (\ref{Classical energy separation}-\ref{Classical heat}),
work and heat might be defined as
\begin{align}
W_{3}(t) & :=\sum_{j}\int_{t_{0}}^{t}ds\,\frac{d\varepsilon_{j}(s)}{ds}\varrho_{j}(s),\label{Terceira forma de trabalho}\\
Q_{3}(t) & :=\sum_{j}\int_{t_{0}}^{t}ds\,\varepsilon_{j}(s)\frac{d\varrho_{j}(s)}{ds},\label{Terceira forma de calor}
\end{align}
such that $U(t)-U(t_{0})=W_{3}(t)+Q_{3}(t)$. Hence, heat is associated
with changes in the distribution of pure states $\left\{ \frac{d}{dt}\varrho_{j}(t)\right\} _{j}$
while the remaining part, depending on the dynamics of both Hamiltonian
$\hat{H}(t)$ and instantaneous basis $\{|\phi_{j}(t)\rangle\}_{j}$,
is identified as work. Such division is motivated by the common recognition
of the von Neumann entropy $S_{vN}(t):=-\sum_{j}\varrho_{j}(t)ln(\varrho_{j}(t))$
as the natural extension of thermodynamic entropy and the classical
relationship between heat flux and entropy variation. Classically,
heat is the portion of energy exchange accompanied by the flow of
entropy into/from the system, while work is the energetic contribution
that does not generate any entropic changes. Along these lines, since
$\frac{d}{dt}S_{vN}(t):=-\sum_{j}\frac{d\varrho_{j}(t)}{dt}ln(\varrho_{j}(t))$,
heat is associated with the non-unitary part of the system dynamics
and, more specifically, identified by the energetic change that also
functionally depends on $\frac{d\varrho_{j}(t)}{dt}$. Thus, if the
time-evolution is unitary we would automatically have both $\frac{d}{dt}S_{vN}(t)=0$
and $Q_{3}(t)=0$ and, therefore, all energetic exchange would be
due to work, $U(t)-U(t_{0})=W_{3}(t)$. Also, it would guarantee the
agreement between $W_{3}(t)$ and $W_{1}(t)$. Interestingly, this
approach does not fundamentally require a semi-classical picture:
even if the Hamiltonian $\hat{H}(t)$ is kept constant, for general
open system dynamics the basis $|\phi_{j}(t)\rangle$ might change
in such a way that $\frac{d\varepsilon_{j}(t)}{dt}\neq0$ and work
$W_{3}(t)$ is not necessarily null for fully quantum interacting
subsystems. It is worth mentioning that in \cite{Bottosso2019} was presented
a simple example where $Q_{3}(t)=0$ for all $t$ while $\frac{d}{dt}S_{vN}(t)\neq0$,
which motivate the authors to argue that this may constitute a possible
inadequacy between the thermodynamic argument and the definitions
above.

Hence, despite having a similar thermodynamic grounding and interesting individual characteristics, all these proposals do not agree with each other in most cases. Naturally, different definitions may predict radically distinct thermodynamic scenarios. In particular, notice that the root of their divergence is due to the initial arbitrary basis representation choice for $U(t)$, which affects the contributions splitting and identification. Besides, it is also clear that coherence must be taken into account, i.e., in general
$[\hat{\rho}(t),\hat{H}(t)]\neq0$ and the quantum coherence - in
any basis choice - plays an indispensable role in the energetic changes.
For instance, even though non-diagonal elements of $\hat{\rho}(t_{0})$
in the Hamiltonian basis at a given instant $t_{0}$, written as $\langle\epsilon_{j}(t_{0})|\hat{\rho}(t_{0})|\epsilon_{k}(t_{0})\rangle$
with $j\neq k$, do not instantaneously contribute to the internal
energy $U(t_{0})$\footnote{Since $\hat{H}(t)=\sum_{j}\epsilon_{j}(t)|\epsilon_{j}(t)\rangle\langle\epsilon_{j}(t)|$,
it is easy to see that 
\begin{equation*}
U(t)=tr\{\hat{H}(t)\hat{\rho}(t)\}=\sum_{j}\epsilon_{j}(t)\langle\epsilon_{j}(t)|\hat{\rho}(t)|\epsilon_{j}(t)\rangle.
\end{equation*}}, their dynamics are - in general - coupled with the populations $\langle\epsilon_{j}(t)|\hat{\rho}(t)|\epsilon_{j}(t)\rangle$\footnote{For some specific situations, their dynamics might be decoupled, e.g.,
Davies maps.}, which means that they are relevant for $U(t)$ at latter times ($t\geq t_{0}$).
In other words, all density matrix elements are essential for accounting
for the energetic change of any given quantum system. Interestingly, in
the classical limit, such contributions become negligible in a way
that these ambiguities disappear and the previous definitions converge
to compatibility. In this sense, it is not obvious if such additional
quantities should be interpreted as work-like or heat-like variations
and, since there is no classical analogous of coherence, classical
thermodynamics does not provide any direct instruction on how to deal
with them. In fact, it is not even clear if coherence should be tied
to such roles. Most of the literature implicitly assumes that either
coherence is shared between both quantities or belongs to one of them,
i.e., is commonly assumed a priori the classical thermodynamic structure
of complementary ingredients. The classical first law states that
$dU\equiv\delta W+\delta Q$, in a way that once defined work or heat,
it is automatically established the remaining part. Thus, it is important
to highlight that there is still room for disputing this assumption
at the quantum level. Along these lines, as an alternative approach,
\cite{Bernardo2020} proposed a redefinition of the first law where the coherence
energetic contribution, given by $\delta C$, is completely separated
from the notions of work and heat, such that $U\equiv\delta W+\delta Q+\delta C$.

In summary, discussions on quantum counterparts of thermodynamic quantities
are still in their early stages and far from being settled. This,
of course, corresponds to a core conceptual issue for the development
of a fully quantum thermodynamics theory and has a direct impact on
the design and operation of effective quantum devices. The most popular
routes offer either an operational framework or a semi-classical description,
which are - essentially - phenomenological approaches in spirit and
do not consistently apply for several scenarios of interest. Especially
those that demand a more symmetrical thermodynamic treatment of all
considered parties.

\subsubsection{Road to effective work and heat\label{subsec:Road-to-effective}}

As mentioned earlier, in addition to not being clear which thermodynamic
role is played by genuine quantum phenomena, it is also mandatory
a previous identification of internal energy in order to properly
recognize quantum versions of work and heat. Such fundamental ambiguities
emphasize the urgency of careful examination of these quantities at
the conceptual level. In particular, it is imperative the identification
of the essential features of what exactly characterizes work and heat.
Along these lines, in Section (\ref{subsec:Local-effective-internal}),
we argued that the local effective Hamiltonians $\tilde{H}^{(1,2)}(t)$
provide a promising and suitable candidate for quantifying the subsystems
internal energy $U^{(k)}(t):=\langle\tilde{H}^{(k)}(t)\rangle$, in
a way that is local, additive ( $\frac{d}{dt}U^{(1)}(t)=-\frac{d}{dt}U^{(2)}(t)$)
and applicable to \textit{arbitrary} scenarios. Hence, it also provides
a starting point for discussing a general, exact and symmetrical understanding
of work and heat along with any dynamical processes. This approach
naturally allows the energetic characterization that transcends the
usual asymmetric thermodynamic description and restrictive regimes,
such as weakly coupled systems and markovian dynamics. Here, nevertheless,
we do not propose fixed definitions for these quantities. Instead,
we advocate that the previously presented work and heat forms from
a dynamical point of view might be well adapted for our local internal
energy identification.

The local internal energy dynamics for each subsystem is simply $\frac{d}{dt}U^{(k)}(t)=\frac{d}{dt}tr_{k}\{\tilde{H}^{(k)}(t)\hat{\rho}^{(k)}(t)\}$.
From previous discussions, it becomes clear that this kind of expression
can be consistently divided into two components in several distinct
ways that still correspond to the first law structure\footnote{Here we will assume a priori the usual first law structure.}.
Then, we are always allowed to write the following expression
\begin{equation}
\frac{d}{dt}U^{(k)}(t)=\frac{d}{dt}\mathbb{W}^{(k)}(t)+\frac{d}{dt}\mathbb{Q}^{(k)}(t),
\end{equation}
where $\mathbb{W}^{(k)}(t)$ and $\mathbb{Q}^{(k)}(t)$ are identified
as the effective work and effective heat flowing from/to subsystem
$(k)$, respectively. Furthermore, the symmetrical treatment of both
parts and internal energy additivity imply that all net energetic
exchange involved during the whole system dynamics should sum up to
zero, i.e.,
\begin{equation}
\frac{d}{dt}\mathbb{W}(t)+\frac{d}{dt}\mathbb{Q}(t)=0,\label{Net work + Net heat}
\end{equation}
where $\mathbb{W}(t)=\mathbb{W}^{(1)}(t)+\mathbb{W}^{(2)}(t)$ and
$\mathbb{Q}(t)=\mathbb{Q}^{(1)}(t)+\mathbb{Q}^{(2)}(t)$ are the respective
net work and net heat transferred throughout the described process.

From Eqs. (\ref{Alicki work}, \ref{Alicki heat}) one might approach
these quantities according to Alicki's proposal in the following straightforward
way
\begin{align}
\mathbb{W}_{1}^{(k)}(t) & :=\int_{t_{0}}^{t}ds\,tr_{k}\left\{ \frac{d}{ds}\tilde{H}^{(k)}(s)\hat{\rho}^{(k)}(s)\right\} =\sum_{j=1}^{d^{(1)}}\int_{t_{0}}^{t}ds\,\lambda_{j}^{2}(s)\langle\varphi_{j}^{(k)}(s)|\frac{d}{ds}\tilde{H}^{(k)}(s)|\varphi_{j}^{(k)}(s)\rangle,\label{workAlickiLike}\\
\mathbb{Q}_{1}^{(k)}(t) & :=\int_{t_{0}}^{t}ds\,tr_{k}\left\{ \tilde{H}^{(k)}(s)\frac{d}{ds}\hat{\rho}^{(k)}(s)\right\} =\sum_{j=1}^{d^{(1)}}\int_{t_{0}}^{t}ds\,\frac{d\lambda_{j}^{2}(s)}{ds}\langle\varphi_{j}^{(k)}(s)|\tilde{H}^{(k)}(s)|\varphi_{j}^{(k)}(s)\rangle,\label{HeatAlickiLike}
\end{align}
where work is the energy exchange associated with the local effective
Hamiltonian dynamics and heat is related to the local state change.
Given Eq. (\ref{Effective Hamiltonian 2}), it is clear that work
$\mathbb{W}_{1}^{(k)}(t)$ is a direct outcome of the time-dependency
induced by the interaction between the subsystems and, therefore,
does not require any ad hoc, external agent for describing it. In
fact, the bare Hamiltonian $\hat{H}^{(k)}$ plays no role in this
energetic exchange since $\hat{H}^{(k)}=cte$ and $\mathbb{W}_{1}^{(k)}(t):=\int_{t_{0}}^{t}ds\,tr_{k}\left\{ \frac{d}{ds}\left(\hat{H}_{LS}^{(k)}(s)+\hat{H}_{X}^{(k)}(s)\right)\hat{\rho}^{(k)}(s)\right\} $.
Also, it intrinsically bound heat $\mathbb{Q}_{1}^{(k)}(t)$ with
entanglement variation due to its explicitly functional dependency
on the population dynamics $\left\{ \frac{d\lambda_{j}^{2}(t)}{dt}\right\} _{j}$.
Additionally, in contrast with the classical equilibrium thermodynamic
scenario, it is also clear that the net work $\mathbb{W}_{1}(t)$
and net heat $\mathbb{Q}_{1}(t)$ are not individually null for general
situations. Thus, for instance, the heat flowing from subsystem $(1)$
is not necessarily translated into the heat absorbed by subsystem
$(2)$, i.e., $\mathbb{Q}_{1}(t)=\mathbb{Q}_{1}^{(1)}(t)+\mathbb{Q}_{1}^{(2)}(t)\neq0$.
Finally, it is easy to show that
\begin{equation}
\langle\varphi_{j}^{(k)}(t)|\frac{d}{dt}\tilde{H}^{(k)}(t)|\varphi_{j}^{(k)}(t)\rangle=\frac{d}{dt}\left(\langle\varphi_{j}^{(k)}(t)|\tilde{H}^{(k)}(t)|\varphi_{j}^{(k)}(t)\rangle\right),
\end{equation}
for all $t$ and, therefore, that there is an equality between Alicki's
proposal and the form presented in Eqs. (\ref{Terceira forma de trabalho},
\ref{Terceira forma de calor}). In such a case, the heat $\mathbb{Q}_{3}^{(k)}(t)$
experienced by subsystem $(k)$ is directly related to changes in
the populations of the local state $\hat{\rho}^{(k)}(t)$, and work
is assigned to the remaining energetic contribution due to the local
effective Hamiltonian dynamics and instantaneous basis $\{|\varphi_{j}^{(k)}(t)\rangle\}_{j}$
variation. Hence
\begin{align}
\mathbb{W}_{1}^{(k)}(t) & =\sum_{j=1}^{d^{(1)}}\int_{t_{0}}^{t}ds\,\lambda_{j}^{2}(s)\frac{d}{dt}\left(\langle\varphi_{j}^{(k)}(s)|\tilde{H}^{(k)}(s)|\varphi_{j}^{(k)}(s)\rangle\right)=\mathbb{W}_{3}^{(k)}(t),\\
\mathbb{Q}_{1}^{(k)}(t) & =\sum_{j=1}^{d^{(1)}}\int_{t_{0}}^{t}ds\,\frac{d\lambda_{j}^{2}(s)}{ds}\langle\varphi_{j}^{(k)}(s)|\tilde{H}^{(k)}(s)|\varphi_{j}^{(k)}(s)\rangle=\mathbb{Q}_{3}^{(k)}(t),
\end{align}
in contrast with what we would have obtained if we just had considered
the local bare Hamiltonians\footnote{By hypothesis, we consider constant local bare Hamiltonians $\hat{H}^{(k)}$.
Thus, the previous equality would not be satisfied, i.e., $0=\langle\varphi_{j}^{(k)}(t)|\frac{d}{dt}\hat{H}^{(k)}|\varphi_{j}^{(k)}(t)\rangle\neq\frac{d}{dt}\left(\langle\varphi_{j}^{(k)}(t)|\hat{H}^{(k)}|\varphi_{j}^{(k)}(t)\rangle\right)$.}.

Alternatively, one might define work and heat in analogy with the
procedure presented along with Eqs. (\ref{Segunda forma de trabalho},
\ref{Segunda forma de calor}). If
\begin{equation}
\tilde{H}^{(k)}(t)\equiv\sum_{j=1}^{d^{(k)}}\tilde{\epsilon}_{j}^{(k)}(t)|\tilde{\epsilon}_{j}^{(k)}(t)\rangle\langle\tilde{\epsilon}_{j}^{(k)}(t)|
\end{equation}
is the instantaneous local effective Hamiltonian spectral decomposition,
where $\{\tilde{\epsilon}_{j}^{(k)}(t)\}_{j}$ are the time-local
effective eigenenergies and $\{|\tilde{\epsilon}_{j}^{(k)}(t)\rangle\}_{j}$
are their eigenstates, it is clear that the effective internal energy
can be cast as
\begin{equation}
U^{(k)}(t)=\langle\tilde{H}^{(k)}(t)\rangle=\sum_{j=1}^{d^{(k)}}\tilde{\epsilon}_{j}^{(k)}(t)\langle\tilde{\epsilon}_{j}^{(k)}(t)|\hat{\rho}^{(k)}(t)|\tilde{\epsilon}_{j}^{(k)}(t)\rangle.
\end{equation}
Then, work $\mathbb{W}_{2}^{(k)}(t)$ and heat $\mathbb{Q}_{2}^{(k)}(t)$
could be identified as
\begin{align}
\mathbb{W}_{2}^{(k)}(t) & =\sum_{j=1}^{d^{(k)}}\int_{t_{0}}^{t}ds\,\frac{d\tilde{\epsilon}_{j}^{(k)}(s)}{ds}\langle\tilde{\epsilon}_{j}^{(k)}(s)|\hat{\rho}^{(k)}(s)|\tilde{\epsilon}_{j}^{(k)}(s)\rangle,\label{WorkForma2like}\\
\mathbb{Q}_{2}^{(k)}(t) & =\sum_{j=1}^{d^{(k)}}\int_{t_{0}}^{t}ds\,\tilde{\epsilon}_{j}^{(k)}(s)\frac{d}{ds}\left(\langle\tilde{\epsilon}_{j}^{(k)}(s)|\hat{\rho}^{(k)}(s)|\tilde{\epsilon}_{j}^{(k)}(s)\rangle\right),\label{HeatForma2like}
\end{align}
such that the former is only associated with changes in the local effective
Hamiltonian spectrum, instead of the whole operator, and the latter
is related to the dynamics of its respective local state $\hat{\rho}^{(k)}(t)$
and basis $\{|\tilde{\epsilon}_{j}^{(k)}(t)\rangle\}_{j}$. Again,
the equality presented in Eq. (\ref{Net work + Net heat}) above is
guarateed by construction and there is no reason why the net work
$\mathbb{W}_{2}(t)$ and net heat $\mathbb{Q}_{2}(t)$ should be individually
null for general cases.

Thus, as presented above, the definition of the local effective internal
energies $U^{(k)}(t)=\langle\tilde{H}^{(k)}(t)\rangle$ provides a
consistent foundation for discussing work and heat and the first law
in an exact, symmetrical and general manner. This, however, is not
enough. As mentioned before, it is imperative to encourage in-depth
investigations at the conceptual level. In particular, a focus on
questions concerning the identification of what are the essential
features that fundamentally characterizes work and heat is particularly
needed. Otherwise, the energetic contribution of coherence will remain
elusive and arbitrarily considered into potential candidates.

%% file: Chapters/Chapter3/Section7.tex
\section{Proof of principle\label{sec:Proof-of-principle}}

To illustrate the concepts presented in this chapter, let us now apply
our formalism for describing the energetic exchange of a simple -
but paradigmatic - example of two interacting qubits as a proof of
principle. Of course, a priori, this method applies to any quantum
system, as long we perform a bipartition.

Suppose the system is described by the following Hamiltonian
\begin{equation}
\hat{H}^{(0)}:=\hbar\frac{\omega_{1}}{2}\hat{\sigma}_{z}^{(1)}\otimes\hat{1}^{(2)}+\hat{1}^{(1)}\otimes\hbar\frac{\omega_{2}}{2}\hat{\sigma}_{y}^{(2)}+\hbar g\left(\hat{\sigma}_{x}^{(1)}\otimes\hat{\sigma}_{y}^{(2)}\right)\in\mathcal{L}(\mathcal{H}^{(0)}),
\end{equation}
where $\hat{H}^{(1)}\equiv\hbar\frac{\omega_{1}}{2}\hat{\sigma}_{z}^{(1)}$
and $\hat{H}^{(2)}\equiv\hbar\frac{\omega_{2}}{2}\hat{\sigma}_{y}^{(2)}$
are the local bare Hamiltonians, $\hat{H}_{int}\equiv\hbar g\left(\hat{\sigma}_{x}^{(1)}\otimes\hat{\sigma}_{y}^{(2)}\right)$
is the interaction term, $\hat{\sigma}_{x,y,z}$ are the usual Pauli
matrices, $g$ is the coupling constant, and $\hbar\omega_{1,2}$
are the energy gaps for $\hat{H}^{(1,2)}$. Since we are dealing with
two qubits, we have $d^{(0)}:=dim(\mathcal{H}^{(0)})=4$ and $d^{(1,2)}:=dim(\mathcal{H}^{(1,2)})=2$.
Hence, for every time $t$, the whole system state will be written - in general - as
\begin{equation}
|\Psi(t)\rangle=\lambda_{1}(t)|\varphi_{1}^{(1)}(t)\rangle\otimes|\varphi_{1}^{(2)}(t)\rangle+\lambda_{2}(t)|\varphi_{2}^{(1)}(t)\rangle\otimes|\varphi_{2}^{(2)}(t)\rangle.
\end{equation}
Given an initial state $|\Psi(0)\rangle$, it will evolve in time
according to $|\Psi(t)\rangle=\hat{\mathcal{U}}(t)|\Psi(0)\rangle$,
where $\hat{\mathcal{U}}(t)=e^{-\frac{i}{\hbar}\hat{H}^{(0)}t}\in\mathcal{L}(\mathcal{H}^{(0)})$
is the time-evolution operator. Conveniently, since the bare Hamiltonian
of subsystem $(2)$ commutes with the interaction term, $[\hat{H}^{(2)},\hat{H}_{int}]=0$,
$\hat{\mathcal{U}}(t)$ simplifies to
\begin{equation}
\hat{\mathcal{U}}(t)=e^{-i\left(\frac{\omega_{1}}{2}\hat{\sigma}_{z}^{(1)}\otimes\hat{1}^{(2)}+g\left(\hat{\sigma}_{x}^{(1)}\otimes\hat{\sigma}_{y}^{(2)}\right)\right)t}e^{-i\frac{\omega_{2}}{2}\hat{\sigma}_{y}^{(2)}t},\label{eq:time evolution operator example}
\end{equation}
which also implies that $[\hat{H}^{(2)},\hat{\mathcal{U}}(t)]=0$\footnote{If $\hat{A}$ and $\hat{B}$ commute, then $e^{\hat{A}+\hat{B}}=e^{\hat{A}}e^{\hat{B}}$.}.
Thus, if $\hat{\sigma}_{y}^{(2)}|\pm_{y}^{(2)}\rangle=\pm|\pm_{y}^{(2)}\rangle$
and $\hat{V}(t):=e^{-i\left(\frac{\omega_{1}}{2}\hat{\sigma}_{z}^{(1)}\otimes\hat{1}^{(2)}+g\left(\hat{\sigma}_{x}^{(1)}\otimes\hat{\sigma}_{y}^{(2)}\right)\right)t}$,
then
\begin{equation}
\hat{\rho}^{(2)}(t)=e^{i\frac{\omega_{2}}{2}\hat{\sigma}_{y}^{(2)}t}tr_{1}\{\hat{V}^{\dagger}(t)|\Psi(0)\rangle\langle\Psi(0)|\hat{V}(t)\}e^{-i\frac{\omega_{2}}{2}\hat{\sigma}_{y}^{(2)}t}
\end{equation}
and, therefore,
\begin{equation}
\langle\pm_{y}^{(2)}|\hat{\rho}^{(2)}(t)|\pm_{y}^{(2)}\rangle=\langle\pm_{y}^{(2)}|\hat{\rho}^{(2)}(0)|\pm_{y}^{(2)}\rangle
\end{equation}
for all $t$, i.e., the populations of $\hat{\rho}^{(2)}(t)$ in the
$\hat{\sigma}_{y}^{(2)}$ basis, $\{|\pm_{y}^{(2)}\rangle\}$, are
constant during the whole dynamics, while the non-diagonal elements
$\langle\mp_{y}^{(2)}|\hat{\rho}^{(2)}(t)|\pm_{y}^{(2)}\rangle$ may
evolve independently. Notice this is not true for qubit $(1)$, since $[\hat{H}^{(1)},\hat{H}_{int}]\neq0$, a priori, the change of all of its density matrix elements are coupled intrinsically with one another. Interestingly, despite their different dynamics, both qubits are guaranteed to maintain their equal purities during the whole time evolution. If $\mathbb{P}[\hat{\sigma}]\equiv tr\{\hat{\sigma}^{2}\}$ is the purity of a state $\hat{\sigma}$, then it is clear that $\mathbb{P}[\hat{\rho}^{(0)}(t)]=1$ - given $\hat{\rho}^{(0)}(t)\equiv|\Psi(t)\rangle\langle\Psi(t)|$ is a pure state for all $t$ - while $\mathbb{P}[\hat{\rho}^{(1)}(t)]=\mathbb{P}[\hat{\rho}^{(2)}(t)]=\lambda_{1}^{4}(t)+\lambda_{2}^{4}(t)\leq1$. The local purity changes also reflect the modification in the entanglement degree between the qubits. As mentioned in Section (\ref{Issues of the von Neumann entropy}), in this context, the von Neumann entropy (or entanglement entropy) represents a direct measure of the entanglement within the bipartition. Under these circumstances, while the whole system's von Neumann entropy is null throughout the dynamics (since it is pure), both subsystems are guaranteed to possess the same value for their von Neumann entropy, i.e., $S_{vN}[\hat{\rho}^{(0)}(t)]=0$ and $S_{vN}[\hat{\rho}^{(1)}(t)]=S_{vN}[\hat{\rho}^{(2)}(t)]=-\lambda_{1}^{2}(t)ln\left(\lambda_{1}^{2}(t)\right)-\lambda_{2}^{2}(t)ln\left(\lambda_{2}^{2}(t)\right)$ for all $t$.

\begin{figure}[h!]
     \centering
     \begin{subfigure}[b]{0.49\textwidth}
         \centering
         \includegraphics[width=\textwidth]{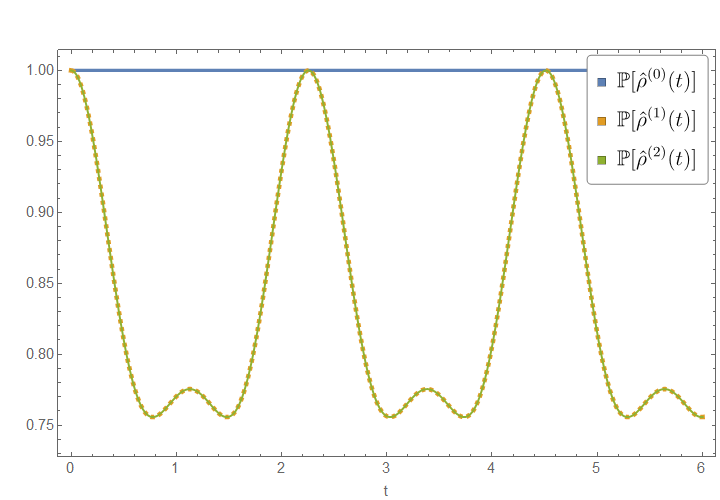}
         \caption{Purity}
         \label{Purity}
     \end{subfigure}
     \begin{subfigure}[b]{0.49\textwidth}
         \centering
         \includegraphics[width=\textwidth]{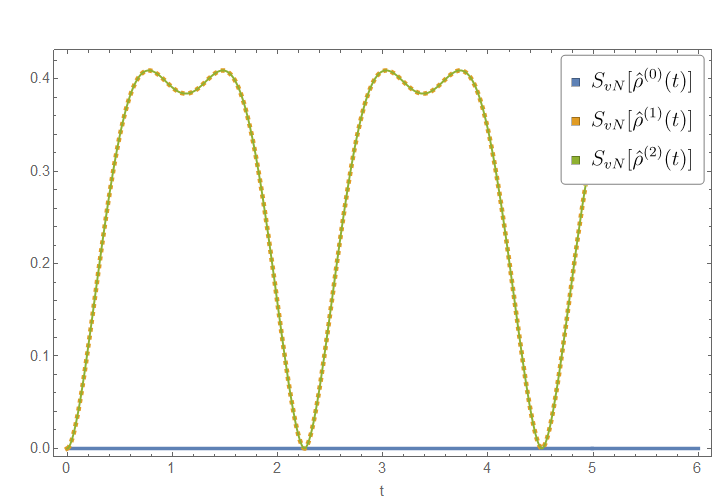}
         \caption{von Neumann entropy}
         \label{vN Entropy}
     \end{subfigure}
        \caption{\textbf{(a)} Purity dynamics for two interacting qubits. While the whole system maintains pure for all t, both qubits oscillate continuously between pure and mixed states with equal purity; \textbf{(b)} Dynamical behaviour of the von Neumann entropies. On the one hand, given $\hat{\rho}^{(0)}(t)\equiv|\Psi(t)\rangle\langle\Psi(t)|$, the whole system's von Neumann entropy maintains null throughout the unitary dynamics. On the other hand, the equal qubits' von Neumann entropies illustrate the oscillation between the degree of entanglement within the bipartition. For the computation of all the plots, it was assumed $a^{(1)}=(5)^{-1/2}, b^{(1)}=2(5)^{-1/2}$, $a^{(2)}=(10)^{-1/2}$ and $b^{(2)}=3(10)^{-1/2}$ for the initial states, $\hbar\equiv1$, $\omega_{1}=1$, $\omega_{2}=5$ and $g=1.3$.}
        \label{fig0}
        Source: By the author.
\end{figure}

The temporal evolution of this simple physical system can be easily checked by numerical analysis. Let us suppose initial uncorrelated states, such that
\begin{equation}
|\Psi(0)\rangle=|\varphi^{(1)}(0)\rangle\otimes|\varphi^{(2)}(0)\rangle
\end{equation}
with $|\varphi^{(k)}(0)\rangle=a^{(k)}|+_{z}^{(k)}\rangle+b^{(k)}|-_{z}^{(k)}\rangle$\footnote{It is worth mentioning this is a convenient simplifying hypothesis
that does not limit the computational analysis or the conclusions.}. Thus, Figure \ref{Purity} shows how the qubits purity oscillates between $1$ and $<1$, which indicates their continuously changing from pure to mixed states. Additionally, it is clear that the whole system's state also oscillates from the initial product to entangled states. In this sense, Figure \ref{vN Entropy} depicts the dynamical behaviour of the von Neumann entropies and illustrates that both quantities are, indeed, correlated. As expected, whenever the local density matrices are pure, their von Neumann entropies are equally null, while the maximum entanglement is obtained when their purities reach their minimum values.

\begin{figure}[h!]
    \centering
    \includegraphics[width=0.7\textwidth]{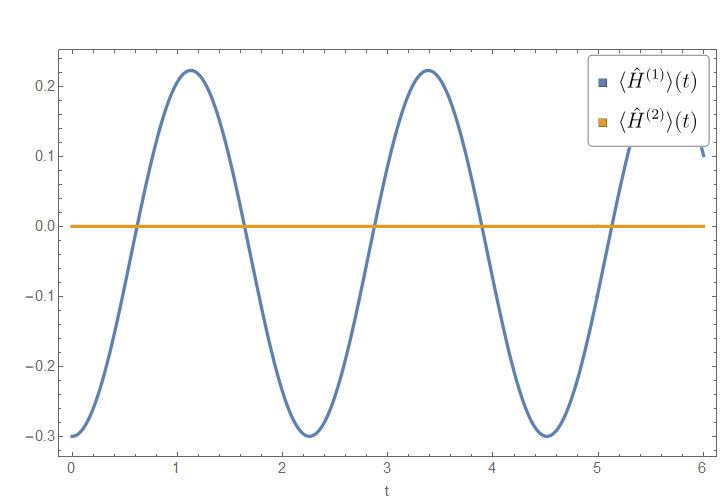}
    \caption{Dynamical behaviour of the expectation values of the local bare Hamiltonians. On the one hand, since $[\hat{H}^{(2)},\hat{H}_{int}]=0$, the populations of $\hat{\rho}^{(2)}$ in the $\hat{H}^{(2)}$ basis are time-invariant and, therefore, $\frac{d}{dt}\langle\hat{H}^{(2)}\rangle(t)=0$ for all $t$. On the other, $\frac{d}{dt}\langle\hat{H}^{(1)}\rangle(t)\neq0$ and $\langle\hat{H}^{(1)}\rangle(t)$ continuously oscillates in time.}
    \label{Fig1}
    Source: By the author.
\end{figure}

More importantly, given that the populations of $\hat{\rho}^{(2)}$ in the $\hat{H}^{(2)}$ basis are time-invariant, the expectation value of the bare Hamiltonian, $\langle\hat{H}^{(2)}\rangle(t)\equiv tr_{2}\{\hat{H}^{(2)}\hat{\rho}^{(2)}(t)\}=\langle\Psi(t)|\hat{H}^{(2)}|\Psi(t)\rangle$, is also guaranteed to be constant in time, i.e.,
\begin{equation}
\langle\hat{H}^{(2)}\rangle=\langle\Psi(0)|\hat{H}^{(2)}|\Psi(0)\rangle\,\Rightarrow\,\frac{d}{dt}\langle\hat{H}^{(2)}\rangle=0.\label{eq:condition}
\end{equation}
As expected, there is no such constraint for the mean value of the bare Hamiltonian of qubit $(1)$, $\langle\hat{H}^{(1)}\rangle(t)$, and it is free to evolve in time. Notice that this conclusion is a very general statement and does not
depend on our particular physical system. It is a simple consequence
of the previous commutation relations\footnote{That is what happens in the usual dephasing model, for instance.}. Figure \ref{Fig1} illustrates this behaviour for the initially uncorrelated qubits: while $\langle\hat{H}^{(1)}\rangle(t)$ continuously oscillates in time, $\langle\hat{H}^{(2)}\rangle$ maintains its initially null value and keeps constant throughout the whole dynamics. This happens because the whole system's internal energy, $U^{(0)}$, is shared between the local bare Hamiltonians and the interaction term, such that 
\begin{equation}
U^{(0)}=\langle\hat{H}^{(1)}\rangle(t)+\langle\hat{H}^{(2)}\rangle(t)+\langle\hat{H}_{int}\rangle(t).
\end{equation}
Thus, if Eq. (\ref{eq:condition}) above is satisfied and given that $\frac{d}{dt}U^{(0)}=0$, we automatically have
\begin{equation}
\frac{d}{dt}\langle\hat{H}^{(1)}\rangle(t)=-\frac{d}{dt}\langle\hat{H}_{int}\rangle(t).
\end{equation}
Along these lines, if the local internal energies are solely associated with the bare Hamiltonian's expectation values, $\langle\hat{H}^{(1,2)}\rangle(t)$, we would be led to conclude that only qubit $(1)$ exchanges energy, even though both qubits continuously evolves in time. It would also imply that the interaction term - essentially - works as an energetic source/sink for this particular qubit since all the exchanges would be attributed only to the energy trapped within its expectation value.

\begin{figure}[h!]
     \centering
     \begin{subfigure}[b]{0.49\textwidth}
         \centering
         \includegraphics[width=\textwidth]{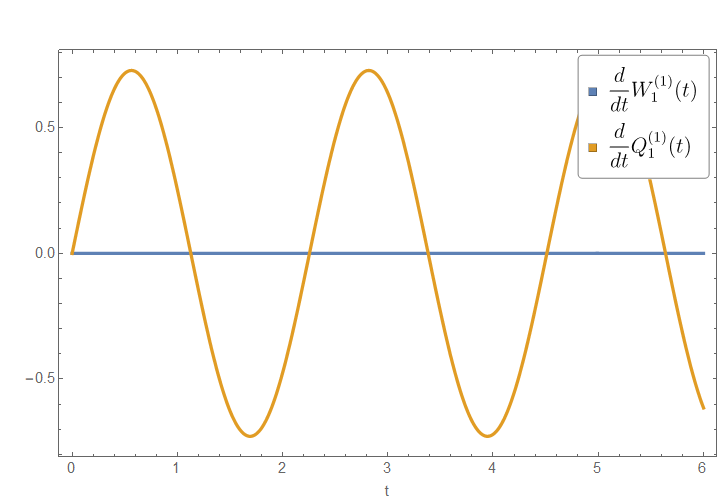}
         \caption{Alicki's proposal for qubit $(1)$}
         \label{AlickiQubit1}
     \end{subfigure}
     \begin{subfigure}[b]{0.49\textwidth}
         \centering
         \includegraphics[width=\textwidth]{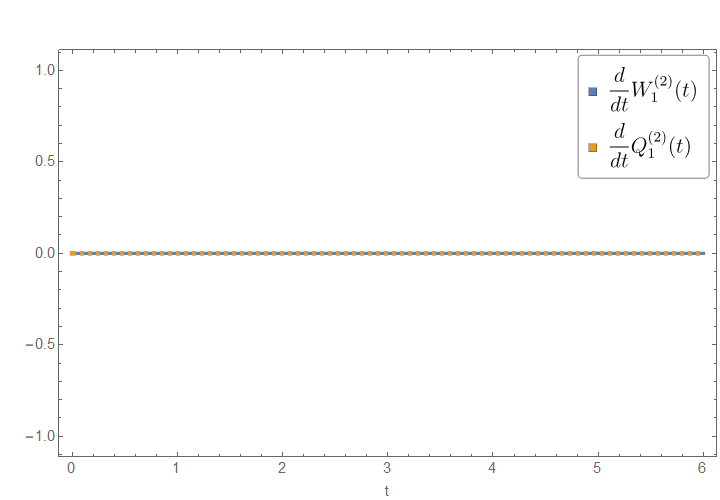}
         \caption{Alicki's proposal for qubit $(2)$}
         \label{AlickiQubit2}
     \end{subfigure}
        \caption{According to Alicki's form of work and heat fluxes (see Eqs. (\ref{Alicki work},\ref{Alicki heat})): \textbf{(a)} Qubit $(1)$ does not perform work, and all energy exchange with the interaction term is interpreted as heat, i.e., $\frac{d}{dt}W_{1}^{(1,2)}(t)=0$ and $\frac{d}{dt}Q_{1}^{(1)}(t) =-\frac{d}{dt}\langle\hat{H}_{int}\rangle(t)$; \textbf{(b)} The energy of qubit $(2)$ is neither submitted to work-like nor heat-like energy changes.}
        \label{fig2}
        Source: By the author.
\end{figure}

Besides, considering the discussions presented in Section (\ref{Dynamical approach}), this simple example also provides an interesting scenario to analyze and compare the behaviours of potential definitions of quantum work and heat. In this sense, given Alicki's proposal, shown in Eqs. (\ref{Alicki work}, \ref{Alicki heat}), work is associated with changes in the local bare Hamiltonian, while heat is linked with the time evolution of the density matrix. Thus, according to these expressions, it is clear that both qubits do not perform work since $\frac{d}{dt}\hat{H}^{(1,2)}=0$, which implies that the all energy exchange of qubit $(1)$ with the interaction term is entirely interpreted as heat, while for qubit $(2)$, this quantity is absent, i.e., $\frac{d}{dt}W_{1}^{(1,2)}(t)=0$, $\frac{d}{dt}Q_{1}^{(1)}(t)=tr\left\{ \hat{H}^{(1)}\frac{d}{dt}\hat{\rho}^{(1)}(t)\right\} =-\frac{d}{dt}\langle\hat{H}_{int}\rangle(t)$ and $\frac{d}{dt}Q_{1}^{(2)}(t)=0$. Figure \ref{fig2} reproduces the fluxes of Alicki's work and heat during the dynamics for both subsystems. Additionally, we might perform the same analyses for the work and heat forms depicted by Eqs. (\ref{Segunda forma de trabalho}, \ref{Segunda forma de calor}), where work is related to modifications in the bare Hamiltonian spectrum while heat depends on the dynamics of both the density matrix and the bare Hamiltonian eigenstates. Nevertheless, since, by hypothesis, $\hat{H}^{(1,2)}$ are constant in time, we can show that these expressions and Alicki's proposal agree with each other for this particular scenario (see Eqs. (\ref{Diferença work}, \ref{Diferença heat})), i.e., $Q_{2}^{(1,2)}(t)=Q_{1}^{(1,2)}(t)$ and $W_{2}^{(1,2)}(t)=W_{1}^{(1,2)}(t)$ for all $t$.

\begin{figure}[t]
     \centering
     \begin{subfigure}[b]{0.49\textwidth}
         \centering
         \includegraphics[width=\textwidth]{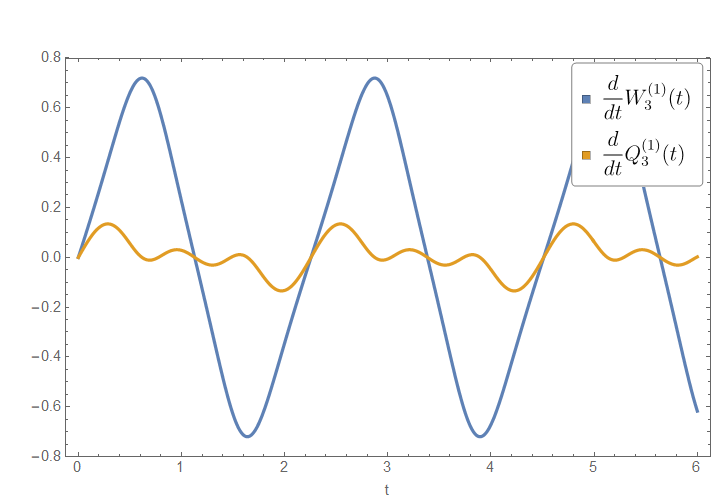}
         \caption{\cite{Ahmadi2019a,Alipour2019b} proposal for qubit $(1)$}
         \label{delCampoQubit1}
     \end{subfigure}
     \begin{subfigure}[b]{0.49\textwidth}
         \centering
         \includegraphics[width=\textwidth]{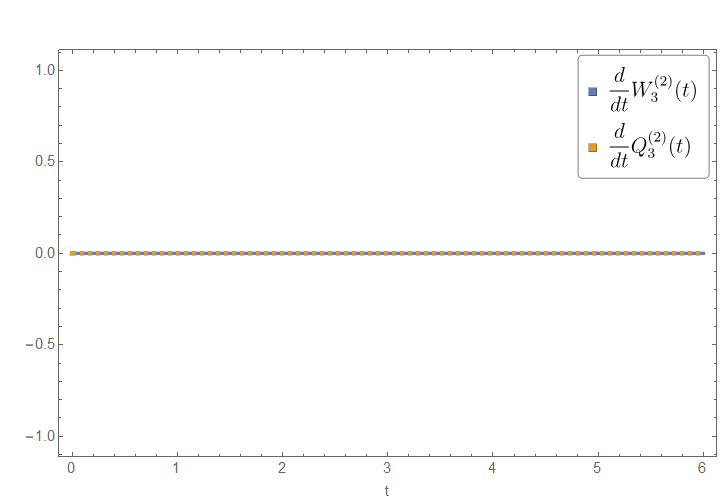}
         \caption{\cite{Ahmadi2019a,Alipour2019b} proposal for qubit $(2)$}
         \label{delCampoQubit2}
     \end{subfigure}
        \caption{According to Eqs. (\ref{Terceira forma de trabalho}, \ref{Terceira forma de calor}): \textbf{(a)} The energetics between qubit $(1)$ and the interaction term has separated contribution from both work and heat, i.e., $\frac{d}{dt}\langle\hat{H}^{(1)}\rangle(t)=\frac{d}{dt}W_{3}^{(1)}(t)+\frac{d}{dt}Q_{3}^{(1)}(t)=-\frac{d}{dt}\langle\hat{H}_{int}\rangle(t)$; \textbf{(b)} Qubit $(2)$ neither performs work nor exchanges heat. Note this happens despite its dynamics and the clear oscillation of its von Neumann entropy $S_{vN}[\hat{\rho}^{(2)}(t)]$ (see Figure \ref{vN Entropy}).}
        \label{fig3}
        Source: By the author.
\end{figure}

Finally, we can also consider the expressions proposed by \cite{Ahmadi2019a,Alipour2019b} and shown in Eqs. (\ref{Terceira forma de trabalho}, \ref{Terceira forma de calor}). According to this proposal, heat is bound to modifications in the populations of the density matrix, while work depends on the dynamics of both the local bare Hamiltonian and the instantaneous basis of $\hat{\rho}^{(1,2)}$, i.e., heat is directly associated with $\left\{ \frac{d}{dt}\lambda_{j}^{2}(t)\right\} _{j}$ due to its relationship with $\frac{d}{dt}S_{vN}[\hat{\rho}^{(1,2)}(t)]$, and work is linked to the remaining terms $\left\{ \frac{d}{dt}\left(\langle\varphi_{j}^{(1,2)}(t)|\hat{H}^{(1,2)}|\varphi_{j}^{(1,2)}(t)\rangle\right)\right\} _{j}$. Figure \ref{fig3} shows the behaviour of the work and heat fluxes for both qubits. In comparison with the previous proposals, despite $\hat{H}^{(1)}$ being time-independent, part of the energy exchanged between qubit $(1)$ and the interaction term is identified as work performed due to the dynamics of the basis $\{|\varphi_{1,2}^{(1)}(t)\rangle\}$, i.e., both work and heat are present such that $\frac{d}{dt}\langle\hat{H}^{(1)}\rangle(t)=\frac{d}{dt}W_{3}^{(1)}(t)+\frac{d}{dt}Q_{3}^{(1)}(t)=-\frac{d}{dt}\langle\hat{H}_{int}\rangle(t)$. Interestingly, for qubit $(2)$, the structure of the dynamics guarantees neither work nor heat exchange during the interaction, $\frac{d}{dt}W_{3}^{(2)}(t)=\frac{d}{dt}Q_{3}^{(2)}(t)=0$ for all $t$. It is worth emphasizing that this occurs even though its von Neumann entropy $S_{vN}[\hat{\rho}^{(2)}(t)]$ oscillates in time (see Figure \ref{vN Entropy}), which is in clear contrast with the original thermodynamic motivation of associating heat with energy modification accompanied by entropic changes. This context is similar to the situation presented in. \cite{Bottosso2019}

\begin{figure}[h!]
     \centering
     \begin{subfigure}[b]{0.49\textwidth}
         \centering
         \includegraphics[width=\textwidth]{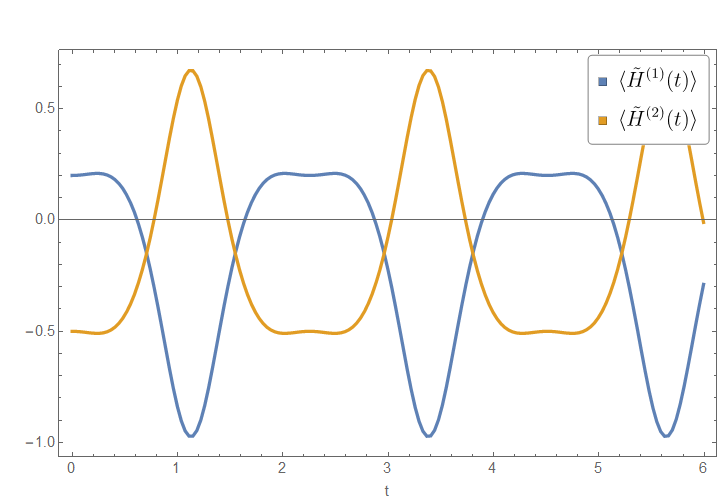}
         \caption{Local effective Hamiltonians}
         \label{LEH}
     \end{subfigure}
     \begin{subfigure}[b]{0.49\textwidth}
         \centering
         \includegraphics[width=\textwidth]{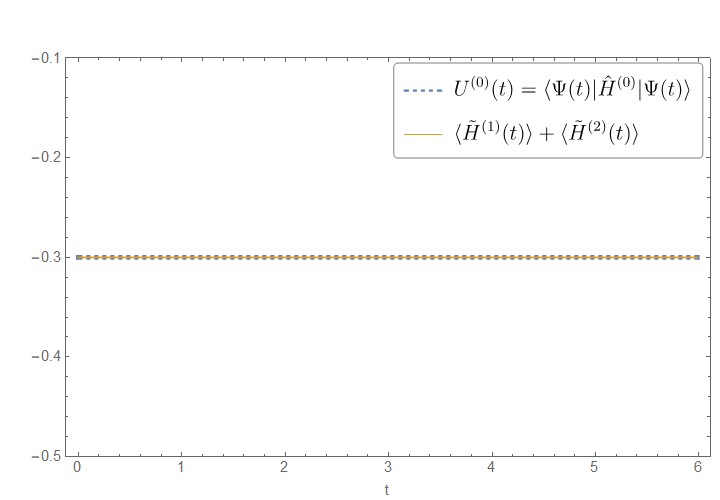}
         \caption{Whole's internal energy}
         \label{WholeInternalEnergy}
     \end{subfigure}
        \caption{Dynamical behaviour of the expectation values of the local effective Hamiltonians and its relationship with the whole's internal energy. \textbf{(a)} Interplay between $\langle\tilde{H}^{(1)}(t)\rangle$ and $\langle\tilde{H}^{(2)}(t)\rangle$. If $\tilde{H}^{(1,2)}(t)$ are interpreted as the representative operators for quantifying local internal energies, then all energy lost from qubit $(1)$ is obtained by qubit $(2)$, and vice versa; \textbf{(b)} The sum of the expectation values of the local effective Hamiltonians is equal to the whole's internal energy. This plot illustrates the energy additivity.}
        \label{fig4}
        Source: By the author.
\end{figure}

\begin{figure}[h!]
     \centering
     \begin{subfigure}[b]{0.49\textwidth}
         \centering
         \includegraphics[width=\textwidth]{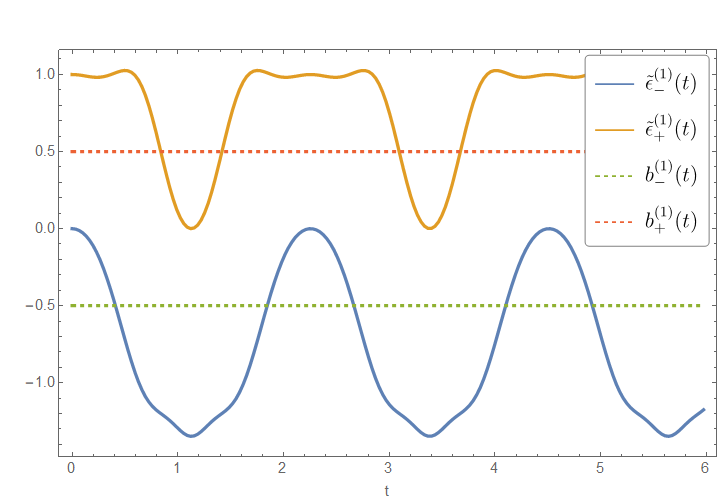}
         \caption{Spectrum - Qubit $(1)$}
         \label{Spectrum1}
     \end{subfigure}
     \begin{subfigure}[b]{0.49\textwidth}
         \centering
         \includegraphics[width=\textwidth]{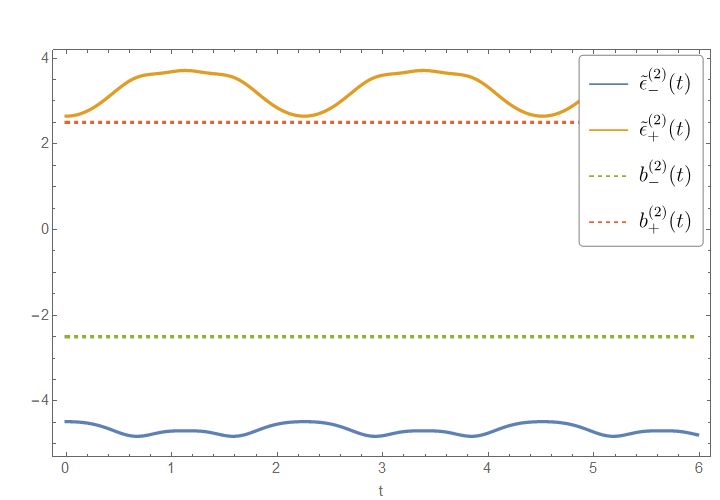}
         \caption{Spectrum - Qubit $(2)$}
         \label{Spectrum2}
     \end{subfigure}
        \caption{Dynamical behaviour of the spectrum of the bare and local effective Hamiltonians, for both qubits. The bare eigenvalues $\{b_{\pm}^{(1,2)}\}$ are constant, while the local effective ones $\{\tilde{\epsilon}_{\pm}^{(1,2)}(t)\}$ clearly change in time. \textbf{(a)} Qubit $(1)$; \textbf{(b)} Qubit $(2)$.}
        \label{fig5}
        Source: By the author.
\end{figure}

\begin{figure}[h!]
     \centering
     \begin{subfigure}[b]{0.49\textwidth}
         \centering
         \includegraphics[width=\textwidth]{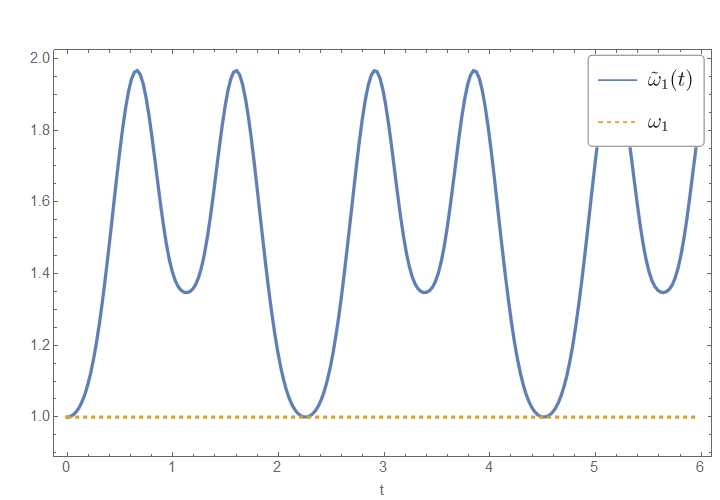}
         \caption{Spectral gaps - Qubit $(1)$}
         \label{Gaps1}
     \end{subfigure}
     \begin{subfigure}[b]{0.49\textwidth}
         \centering
         \includegraphics[width=\textwidth]{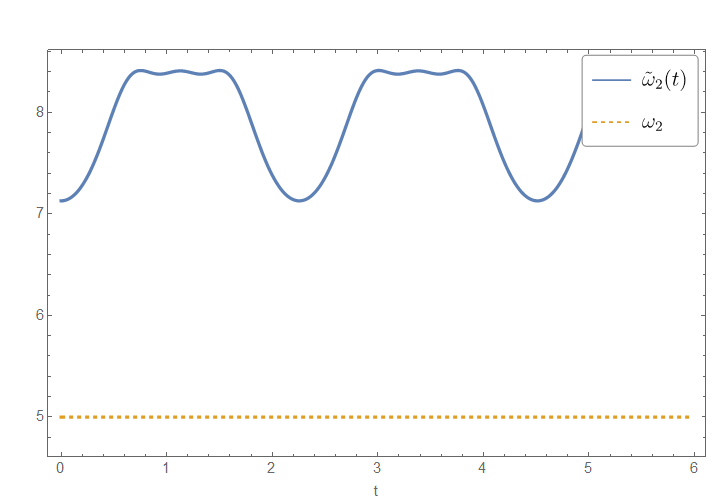}
         \caption{Spectral gaps - Qubit $(2)$}
         \label{Gaps2}
     \end{subfigure}
        \caption{Dynamical behaviour of the spectral gaps of the bare and local effective Hamiltonians, for both qubits. The changes in the spectrum modulate the local gaps $\tilde{\omega}_{1,2}(t)=\tilde{\epsilon}_{+}^{(1,2)}(t)-\tilde{\epsilon}_{-}^{(1,2)}(t)$ between the effective energy levels while $\hbar\omega_{1,2}=b_{+}^{(1,2)}-b_{-}^{(1,2)}$ is maintained fixed. \textbf{(a)} Qubit $(1)$; \textbf{(b)} Qubit $(2)$.}
        \label{fig6}
        Source: By the author.
\end{figure}

As we saw earlier, instead of focusing on the local bare Hamiltonians $\hat{H}^{(1,2)}$, if we use the local effective ones, $\tilde{H}^{(1,2)}(t)$, as the representative operators for quantifying local internal energies, the energy becomes additive and, therefore, the whole internal energy is simply written as
\begin{equation}
U^{(0)}=\langle\tilde{H}^{(1)}(t)\rangle+\langle\tilde{H}^{(2)}(t)\rangle.\label{eq:Soma}
\end{equation}
According to this formulation, both qubits exchange energy continuously, such that 
\begin{equation}
\frac{d}{dt}\langle\tilde{H}^{(1)}(t)\rangle=-\frac{d}{dt}\langle\tilde{H}^{(2)}(t)\rangle,
\end{equation}
and there is no space for additional elements working as energetic sources or sinks, i.e., the qubits are the only entities required for characterizing the energetics within $|\Psi(t)\rangle$. The dynamics of the expectation values of the local effective Hamiltonians of our example are depicted in Figure \ref{LEH}. Thus, it is clear that during the interaction, all energy flowing outside qubit $(1)$ is entirely acquired by qubit $(2)$ and vice versa. Besides, Figure \ref{WholeInternalEnergy} shows that the sum of both local quantities is constant and equal to the whole system's internal energy, which illustrates the additivity property presented in Eq. (\ref{eq:Soma}) above. 
Notice that, while the time-dependency of $\langle\hat{H}^{(1)}\rangle(t)$ is only due to the state $\hat{\rho}^{(1)}(t)$ dynamics, the $\tilde{H}^{(1,2)}(t)$ are explicitly time-dependent and play a significant role in the changes of their mean values $\langle\tilde{H}^{(1,2)}(t)\rangle$. In this sense, it is interesting to contrast the static nature of the bare Hamiltonians with the dynamic behaviour of the local effective operators. In order to do that, let us cast their following spectral decompositions
\begin{equation}
\hat{H}^{(1,2)}\equiv\sum_{j=\pm}b_{j}^{(1,2)}|b_{j}^{(1,2)}\rangle\langle b_{j}^{(1,2)}|
\end{equation}
and 
\begin{equation}
\tilde{H}^{(1,2)}(t)\equiv\sum_{j=\pm}\tilde{\epsilon}_{j}^{(1,2)}(t)|\tilde{\epsilon}_{j}^{(1,2)}(t)\rangle\langle\tilde{\epsilon}_{j}^{(1,2)}(t)|
\end{equation}
where $\{b_{\pm}^{(1,2)}=\pm\hbar\frac{\omega_{1,2}}{2}\}$ and $\{\tilde{\epsilon}_{\pm}^{(1,2)}(t)\}$ are the respective eigenvalues of $\hat{H}^{(1,2)}$ and $\tilde{H}^{(1,2)}(t)$, while $\{|b_{\pm}^{(1)}\rangle=|\pm_{z}^{(1)}\rangle\}$, $\{|b_{\pm}^{(2)}\rangle=|\pm_{y}^{(2)}\rangle\}$ and $\{|\tilde{\epsilon}_{\pm}^{(1,2)}(t)\rangle\}$ are the eigenbasis of $\hat{H}^{(1)}$, $\hat{H}^{(2)}$ and $\tilde{H}^{(1,2)}(t)$, respectively. Figures \ref{Spectrum1} and \ref{Spectrum2} portray the dynamics of their energy levels. The time dependency of $\{\tilde{\epsilon}_{\pm}^{(1,2)}(t)\}$ is a direct consequence of the interaction term that is automatically comprised within the local effective operators (see Eq. (\ref{Effective Hamiltonian 2})). Interestingly, not only these eigenvalues are time-dependent, but also the energy gaps are modulated in time. Given $\hbar\omega_{1,2}=b_{+}^{(1,2)}-b_{-}^{(1,2)}$ and $\hbar\tilde{\omega}_{1,2}(t)=\tilde{\epsilon}_{+}^{(1,2)}(t)-\tilde{\epsilon}_{-}^{(1,2)}(t)$, Figures \ref{Gaps1} and \ref{Gaps2} illustrate how the local gaps $\hbar\tilde{\omega}_{1,2}(t)$ change, while $\hbar\omega_{1,2}$ is maintained fixed during the whole dynamics.

\begin{figure}[h!]
     \centering
     \begin{subfigure}[b]{0.49\textwidth}
         \centering
         \includegraphics[width=\textwidth]{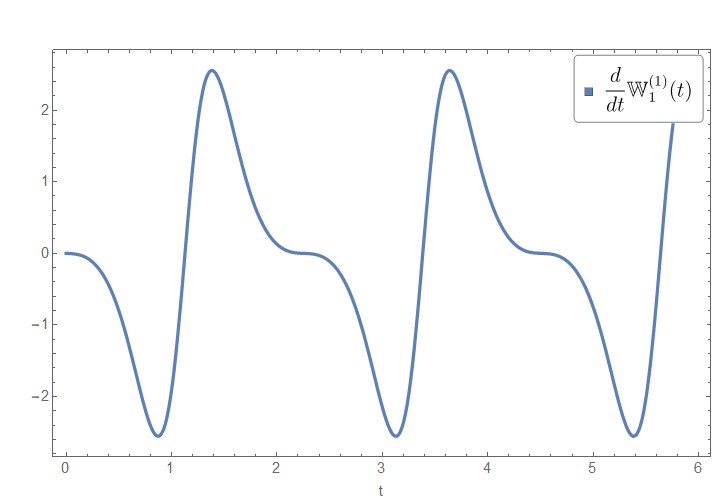}
         \caption{Alicki's form of work flux for qubit $(1)$}
         \label{AlickiformW1}
     \end{subfigure}
     \begin{subfigure}[b]{0.49\textwidth}
         \centering
         \includegraphics[width=\textwidth]{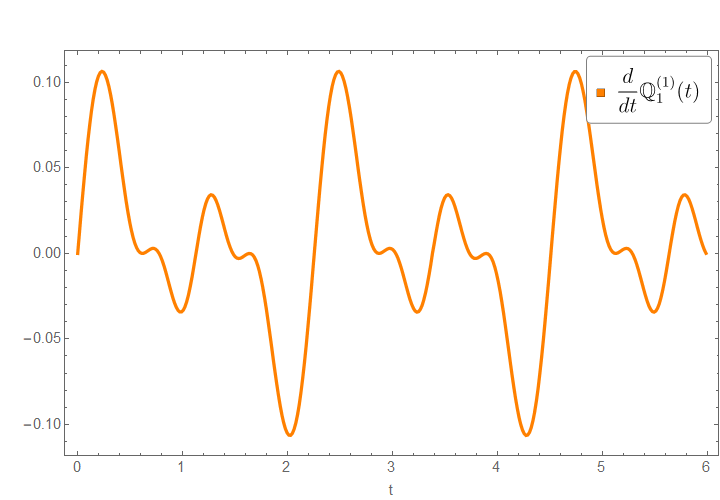}
         \caption{Alicki's form of heat flux for qubit $(1)$}
         \label{AlickiformQ1}
     \end{subfigure}
        \caption{Work and heat fluxes according to Eqs. (\ref{workAlickiLike}, \ref{HeatAlickiLike}) for qubit $(1)$. \textbf{(a)} Work flux; \textbf{(b)} Heat flux.}
        \label{fig7}
        Source: By the author.
\end{figure}

\begin{figure}[h!]
     \centering
     \begin{subfigure}[b]{0.49\textwidth}
         \centering
         \includegraphics[width=\textwidth]{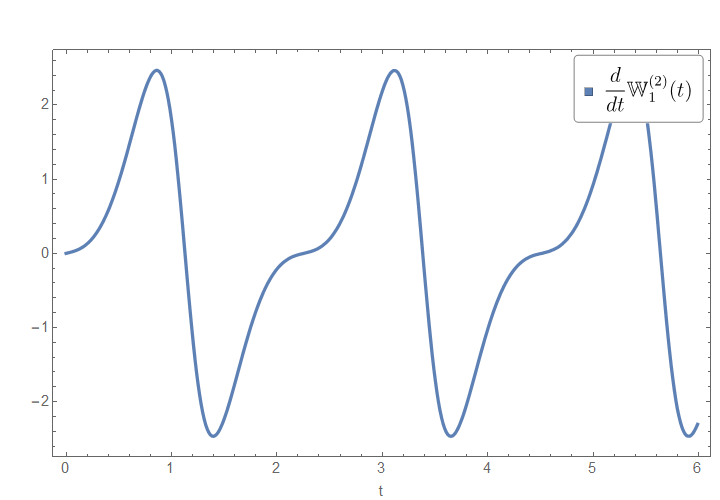}
         \caption{Alicki's form of work flux for qubit $(2)$}
         \label{AlickiformW2}
     \end{subfigure}
     \begin{subfigure}[b]{0.49\textwidth}
         \centering
         \includegraphics[width=\textwidth]{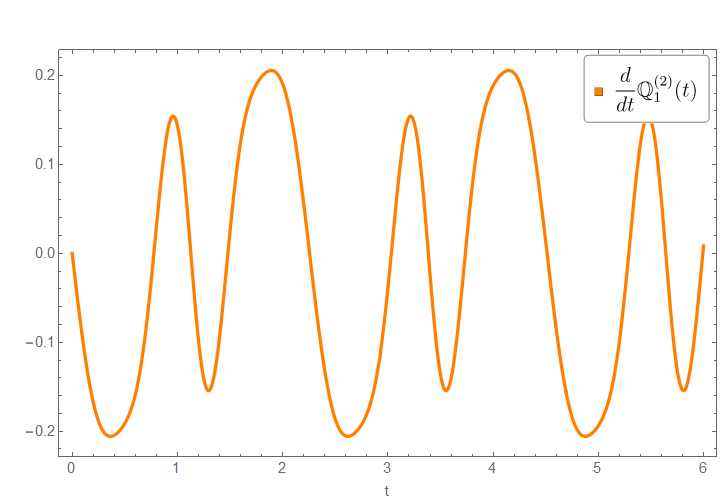}
         \caption{Alicki's form of heat flux for qubit $(2)$}
         \label{AlickiformQ2}
     \end{subfigure}
        \caption{Work and heat fluxes according to Eqs. (\ref{workAlickiLike}, \ref{HeatAlickiLike}) for qubit $(2)$. \textbf{(a)} Work flux; \textbf{(b)} Heat flux.}
        \label{fig8}
        Source: By the author.
\end{figure}

\begin{figure}[h!]
     \centering
     \begin{subfigure}[b]{0.49\textwidth}
         \centering
         \includegraphics[width=\textwidth]{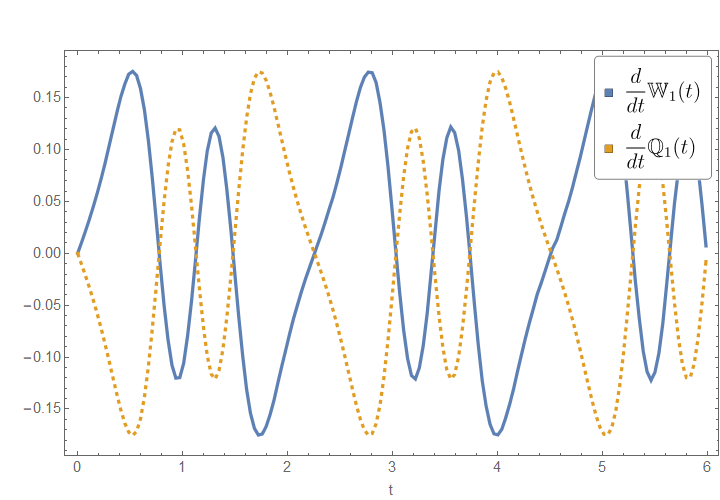}
         \caption{Net work and net heat fluxes}
         \label{NetWorkHeatAlicki}
     \end{subfigure}
     \begin{subfigure}[b]{0.49\textwidth}
         \centering
         \includegraphics[width=\textwidth]{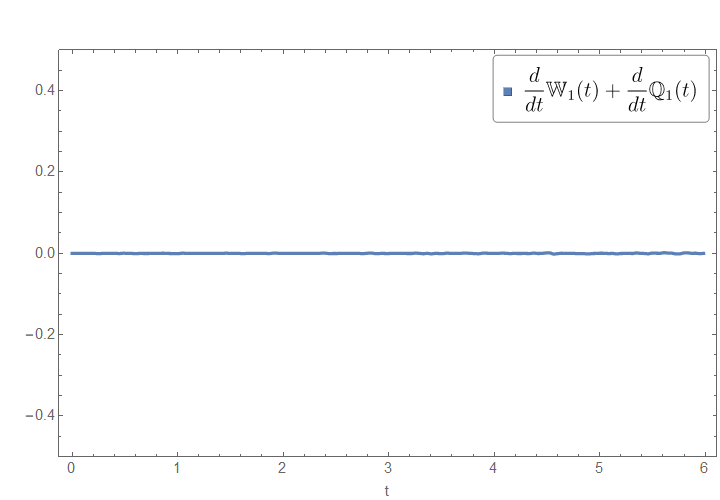}
         \caption{Net energetic exchange}
         \label{SumNetAlicki}
     \end{subfigure}
        \caption{Net energy fluxes according to Eqs. (\ref{workAlickiLike}, \ref{HeatAlickiLike}). \textbf{(a)} The total work flux performed and the total heat flux transferred throughout the dynamics; \textbf{(b)} All net energetic exchanges involved during the whole system dynamics sum up to zero, Eq. (\ref{Net work + Net heat}).}
        \label{fig9}
        Source: By the author.
\end{figure}

Finally, as mentioned earlier, the use of the local effective Hamiltonians also provides an interesting starting point for discussing a general, exact and symmetrical understanding of quantum work and heat. Hence, let us compare the potential approaches for these quantities presented in Section (\ref{subsec:Road-to-effective}). Along these lines, Eqs. (\ref{workAlickiLike}, \ref{HeatAlickiLike}) are similar to Alicki's proposal. However, since the internal energy is computed by $\langle\tilde{H}^{(1,2)}(t)\rangle$, heat is associated with changes in the whole density matrix, and work is the energy exchange related to the local effective Hamiltonian dynamics. Thus, work is a direct outcome of the interactions between the qubits instead of being the result of the addition of classical external control. Figures \ref{fig7} and \ref{fig8} reproduce  the non-null work and heat fluxes, $\frac{d}{dt}\mathbb{W}_{1}^{(1,2)}(t)$ and $\frac{d}{dt}\mathbb{Q}_{1}^{(1,2)}(t)$, during the dynamics for qubits $(1)$ and $(2)$, respectively.
Additionally, Figure \ref{NetWorkHeatAlicki} shows the behaviour of  the non-null net work and net heat fluxes involved during the process, where $\mathbb{W}_{1}(t)=\mathbb{W}_{1}^{(1)}(t)+\mathbb{W}_{1}^{(2)}(t)$ and $\mathbb{Q}_{1}(t)=\mathbb{Q}_{1}^{(1)}(t)+\mathbb{Q}_{1}^{(2)}(t)$, while Figure \ref{SumNetAlicki} confirms that all net energetic exchanges sum up to zero, i.e., $\frac{d}{dt}\mathbb{W}_{1}(t)+\frac{d}{dt}\mathbb{Q}_{1}(t)=0$ (see Eq. (\ref{Net work + Net heat})). It is worth mentioning that earlier was shown that the analogue forms of Eqs. (\ref{Terceira forma de trabalho}, \ref{Terceira forma de calor}), considering the local effective Hamiltonians, are equivalent to the previous ones, i.e., $\mathbb{W}_{1}^{(1,2)}(t)=\mathbb{W}_{3}^{(1,2)}(t)$ and $\mathbb{Q}_{1}^{(1,2)}(t)=\mathbb{Q}_{3}^{(1,2)}(t)$.

\begin{figure}[t]
     \centering
     \begin{subfigure}[b]{0.49\textwidth}
         \centering
         \includegraphics[width=\textwidth]{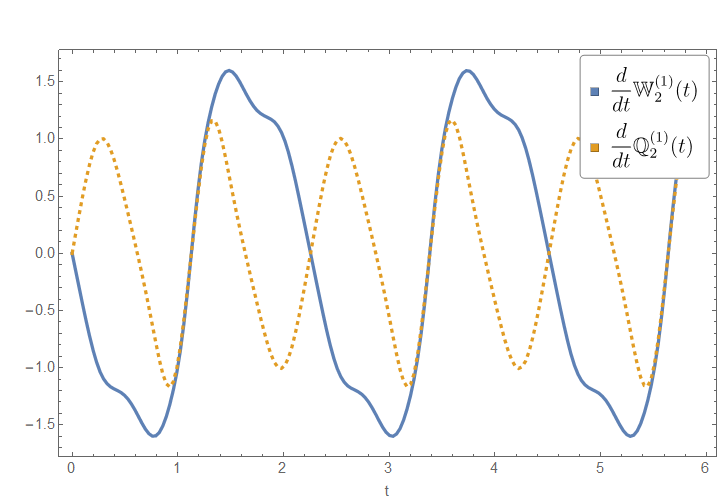}
         \caption{Work and heat fluxes for qubit $(1)$}
         \label{SecondFormQubit1}
     \end{subfigure}
     \begin{subfigure}[b]{0.49\textwidth}
         \centering
         \includegraphics[width=\textwidth]{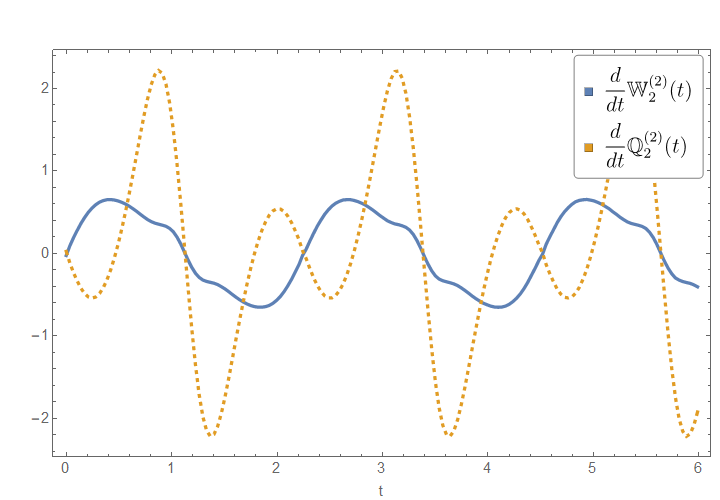}
         \caption{Work and heat fluxes for qubit $(2)$}
         \label{SecondFormQubit2}
     \end{subfigure}
        \caption{Energy fluxes according to Eqs. (\ref{WorkForma2like}, \ref{HeatForma2like}). \textbf{(a)} Qubit $(1)$; \textbf{(b)} Qubit $(2)$.}
        \label{fig10}
        Source: By the author.
\end{figure}

\begin{figure}[h!]
     \centering
     \begin{subfigure}[b]{0.49\textwidth}
         \centering
         \includegraphics[width=\textwidth]{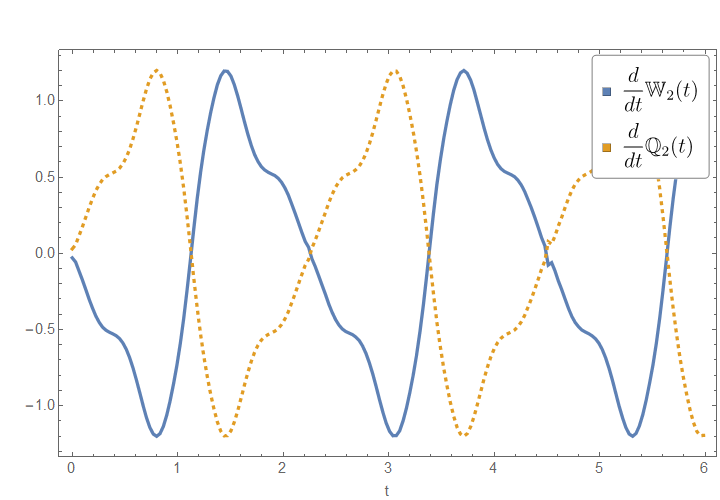}
         \caption{Net work and net heat fluxes}
         \label{NetWorkHeatSecondForm}
     \end{subfigure}
     \begin{subfigure}[b]{0.49\textwidth}
         \centering
         \includegraphics[width=\textwidth]{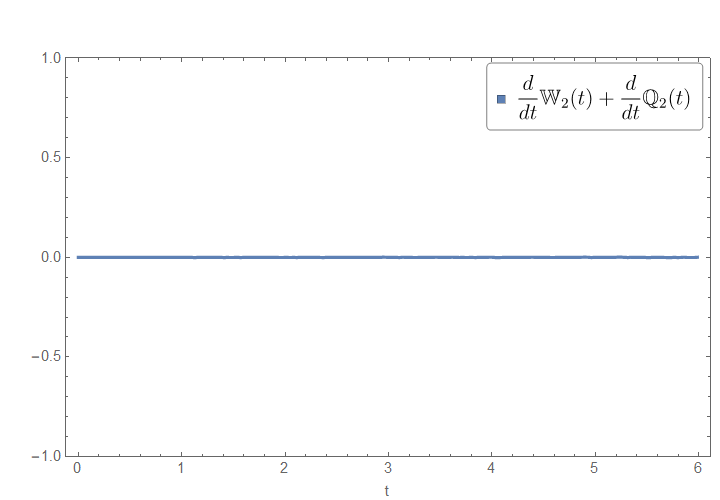}
         \caption{Net energetic exchange}
         \label{SumNetSecondForm}
     \end{subfigure}
        \caption{Net energy fluxes according to Eqs. (\ref{WorkForma2like}, \ref{HeatForma2like}). \textbf{(a)} The total work flux performed and the total heat flux transferred throughout the dynamics; \textbf{(b)} All net energetic exchanges involved during the whole system dynamics sum up to zero, Eq. (\ref{Net work + Net heat}).}
        \label{fig11}
        Source: By the author.
\end{figure}

Alternatively, we might define these quantities according to Eqs. (\ref{WorkForma2like}, \ref{HeatForma2like}). In this case, work is associated with changes in the local effective Hamiltonian spectrum, $\left\{ \frac{d}{dt}\tilde{\epsilon}_{j}^{(1,2)}(t)\right\} _{j}$, and heat depends both on the dynamics of the density matrix $\hat{\rho}^{(1,2)}(t)$ and the basis $\left\{ |\tilde{\epsilon}_{j}^{(1,2)}(t)\rangle\right\}$. Figures \ref{SecondFormQubit1} and \ref{SecondFormQubit2} show the interplay between work and heat fluxes, $\frac{d}{dt}\mathbb{W}_{2}^{(1,2)}(t)$ and $\frac{d}{dt}\mathbb{Q}_{2}^{(1,2)}(t)$, for qubit $(1)$ and $(2)$, respectively. Similarly to the previous proposal, Figure \ref{NetWorkHeatSecondForm} presents the non-null dynamics of the net work and net heat fluxes during the interaction, where $\mathbb{W}_{2}(t)=\mathbb{W}_{2}^{(1)}(t)+\mathbb{W}_{2}^{(2)}(t)$ and $\mathbb{Q}_{2}(t)=\mathbb{Q}_{2}^{(1)}(t)+\mathbb{Q}_{2}^{(2)}(t)$. Besides, Figure \ref{SumNetSecondForm} shows that all net energetic exchanges within the bipartition computed with these expressions are also in accordance with Eq. (\ref{Net work + Net heat}), i.e., they sum up to zero.

Therefore, it is clear that different work and heat proposals may represent radically distinct thermodynamic scenarios. Still, the use of $\tilde{H}^{(1,2)}(t)$ provides a consistent foundation for advancing such discussions in an absolutely general manner.

In short, it was both illustrated a simple example of the application of our local effective Hamiltonians formalism and highlighted a challenging issue of the usual identification of the bare Hamiltonians as the operators for representing the local internal energies whenever the interaction term $\hat{H}_{int}$ is non-negligible (and the strict energy conservation is not applicable). While the latter identification requires the interpretation of the interaction term as an additional energetic source or sink, the former only attributes local internal energies for the described subsystems. In this particular physical system, the global Hamiltonian structure guarantees that $\langle\hat{H}^{(2)}\rangle$ is a constant of motion, even though $\hat{\rho}^{(2)}(t)$ explicitly evolves in time (Figure \ref{fig0}), and the mean value $\langle\hat{H}^{(1)}\rangle$ changes (Figure \ref{Fig1}). This behaviour, of course, reflects the non-additivity of the expectation values of the bare Hamiltonians and the essential role played by the interaction in the total internal energy computation. Naturally, such a role is also critical for characterizing energy exchanges and must be appreciated, in one way or another, into any \textit{consistent} and \textit{general} definition of work and heat. Along these lines, since Alicki's definition proposal in \cite{Alicki19792}, the bare Hamiltonians represent a common starting point for defining these thermodynamic quantities, even when considered alternative forms, such as the presented in. \cite{Ahmadi2019a,Alipour2019b} In contrast, the use of the expectation values of the local effective Hamiltonians $\langle\tilde{H}^{(1,2)}(t)\rangle$ as the representative operators for characterizing physical internal energies satisfies the additivity property and offers an important interpretative advantage for the framework of quantum thermodynamics.

%% file: Chapters/Chapter3/Section8.tex
\section{Generalization for mixed states\label{Section: Generalization - mixed states}}

In this section, we will take our approach one step further by allowing the possibility of describing \textit{mixed} quantum states. During the previous sections, we explicitly assumed a finite, isolated
and nondegenerate pure quantum system, depicted by $|\Psi(t)\rangle\langle\Psi(t)|$,
composed of two arbitrary smaller interacting subsystems, $(1)$ and
$(2)$, with dimensions $d^{(k)}:=dim(\mathcal{H}^{(k)})$, such that
$d^{(0)}=(d^{(1)}+d^{(2)})$ and - without any loss of generality
- $d^{(1)}\leq d^{(2)}$. The fully quantum autonomous object is described
by the time-independent Hamiltonian $\hat{H}^{(0)}$ that generates
the whole system dynamics, such that 
\begin{equation}
\hat{H}^{(0)}=\hat{H}^{(1)}\otimes\hat{1}^{(2)}+\hat{1}^{(1)}\otimes\hat{H}^{(2)}+\hat{H}_{int},\label{Total hamiltonian, mixed state}
\end{equation}
where $\hat{H}^{(1,2)}\in\mathcal{L}(\mathcal{H}^{(1,2)})$ are the
local bare Hamiltonians of each subsystem and $\hat{H}_{int}\in\mathcal{L}(\mathcal{H}^{(0)})$
is the interaction between them. Let us now generalize and expand
our formalism to include a more comprehensive and realistic experimental
description\footnote{Many of the steps presented here - from now on - will be very similar
to those shown in Section (\ref{Section: Setup}). For completeness
and for establishing the new notation, the essential features will
be repeated. However, to avoid redundancy, some discussions will be
purposely suppressed.}. From a pragmatical point of view, this setting only characterizes
specific situations. In real-world scenarios, the quantum system must
be somehow experimentally prepared, in a way that it is inevitably
distributed across an ensemble of possible states. This, of course,
corresponds to a classical lack of information due to the preparation
procedure itself. In these cases, the system in question is better
described by a statistical mixture of pure states $\{P_{\eta},|\Psi_{\eta}(t)\rangle\in\mathcal{H}^{(0)}=\mathcal{H}^{(1)}\otimes\mathcal{H}^{(2)}\}_{\eta=1,...,d^{(0)}}$,
such that 
\begin{equation}
\hat{\rho}^{(0)}(t)\equiv\sum_{\eta=1}^{d^{(0)}}P_{\eta}\hat{\sigma}_{\eta}^{(0)}(t),\label{MATRIZ DENSIDADE GLOBAL: ESTADO MISTO}
\end{equation}
where $\{\hat{\sigma}_{\eta}^{(0)}(t)\equiv|\Psi_{\eta}(t)\rangle\langle\Psi_{\eta}(t)|\}_{\eta}$
are pure states, $\langle\Psi_{\alpha}(t)|\Psi_{\beta}(t)\rangle=\delta_{\alpha\beta}$
and $tr\{\hat{\rho}^{(0)}(t)\}=\sum_{\eta=1}^{d^{(0)}}P_{\eta}=1$. Note that such description also provides us with the means to represent and characterize systems initially prepared at thermal states, i.e., if $\hat{H}^{(0)}\equiv\sum_{\eta=1}^{d^{(0)}}b_{\eta}^{(0)}|b_{\eta}^{(0)}\rangle\langle b_{\eta}^{(0)}|$, then $P_{\eta}\equiv\frac{e^{-\beta b_{\eta}^{(0)}}}{Z^{(0)}}$ and $|\Psi_{\eta}(t_{0})\rangle\equiv|b_{\eta}^{(0)}\rangle$ in a way that $\hat{\rho}^{(0)}(t_{0})=\sum_{\eta=1}^{d^{(0)}}\frac{e^{-\beta b_{\eta}^{(0)}}}{Z^{(0)}}|b_{\eta}^{(0)}\rangle\langle b_{\eta}^{(0)}|$.
Also, since the whole system is isolated, the populations
$\{P_{\eta}\}_{\eta}$ remains constant throughout the unitary dynamics.
Thus, it is clear that a single pure state, instead of the convex
sum above, is just a particular case of a much more broad representation,
where $P_{\eta}=\delta_{0\eta}$ and $\hat{\rho}^{(0)}(t)=\hat{\sigma}_{0}^{(0)}(t)$.
Locally, as will be shown below, such a mixed state implies that both
subsystems will also be portrayed as ensembles themselves, although
not of pure states. It is worth mentioning that one may argue that
there is a clear and noticeable ambiguity in such a process: it is
well-known that there are infinite possible ways to express the same
density matrix $\hat{\rho}^{(0)}(t)$ as a convex sum of pure - and
not necessarily orthogonal - states. However, on the one hand, from
a physical perspective, all possible representations are indistinguishable,
thus any observable or internal thermodynamic description should not
depend on this choice. On the other, there is a unique description
in terms of an orthonormal basis which is given by the spectral decomposition
above. Thus, Eq. (\ref{MATRIZ DENSIDADE GLOBAL: ESTADO MISTO}) represents
the most natural depiction choice for our purposes. Finally, from
now on, the addition of a label $\eta$ will be necessary to identify
which element of Eq. (\ref{MATRIZ DENSIDADE GLOBAL: ESTADO MISTO})
we are dealing with and to distinguish an ensemble or a pure-state-level
treatment.

Now, let us proceed in complete analogy with what was previously shown
in Section (\ref{Section: Setup}). Every pure state dynamics is governed
by the usual Schrödinger equation $i\hbar\frac{d}{dt}|\Psi_{\eta}(t)\rangle=\hat{H}^{(0)}|\Psi_{\eta}(t)\rangle$,
such that for any initial state $|\Psi_{\eta}(t_{0})\rangle$ and
$t\geq t_{0}$ we have
\begin{equation}
|\Psi_{\eta}(t)\rangle=\hat{\mathcal{U}}(t,t_{0})|\Psi_{\eta}(t_{0})\rangle,\label{Mixe universe - pure state time evolution}
\end{equation}
where $\hat{\mathcal{U}}(t,t_{0})=e^{-\frac{i}{\hbar}\hat{H}^{(0)}(t-t_{0})}\in\mathcal{L}(\mathcal{H}^{(0)})$
is the time-evolution operator of the whole bipartite system. Then,
let us represent them according to their respective Schmidt decomposition
form
\begin{equation}
|\Psi_{\eta}(t)\rangle=\sum_{j=1}^{d^{(1)}}\lambda_{\eta j}(t)|\varphi_{\eta j}^{(1)}(t)\rangle\otimes|\varphi_{\eta j}^{(2)}(t)\rangle,\label{Schmidt; Estado misto}
\end{equation}
for every time $t$, where $\{\lambda_{\eta j}(t)\geq0\}_{j}$ and
$\{|\varphi_{\eta j}^{(k)}(t)\rangle\}_{j}\in\mathcal{H}^{(k)}$ are
the $\eta$th time-local Schmidt coefficients and local Schmidt basis
of subsystem $(k)$, respectively. The pure states $\{|\Psi_{\eta}(t)\rangle\}_{\eta}$
orthonormality implies that $\sum_{q,j=1}^{d^{(1)}}\lambda_{\beta j}(t)\lambda_{\alpha q}(t)\langle\varphi_{\alpha q}^{(1)}(t)|\varphi_{\beta j}^{(1)}(t)\rangle\langle\varphi_{\alpha q}^{(2)}(t)|\varphi_{\beta j}^{(2)}(t)\rangle=\delta_{\alpha\beta}$
and, therefore, the normalization $\sum_{j=1}^{d^{(1)}}\lambda_{\eta j}^{2}(t)=1$
for all $\eta$, also the orthonormality of the local basis elements
assure that $\langle\varphi_{\eta\alpha}^{(k)}(t)|\varphi_{\eta\beta}^{(l)}(t)\rangle=\delta_{kl}\delta_{\alpha\beta}$.
Besides, since each $\eta$th set $\{|\varphi_{\eta j}^{(k)}(t)\rangle\}_{j}$
constitute a possible basis for the same Hilbert space $\mathcal{H}^{(k)}$,
any pair of Schmidt basis for a given subsystem should be unitarily
related, such that
\begin{equation}
|\varphi_{\eta j}^{(k)}(t)\rangle=\hat{T}_{\eta\alpha}^{(k)}(t)|\varphi_{\alpha j}^{(k)}(t)\rangle,\label{Unitary relation}
\end{equation}
where $\hat{T}_{\eta\alpha}^{(k)}(t)=\sum_{m=1}^{d^{(k)}}|\varphi_{\eta m}^{(k)}(t)\rangle\langle\varphi_{\alpha m}^{(k)}(t)|$,
$\hat{T}_{\alpha\eta}^{(k)}(t)=\hat{T}_{\eta\alpha}^{(k)\dagger}(t)$
and $\hat{T}_{\eta\alpha}^{(k)}(t)\hat{T}_{\eta\alpha}^{(k)\dagger}(t)=\hat{1}^{(k)}$.

Concerning the subsystems representations, for each possible pure
state $|\Psi_{\eta}(t)\rangle$ of the whole, one can find the local
states by the usual procedure of partial tracing their complementary
degrees of freedom, such that $\hat{\sigma}_{\eta}^{(1,2)}(t)\equiv tr_{2,1}\{\hat{\sigma}_{\eta}^{(0)}(t)\}$.
Thus, as expected,
\begin{align}
\hat{\sigma}_{\eta}^{(1)}(t) & =\sum_{j=1}^{d^{(1)}}\lambda_{\eta j}^{2}(t)|\varphi_{\eta j}^{(1)}(t)\rangle\langle\varphi_{\eta j}^{(1)}(t)|,\label{Possible local state 1}\\
\hat{\sigma}_{\eta}^{(2)}(t) & =\sum_{j=1}^{d^{(1)}}\lambda_{\eta j}^{2}(t)|\varphi_{\eta j}^{(2)}(t)\rangle\langle\varphi_{\eta j}^{(2)}(t)|,\label{Possible local state 2}
\end{align}
where the eigenvalues and eigenvectors are given by the Schmidt coefficients
squared $\{\lambda_{\eta j}^{2}(t)\}_{j}$ and Schmidt basis $\{|\varphi_{\eta j}^{(1,2)}(t)\rangle\}_{j}$,
respectively. Notice that Eqs. (\ref{Possible local state 1}, \ref{Possible local state 2})
above do not describe the total local states, in the sense that they
only represent the possible local density operators obtained from
the whole distribution of pure states $\{P_{\eta},|\Psi_{\eta}(t)\rangle\}_{\eta}$.
This also characterizes a \textit{subensemble} of local representations
$\{P_{\eta},\hat{\sigma}_{\eta}^{(1,2)}(t)\}_{\eta}$, such that the
total local states are simply given by the averages
\begin{align}
\hat{\rho}^{(1)}(t) & \equiv tr_{2}\{\hat{\rho}^{(0)}(t)\}=\sum_{\eta=1}^{d^{(0)}}P_{\eta}\hat{\sigma}_{\eta}^{(1)}(t),\label{Entire local state 1}\\
\hat{\rho}^{(2)}(t) & \equiv tr_{1}\{\hat{\rho}^{(0)}(t)\}=\sum_{\eta=1}^{d^{(0)}}P_{\eta}\hat{\sigma}_{\eta}^{(2)}(t).\label{Entire local state 2}
\end{align}
In short, every possible pure bipartite state from the mixed ensemble
presented in Eq. (\ref{MATRIZ DENSIDADE GLOBAL: ESTADO MISTO}), give
rise to a pair of local density matrices $\hat{\sigma}_{\eta}^{(1,2)}(t)$
in such a way that the entire local states are given by the distribution
portrayed in Eqs. (\ref{Entire local state 1}, \ref{Entire local state 2})
above.

\subsection{Schmidt basis dynamics and local effective Hamiltonians}

Next, following the same reasoning used in Section (\ref{SECTION: Schmidt-basis-dynamics}),
we are interested in a local dynamical description for the set of
Schmidt basis $\{|\varphi_{\eta j}^{(1,2)}(t)\rangle\}_{\eta,j}$.
At any time interval $[t_{0},t_{1}]$, instead of having a single
curve associated with the whole system time-evolution, now we have
an ensemble of possible trajectories $\{\mathcal{P}_{\eta}^{(0)}:|\Psi_{\eta}(t)\rangle,\:t\in[t_{0},t_{1}]\}_{\eta}$
in the total Hilbert space $\mathcal{H}^{(0)}$. Thus, just like before,
each path can be mapped into two coupled trajectories $\mathcal{P}_{\eta j}^{(1,2)}:|\varphi_{\eta j}^{(1,2)}(t)\rangle,\:t\in[t_{0},t_{1}]$
followed by the Schmidt basis in their own Hilbert spaces $\mathcal{H}^{(1,2)}$,
and the paths of the Schmidt coefficients, $\mathcal{\mathcal{P}}_{\eta j}^{\lambda}:\lambda_{\eta j}(t)\geq0,\:t\in[t_{0},t_{1}]$,
such that $\lambda_{\eta j}^{2}(t)\in[0,1]$ for all $j$ and $\eta$,
and $\sum_{j=1}^{d^{(1)}}\lambda_{\eta j}^{2}(t)=1$. Clearly, the
whole time-evolution is fully characterized by the initial states
$\{|\Psi_{\eta}(t_{0})\rangle\}_{\eta}$ and the unitary operator
$\hat{\mathcal{U}}(t,t_{0})$ through Eq. (\ref{Mixe universe - pure state time evolution}),
i.e., every possible initial condition will be time-evolved under
the same generator. Nevertheless, the local dynamical behaviours will
definitely depend on the whole pure state in question.

Let us define the local dynamical maps $\tilde{\mathcal{U}}_{\eta}^{(k)}:\mathcal{H}^{(k)}\rightarrow\mathcal{H}^{(k)}$
($k=1,2$) associated with the time-evolution of each $\eta$th path
from $\{\mathcal{P}_{\eta j}^{(k)}\}_{j}$, in a way that every Schmidt
basis ket continuously follows
\begin{equation}
|\varphi_{\eta j}^{(k)}(t)\rangle=\tilde{\mathcal{U}}_{\eta}^{(k)}(t,t_{0})|\varphi_{\eta j}^{(k)}(t_{0})\rangle,\label{Estado misto - Schmidt time evolution}
\end{equation}
for any $t\geq t_{0}$, with $\underset{t\rightarrow t_{0}}{lim}|\varphi_{\eta j}^{(k)}(t)\rangle=|\varphi_{\eta j}^{(k)}(t_{0})\rangle$
or $\underset{t\rightarrow t_{0}}{lim}\,\tilde{\mathcal{U}}_{\eta}^{(k)}(t,t_{0})=\hat{1}^{(k)}$.
This time-evolution operator is unitary, $\tilde{\mathcal{U}}_{\eta}^{(k)\dagger}(t,t_{0})\tilde{\mathcal{U}}_{\eta}^{(k)}(t,t_{0})=\hat{1}^{(k)}$,
and should satisfy\footnote{See Section (\ref{SECTION: Schmidt-basis-dynamics}) for more details.}
\begin{equation}
i\hbar\frac{d}{dt}\tilde{\mathcal{U}}_{\eta}^{(k)}(t,t_{0})=\tilde{H}_{\eta}^{(k)}(t)\tilde{\mathcal{U}}_{\eta}^{(k)}(t,t_{0}),\label{Estado misto - Time evolution operator}
\end{equation}
where $\tilde{H}_{\eta}^{(k)}(t)=\tilde{H}_{\eta}^{(k)\dagger}(t)\in\mathcal{L}(\mathcal{H}^{(k)})$
is a hermitian and possibly time-dependent operator. Then, from Eqs.
(\ref{Estado misto - Schmidt time evolution}, \ref{Estado misto - Time evolution operator})
we have
\begin{equation}
i\hbar\frac{d}{dt}|\varphi_{\eta j}^{(k)}(t)\rangle=\tilde{H}_{\eta}^{(k)}(t)|\varphi_{\eta j}^{(k)}(t)\rangle,
\end{equation}
for all $j$ and $\eta$. It is clear that $\tilde{H}_{\eta}^{(k)}(t)$
is the time-translation generator of the Schmidt basis $\{|\varphi_{\eta j}^{(k)}(t)\rangle\}_{\eta,j}$
and the $\eta$th local effective Hamiltonian for subsystem $(k)$
linked with the state $\hat{\sigma}_{\eta}^{(k)}(t)$. As before,
this operator can be simply cast as
\begin{equation}
\tilde{H}_{\eta}^{(k)}(t)\equiv i\hbar\sum_{j=1}^{d^{(k)}}\frac{d}{dt}|\varphi_{\eta j}^{(k)}(t)\rangle\langle\varphi_{\eta j}^{(k)}(t)|.\label{Estado misto - local effective Hamiltonian}
\end{equation}
Thus, essentially, we have a collection of local operators $\{\tilde{H}_{\eta}^{(k)}(t)\}_{\eta}$
associated with the subensemble of local states $\{\hat{\sigma}_{\eta}^{(k)}(t)\}_{\eta}$.
Interestingly, from Eq. (\ref{Unitary relation}), one can also show
that all possible local effective Hamiltonians can be directly related
as follows
\begin{equation}
\tilde{H}_{\eta}^{(k)}(t)=\hat{T}_{\eta\alpha}(t)\tilde{H}_{\alpha}^{(k)}(t)\hat{T}_{\eta\alpha}^{(k)\dagger}(t)-i\hbar\hat{T}_{\eta\alpha}(t)\frac{d}{dt}\hat{T}_{\eta\alpha}^{(k)\dagger}(t).
\end{equation}

Also, for every $\eta$, one can break down Eq. (\ref{Estado misto - local effective Hamiltonian})
into the contributions of the local bare Hamiltonian $\hat{H}^{(k)}$
and the by-products of the interaction term $\hat{H}_{int}$. Using
the spectral decomposition from Eq. (\ref{local bare hamiltonian})
and defining the projection $\langle b_{j}^{(k)}|\varphi_{\eta l}^{(k)}(t)\rangle:=r_{j\eta l}^{(k)}(t)e^{-\frac{i}{\hbar}b_{j}^{(k)}t}$,
one can rewrite the local effective Hamiltonians as
\begin{equation}
\tilde{H}_{\eta}^{(k)}(t)=\hat{H}^{(k)}+\hat{H}_{LS;\eta}^{(k)}(t)+\hat{H}_{X;\eta}^{(k)}(t),\label{Estado misto hamiltoniano efetivo}
\end{equation}
where
\begin{align}
\hat{H}_{LS;\eta}^{(k)}(t) & :=i\hbar\sum_{j=1}^{d^{(k)}}\left(\sum_{l=1}^{d^{(k)}}\frac{d}{dt}r_{j\eta l}^{(k)}(t)r_{j\eta l}^{(k)*}(t)\right)|b_{j}^{(k)}\rangle\langle b_{j}^{(k)}|,\\
\hat{H}_{X;\eta}^{(k)}(t) & :=i\hbar\sum_{j=1}^{d^{(k)}}\sum_{m\neq j}^{d^{(k)}}\left(\sum_{l=1}^{d^{(k)}}\frac{d}{dt}r_{m\eta l}^{(k)}(t)r_{j\eta l}^{(k)*}(t)\right)e^{\frac{i}{\hbar}\left(b_{j}^{(k)}-b_{m}^{(k)}\right)t}|b_{m}^{(k)}\rangle\langle b_{j}^{(k)}|.
\end{align}
Obviously, these expressions above have the same structure as the
ones previously discussed. However, it is worth mentioning again that
the local Hamiltonian time-dependency is induced by $\hat{H}_{LS;\eta}^{(k)}(t)$
and $\hat{H}_{X;\eta}^{(k)}(t)$, where the former is a general Lamb-shift-like
term such that $[\hat{H}_{LS:\eta}^{(k)}(t),\hat{H}^{(k)}]=0$ for
all $t$ and $\eta$, and the latter contain only non-diagonal elements
in the bare Hamiltonian basis $\{|b_{j}^{(k)}\rangle\}_{j}$.

\subsection{Local states dynamics}

Although mixed, the whole bipartite system is still assumed to be
isolated. Thus, its dynamic is unitary and expressed by the usual
Liouville-von Neumann equation below
\begin{equation}
i\hbar\frac{d}{dt}\hat{\rho}^{(0)}(t)=[\hat{H}^{(0)},\hat{\rho}^{(0)}(t)],
\end{equation}
which also implies - as already mentioned - that every pure state
from the ensemble evolves unitarily (Eq. (\ref{Mixe universe - pure state time evolution}))
and $\frac{d}{dt}P_{\eta}=0$ for all $\eta$. Locally, the subsystem's
dynamics is not unitary and can be simply obtained by partial tracing
the equation above, such that
\begin{equation}
i\hbar\frac{d}{dt}\hat{\rho}^{(k)}(t)=[\hat{H}^{(k)},\hat{\rho}^{(k)}(t)]+tr_{\bar{k}}\{[\hat{H}_{int},\hat{\rho}^{(0)}(t)]\}.
\end{equation}
where it clearly depends on unitary and non-unitary parts. Nevertheless,
given Eqs. (\ref{MATRIZ DENSIDADE GLOBAL: ESTADO MISTO}, \ref{Entire local state 1},
\ref{Entire local state 2}) we might write similar expressions for
each element of the local subensembles $\{\hat{\sigma}_{\eta}^{(k)}(t)\}_{\eta}$,
i.e.,
\begin{equation}
i\hbar\frac{d}{dt}\hat{\sigma}_{\eta}^{(k)}(t)=[\hat{H}^{(k)},\hat{\sigma}_{\eta}^{(k)}(t)]+tr_{\bar{k}}\{[\hat{H}_{int},\hat{\sigma}_{\eta}^{(0)}(t)]\},\label{Subensemble master equation 1}
\end{equation}
which is similar to Eq. (\ref{Master Eq. 1}). Instead, the expression
above might be analogously written as Eq. (\ref{Master Eq. 2}), such
that
\begin{equation}
i\hbar\frac{d}{dt}\hat{\sigma}_{\eta}^{(k)}(t)=[\tilde{H}_{\eta}^{(k)}(t),\hat{\sigma}_{\eta}^{(k)}(t)]+i\hbar\sum_{j=1}^{d^{(1)}}\frac{d}{dt}\lambda_{\eta j}^{2}(t)|\varphi_{\eta j}^{(k)}(t)\rangle\langle\varphi_{\eta j}^{(k)}(t)|,\label{Subensemble master equation 2}
\end{equation}
where the commutator is the unitary part, in terms of the effective
Hamiltonian - instead of the bare one - and the remaining element
is the non-unitary contribution represented by the population changes.
Along these lines, now, there is a set of parametric curves $\{\mathcal{C}_{\eta}^{(k)}:\hat{\sigma}_{\eta}^{(k)}(t),\:t\in[t_{0},t_{1}]\}_{\eta}$
followed by the subsystem $(k)$ in its respective density operator
space $\mathscr{D}(\mathcal{H}^{(k)})$, where the unitary contribution
is due to the generators of the Schmidt basis paths $\{\mathcal{P}_{\eta j}^{(k)}\}_{\eta,j}$,
and the non-unitary factor is given by the population's trajectories
$\{\mathcal{P}_{\eta j}^{(\lambda)}\}_{\eta,j}$.

Finally, since both Eq. (\ref{Subensemble master equation 1}) and
Eq. (\ref{Subensemble master equation 2}) describe the same dynamics,
it is clear that they must be directly associated. Hence, given Eq.
(\ref{Estado misto hamiltoniano efetivo}), both unitary contributions
satisfy the following equality
\begin{equation}
[\tilde{H}_{\eta}^{(k)}(t),\hat{\sigma}_{\eta}^{(k)}(t)]=[\hat{H}^{(k)},\hat{\sigma}_{\eta}^{(k)}(t)]+[\hat{H}_{LS;\eta}^{(k)}(t),\hat{\sigma}_{\eta}^{(k)}(t)]+[\hat{H}_{X;\eta}^{(k)}(t),\hat{\sigma}_{\eta}^{(k)}(t)]
\end{equation}
and, therefore,
\begin{equation}
tr_{\bar{k}}\{[\hat{H}_{int},\hat{\sigma}_{\eta}^{(0)}(t)]\}=i\hbar\sum_{j=1}^{d^{(1)}}\frac{d}{dt}\lambda_{\eta j}^{2}(t)|\varphi_{\eta j}^{(k)}(t)\rangle\langle\varphi_{\eta j}^{(k)}(t)|+[\hat{H}_{LS;\eta}^{(k)}(t)+\hat{H}_{X;\eta}^{(k)}(t),\hat{\sigma}_{\eta}^{(k)}(t)],
\end{equation}
for all $\eta$.

\subsection{Local effective internal energy}

Let us now characterize the energetics within this mixed bipartite
quantum system. The whole system is isolated by hypothesis, thus it
does not interact with any other element, classical or quantum. From
an energetic point of view, it means the energy must be conserved
inside the bipartition. Along these lines, the role of total internal
energy is naturally - and unambiguously - attributed to the expectation value of the Hamiltonian $\hat{H}^{(0)}$, i.e.,
\begin{equation}
U^{(0)}\equiv\langle\hat{H}^{(0)}\rangle=tr\{\hat{H}^{(0)}\hat{\rho}^{(0)}(t)\}.
\end{equation}
As expected, the unitary time evolution of $\hat{\rho}^{(0)}(t)$
automatically guarantees that this internal energy is a conserved
quantity, such that
\begin{equation}
\frac{d}{dt}U^{(0)}=0.
\end{equation}
Additionally, given that the whole state is mixed, as shown in Eq.
(\ref{MATRIZ DENSIDADE GLOBAL: ESTADO MISTO}), one can easily cast
the expectation value above as the following average over the ensemble
of pure states
\begin{equation}
U^{(0)}=\sum_{\eta=1}^{d^{(0)}}P_{\eta}U_{\eta}^{(0)},\label{whole state Internal energy, mixed state}
\end{equation}
where
\begin{equation}
U_{\eta}^{(0)}\equiv tr\{\hat{H}^{(0)}\hat{\sigma}_{\eta}^{(0)}(t)\}=\langle\Psi_{\eta}(t)|\hat{H}^{(0)}|\Psi_{\eta}(t)\rangle\label{pure state Internal energy, mixed state}
\end{equation}
is immediately recognized as the whole's internal energy relative
to the $\eta$th individual state $\hat{\sigma}_{\eta}^{(0)}(t)$,
which also satisfy 
\begin{equation}
\frac{d}{dt}U_{\eta}^{(0)}=0\label{eq: mixed state, energy conservation}
\end{equation}
for all $\eta$.

Nevertheless, we are interested to understand how exactly the internal
energy is distributed between the bipartition. In this sense, it is
clear that we can refer to and analyse the whole's internal energy
both from the ensemble ($U^{(0)}$) and pure state ($U_{\eta}^{(0)}$)
perspectives:

\subsubsection{Pure state level}

As discussed earlier, the direct use of the local bare Hamiltonians
$\hat{H}^{(1,2)}$ as the proper operators for this task is not satisfactory
for general scenarios. Thus, following the exact same procedure from
Subsection (\ref{subsec:Local-effective-internal}), we will recognize
the local effective Hamiltonians, $\tilde{H}_{\eta}^{(1,2)}(t)$,
as the representative local operators for characterizing the physical
internal energy of each possible pure state of the whole. Along these
lines, since $\sum_{j=1}^{d^{(1)}}\lambda_{\eta j}^{2}(t)=1$ and
$\sum_{j=1}^{d^{(1)}}\lambda_{\eta j}(t)\frac{d}{dt}\lambda_{\eta j}(t)=\frac{1}{2}\frac{d}{dt}\sum_{j=1}^{d^{(1)}}\lambda_{\eta j}^{2}(t)$,
we can show that
\begin{equation}
\langle\Psi_{\eta}(t)|\hat{H}^{(0)}|\Psi_{\eta}(t)\rangle=\langle\Psi_{\eta}(t)|\tilde{H}_{\eta}^{(1)}(t)|\Psi_{\eta}(t)\rangle+\langle\Psi_{\eta}(t)|\tilde{H}_{\eta}^{(2)}(t)|\Psi_{\eta}(t)\rangle\label{pure state level, total hamiltonian}
\end{equation}
for every $\eta$, where $\langle\Psi_{\eta}(t)|\tilde{H}_{\eta}^{(1,2)}(t)|\Psi_{\eta}(t)\rangle=tr\{\tilde{H}_{\eta}^{(1,2)}(t)\hat{\sigma}_{\eta}^{(0)}(t)\}=tr_{1,2}\{\tilde{H}_{\eta}^{(1,2)}(t)\hat{\sigma}_{\eta}^{(1,2)}(t)\}$
is the expectation value of the local effective Hamiltonian concerning
the $\eta$th pure state. Hence, as long we identify 
\begin{equation}
U_{\eta}^{(1,2)}(t)\equiv\langle\Psi_{\eta}(t)|\tilde{H}_{\eta}^{(1,2)}(t)|\Psi_{\eta}(t)\rangle\label{local internal energy, mixed state}
\end{equation}
as the proper local effective internal energy of subsystem $(1,2)$,
every possible element of the mixed ensemble $\hat{\rho}^{(0)}(t)$
will individually satisfy the energy additivity, such that 
\begin{equation}
U_{\eta}^{(0)}=U_{\eta}^{(1)}(t)+U_{\eta}^{(2)}(t).\label{pure state, Energy additivity mixe states}
\end{equation}
Besides, given the energy conservation property (Eq. (\ref{eq: mixed state, energy conservation})),
the equation above also implies that the energy flowing from one subsystem
is entirely obtained by the other, i.e., 
\begin{equation}
\frac{d}{dt}U_{\eta}^{(1)}(t)=-\frac{d}{dt}U_{\eta}^{(2)}(t).
\end{equation}

Also, since $\hat{H}^{(0)}=\hat{H}^{(1)}\otimes\hat{1}^{(2)}+\hat{1}^{(1)}\otimes\hat{H}^{(2)}+\hat{H}_{int}$
and Eq. (\ref{pure state level, total hamiltonian}), it is straightforward
to show that
\begin{align}
\langle\Psi_{\eta}(t)|\hat{H}_{int}|\Psi_{\eta}(t)\rangle & =\sum_{k=1,2}\langle\Psi_{\eta}(t)|\left(\hat{H}_{LS;\eta}^{(k)}(t)+\hat{H}_{X;\eta}^{(k)}(t)\right)|\Psi_{\eta}(t)\rangle.\label{pure ensemble level, interaction term}
\end{align}

\subsubsection{Ensemble level}

However, from the total ensemble perspective, the whole's internal
energy is characterized by Eq. (\ref{whole state Internal energy, mixed state}).
Thus, along with Eq. (\ref{pure state, Energy additivity mixe states}),
we obtain
\begin{equation}
U^{(0)}=U^{(1)}(t)+U^{(2)}(t),
\end{equation}
where the entire local internal energies $U^{(1,2)}(t)$ are simply
identified as the averages of its possible values
\begin{equation}
U^{(1,2)}(t)\equiv\sum_{\eta=1}^{d^{(0)}}P_{\eta}U_{\eta}^{(1,2)}(t),\label{entire local internal energies, mixed state}
\end{equation}
which is clearly an additive property, such that
\begin{equation}
\frac{d}{dt}U^{(1)}(t)=-\frac{d}{dt}U^{(2)}(t).
\end{equation}
As expected, these quantities automatically contain both the bare
and interaction contributions, i.e.,
\begin{equation}
U^{(k)}(t)=\langle\hat{H}^{(k)}\rangle(t)+\sum_{\eta=1}^{d^{(0)}}P_{\eta}tr_{k}\{(\hat{H}_{LS;\eta}^{(k)}(t)+\hat{H}_{X;\eta}^{(k)}(t))\hat{\sigma}_{\eta}^{(k)}(t)\}
\end{equation}
where $\langle\hat{H}^{(k)}\rangle(t)=tr_{k}\{\hat{H}^{(k)}\hat{\rho}^{(k)}(t)\}$.
Additionally, one can generalize Eq. (\ref{pure ensemble level, interaction term})
and associate the expectation value of the interaction term, $\langle\hat{H}_{int}\rangle(t)\equiv tr\{\hat{H}_{int}\hat{\rho}^{(0)}(t)\}$,
with the local operators $\hat{H}_{LS;\eta}^{(1,2)}(t)$ and $\hat{H}_{X;\eta}^{(1,2)}(t)$,
such that
\begin{equation}
\langle\hat{H}_{int}\rangle(t)=\sum_{\eta=1}^{d^{(0)}}P_{\eta}\sum_{k=1,2}\langle\Psi_{\eta}(t)|\left(\hat{H}_{LS;\eta}^{(k)}(t)+\hat{H}_{X;\eta}^{(k)}(t)\right)|\Psi_{\eta}(t)\rangle.
\end{equation}

Notice that the entire local internal energies $U^{(k)}(t)$ showed
in Eq. (\ref{entire local internal energies, mixed state}) are fundamentally
different from the ones presented in Eq. (\ref{local internal energy, mixed state})
associated with each pure state, $U_{\eta}^{(k)}(t)$. While the latter
is given by the mean values of local observables, the former is
represented by the averages of these quantities over the whole system's
ensemble. Thus, even though the local effective Hamiltonians $\tilde{H}_{\eta}^{(k)}(t)$
represent suitable operators for individually characterizing the energetics
within possible elements of $\{\hat{\sigma}_{\eta}^{(0)}(t)\}_{\eta}$,
we did not yet present appropriate local operators for the entire
subsystem's internal energy. Along these lines, it is desired to define
the operators $\tilde{H}^{(k)}(t)$, such that
\begin{equation}
\langle\tilde{H}^{(k)}(t)\rangle=tr_{k}\{\tilde{H}^{(k)}(t)\hat{\rho}^{(k)}(t)\}\equiv\sum_{\eta=1}^{d^{(0)}}P_{\eta}tr_{k}\{\tilde{H}_{\eta}^{(k)}(t)\hat{\sigma}_{\eta}^{(k)}(t)\},
\end{equation}
or $\sum_{\eta=1}^{d^{(0)}}P_{\eta}tr_{k}\{(\tilde{H}^{(k)}(t)-\tilde{H}_{\eta}^{(k)}(t))\hat{\sigma}_{\eta}^{(k)}(t)\}=0$.
The simplest way to achieve this is to guarantee 
\begin{equation}
\langle\varphi_{\eta j}^{(k)}(t)|\tilde{H}^{(k)}(t)|\varphi_{\eta j}^{(k)}(t)\rangle=\langle\varphi_{\eta j}^{(k)}(t)|\tilde{H}_{\eta}^{(k)}(t)|\varphi_{\eta j}^{(k)}(t)\rangle,
\end{equation}
for all $\eta$ and $j$. As long the equality above is satisfied,
the operators $\tilde{H}^{(1,2)}(t)$ are qualified to be the representative
observables for quantifying the entire local internal energy, such
that
\begin{equation}
U^{(k)}(t)=\langle\tilde{H}^{(k)}(t)\rangle=tr_{k}\{\tilde{H}^{(k)}(t)\hat{\rho}^{(k)}(t)\},
\end{equation}
and $U^{(0)}=U^{(1)}(t)+U^{(2)}(t)$.

In short, by identifying the local effective Hamiltonians as the operators
that characterize the physical local internal energies, we were able
to consistently generalize the expressions presented in Section (\ref{subsec:Internal-energy-and})
and describe the energetics of general bipartite mixed quantum systems,
both at the individual pure state and ensemble levels. As required,
these energies are additive and locally accessible. Finally, it is
worth highlighting that this procedure is exact and applicable to
any setting and regime.

\subsubsection{Brief remarks}

In Section (\ref{sec:Local-phase-gauge}) it was presented the phase
ambiguity inbuilt in the Schmidt decomposition formalism and explained
how this gauge freedom influences our calculations and the local effective
description. Clearly, it would also play a role in the previous expressions.
Nevertheless, considering the discussion shown in Section (\ref{Recap and gauge}), we are implicitly assuming the use of the gauges that maintain physical consistency for all the local effective Hamiltonians. 

Besides, as mentioned earlier, the identification of appropriate candidates
for internal energy is a requirement for defining further thermodynamic
quantities. Along these lines, the discussion presented in Section
(\ref{subsec:Road-to-effective}) can be entirely imported to the
present context of mixed states. Even though we do not introduce work
and heat definitions, it provides a consistent framework for beginning
a discussion concerning general, exact and symmetrical understanding
of work and heat along with any dynamical processes.

%% file: Chapters/Chapter3/Section9.tex
\section{Discussion and summary\label{sec:Discussion-and-summary}}

In this chapter, it was introduced a novel formalism for describing
the energetics within isolated bipartite quantum systems. The formal
procedure is based on the well-known Schmidt decomposition and provides
a promising route for properly defining effective Hamiltonians and
characterizing the subsystem's internal energies in a symmetrical
fashion. In contrast with current methodology, such a framework is
exact and do not rely on any sort of approximations and additional
hypotheses, such as particular coupling regimes, convenient Hamiltonian
structures and specific type of dynamics. Surprisingly, despite such
generality, this description allows the definition of local properties,
i.e., quantities accessible by local observations, that recovers the
usual thermodynamic notion of energy additivity. Besides, these expressions
also establish a new route for defining further general thermodynamic
quantities to the quantum regime, which is also imperative for the
design and development of functional quantum devices.

Additionally, from a conceptual perspective, the most common procedures found in the literature of QT are based on semi-classical approaches that are not entirely suitable for a general thermodynamic description of fully autonomous quantum objects. The implicit assumption of a classical agent to externally control and measure the system of interest - and potentially even process this information - restricts QT to this particular semi-classical picture of coherent control where both the quantum nature of the control fields is unimportant to the system's dynamics and the system itself does not change the control's state. Thus, any QT framework that requires an external agent to perform any task, such as driving or measuring the system, is fundamentally limited and phenomenological in spirit. The quantization of this entity is an essential step toward the generalization of QT and the design of autonomous quantum devices. Along these lines, our formalism does not suffer from this shortcoming and, therefore, contributes to further understanding of foundational aspects concerning the development of a fully quantum thermodynamic theory.

The present work focused on the theoretical aspects and viability of the procedure introduced in this chapter. Despite its mathematical consistency and compelling features, a more rigorous analysis on experimental grounds is necessary to support the use of our proposal. In this sense, a detailed examination of the potential physical observations and experiments that one might perform to investigate this framework is out of the scope of this thesis. However, it is of the author's opinion that such deliberation is crucial for building a robust and meaningful physical theory. Thus, let us briefly remark on this topic. Essentially, we associate
the local effective Hamiltonian, $\tilde{H}^{(k)}(t)\equiv\sum_{j=1}^{d^{(k)}}\tilde{\epsilon}_{j}^{(k)}(t)|\tilde{\epsilon}_{j}^{(k)}(t)\rangle\langle\tilde{\epsilon}_{j}^{(k)}(t)|$,
as the observable for characterizing the physical local internal energy,
$U^{(k)}(t)$, of subsystem $(k)$. These operators, by construction, depend on the dynamics of the local Schmidt basis and - a priori - can be experimentally reconstructed once known the time evolution of the local density matrices. Observe that the emergent energetic
time dependency clearly implies that the energy spectra $\{\tilde{\epsilon}_{j}^{(k)}(t)\}_{j}$
is modified due to the interactions between the subsystems. Notice,
however, that this does not happen if the physical internal energy
is associated with the bare Hamiltonians $\hat{H}^{(k)}$: clearly,
the expectation values might be changing in time, but that would be a
consequence of the local state dynamics since the Hamiltonians are
constant for autonomous systems, i.e., $\langle\hat{H}^{(k)}\rangle(t)=tr\{\hat{H}^{(k)}\hat{\rho}^{(k)}(t)\}$.
In other words, as long as the subsystems are interacting, their eigenenergies
are affected by their interaction $\hat{H}_{int}$. Suppose, for instance,
a qubit interacting with another object\footnote{In principle, it could be as simple as another two-level system, a
single photon, or a complicated body.}: on the one hand, the former's bare Hamiltonian could be written
as $\hat{H}^{(1)}=\frac{1}{2}\hbar\omega_{0}\hat{\sigma}_{z}^{(1)}$,
where $\hat{\sigma}_{z}^{(1)}$ is the usual Pauli matrix, and $\hbar\omega_{0}$
is the gap between the energy levels; on the other, the local effective
Hamiltonian will have the following general structure $\tilde{H}^{(1)}(t)\equiv\frac{1}{2}\hbar\omega_{eff}(t)\tilde{\sigma}_{z}^{(1)}(t)$,
where $\tilde{\sigma}_{z}^{(1)}(t)$ is the Pauli matrix relative to
the time-dependent effective eigenbasis and $\hbar\omega_{eff}(t)$ is the time-dependent
effective gap induced by the interaction. Notice that, while in the
former, the eigenbasis and energy spectrum are static, in the latter, they are dynamic.
Along these lines, as we know, the gap structure dictates the energy
frequency absorption and emission, and - in principle - this corresponds
to a realistic scenario that could be probed: while the qubit's effective
Hamiltonian is changing in time due to the interaction, its absorption
frequency $\omega_{eff}(t)$ is also being modulated by subsystem
$(2)$, in such a way that its interaction capability with a third
party is affected. Similar situations might be constructed considering
qutrits and their energy level structures. Essentially, if such a
spectrum modulation is observed, it would be in accordance with this
work's proposal. Also, it is worth mentioning that this kind of setting and test might be experimentally feasible in optical and solid-state setups and NMR, for instance.

In short, we provided a useful and novel framework for characterizing
the energetics within interacting quantum systems. Hopefully, it became
clear both the importance of this kind of task and how they are not
easy or trivial questions, especially because several subtleties should
be considered once more foundational aspects are being raised. Finally,
it is - optimistically - expected that these results will also have
the potential to motivate the flourishing of new definitions of quantum
work and heat, along with the refinement of the understanding of the
laws of thermodynamics in the quantum realm.

%% file: Chapters/Chapter4/Chapter4.tex
\chapter{Conclusion and outlook\label{part:Chapter-4--}}

Quantum thermodynamics is an exciting and promising research field
that will - certainly - play a pivotal role in the design and development
of future quantum-based technologies. In addition to the purely practical-driven
interests, fundamental theoretical aspects lie at the heart of the
intersection of quantum mechanics and thermodynamics. As mentioned
earlier, despite many efforts aiming to extend the well-known laws
of thermodynamics to the microscopic realm of non-equilibrium and
quantum processes, there is still no unifying and general picture
for the theory. Along these lines, some fundamental questions remain
unanswered. Besides, most current proposals rely on regimes and settings
that, although familiar to the macroscopic description, are restrictive
to our more ambitious purposes of characterizing the thermodynamics
within arbitrary autonomous quantum systems. The present work belongs
to this general context.

In this thesis, we focus on the energetic analysis within isolated
bipartite quantum systems. More specifically, we propose a novel and
general formalism for a dynamic description of the energy exchanges
between interacting subsystems. To this aim, instead of using the
bare Hamiltonians, we introduce a new effective operator as being
the representative element for characterizing the local dynamics and
internal energy, i.e., the Schmidt decomposition approach allows the
identification of effective Hamiltonians whose expectation values satisfy
the desired properties of appropriate definitions of internal energies,
namely being local and additive quantities. Such proposal is independent
of the Hamiltonian structures (including the interaction term), coupling
strengths and other regular constraints, which establishes a promising
route for the thermodynamic analysis of general autonomous quantum
dynamics\footnote{The discussions and results introduced in this thesis were presented in \cite{Malavazi} after its defense.}.

The definition of quantum counterparts of classical thermodynamic
variables is one of the core conceptual issues of the field. The identification
of quantum thermodynamic entropy, internal energy, work and heat,
along with their relationships, is crucial from a foundational point
of view and of extreme practical relevance for designing and operating
functional quantum devices. Along these lines, our proposal opens
up many possibilities for future investigations. A consistent definition
of local internal energy corresponds to the first and fundamental
step toward the definition of other relevant quantities, especially
those directly derived from the energy flow, like work and heat. Consequently,
it also provides the means for helping to establish general quantum
versions of the first and second laws.

Future research will aim at these topics. However, more importantly,
it will also focus on suggesting realistic experimental designs for
assessing these quantities and investigating our proposal. As discussed
earlier, the local effective internal energies are, a priori, experimentally
accessible properties. Still, further research is necessary for identifying
an appropriate physical setup and the corresponding parameters.

In summary, quantum thermodynamics is a young discipline, and its
development is still a work in progress. On the one hand, it implies
that it is a fertile field to explore; On the other, it also means
there is no solid and cohesive foundation yet. In this thesis, we
identify a consistent candidate for quantifying internal energy and
provide a simple framework suitable for the energetic analysis of
autonomous quantum systems. The proposed formalism does not assume
any approximations or restrictive hypotheses, treat the bipartitions
on equal footing and is completely general.